\documentclass{article}
\usepackage{arxiv}
\newcommand{\backmatter}{}

\usepackage{graphicx}%
\usepackage{multirow}%
\usepackage{amsmath,amssymb,amsfonts}%
\usepackage{amsthm}%
\usepackage{mathrsfs}%
\usepackage{mathtools}

\usepackage{tabularx}
\usepackage{makecell}
\usepackage{rotating}
\usepackage[table]{xcolor}
\definecolor{tablealtgray}{gray}{0.9}
\newcommand{\tablealtrows}{\rowcolors{2}{white}{tablealtgray}}

\usepackage{tikz}
\usetikzlibrary{positioning,decorations.pathreplacing,arrows.meta}

\usepackage{booktabs}%
\newcommand{\botrule}{\bottomrule}
\usepackage{listings}%

\usepackage{orcidlink}
\usepackage{comment}
\usepackage[normalem]{ulem} 

\usepackage{comment}

\usepackage{ccicons} 
\usepackage[%
style=base,
singlelinecheck=false,%
format=plain,%
labelformat=simple,%
labelsep=period,%
justification=justified,%
font=none,small,%
labelfont=bf,%
textfont=none%
]{caption}
\usepackage{subcaption}
\usepackage[section]{placeins}

\usepackage{cleveref}
\crefname{figure}{figure}{figures}
\Crefname{figure}{Figure}{Figures}
\crefname{table}{table}{tables}
\Crefname{table}{Table}{Tables}
\usepackage{hyperref}

\usepackage{siunitx}
\AtBeginDocument{\RenewCommandCopy\qty\SI}
\ExplSyntaxOn
\msg_redirect_name:nnn { siunitx } { physics-pkg } { none }
\ExplSyntaxOff
\DeclareSIUnit{\Gb}{\giga\bit}
\DeclareSIUnit{\fJ}{\femto\joule}
\DeclareSIUnit{\sample}{S}
\DeclareSIUnit{\GS}{\giga\sample}
\DeclareSIUnit{\Gbps}{\giga\bit\per\second}
\DeclareSIUnit{\yr}{yr}

\usepackage{physics}
\usepackage[version=4]{mhchem}

\usepackage[english]{babel}
\usepackage{csquotes}
\usepackage[super]{nth}

\usepackage[style=numeric, sorting=none]{biblatex}
\title{Memory in Integrated Photonic Neural Networks: From Physical Mechanisms to Neuromorphic Architectures}

\hypersetup{
pdftitle={Memory in Integrated Photonic Neural Networks: From Physical Mechanisms to Neuromorphic Architectures},
pdfsubject={physics.optics},
pdfauthor={Alessandro Foradori, Ilya Auslender, Stefano Biasi, Stefano Gretter, Alessio Lugnan, Emiliano Staffoli, Lorenzo Pavesi},
pdfkeywords={Neuromorphic Photonics, Optical Memory, Silicon Photonics},
}

\usepackage{authblk}

\newcommand{\fnm}[1]{#1}
\newcommand{\sur}[1]{#1}
\newcommand{\email}[1]{}

\author[1,2]{\orcidlink{0009-0001-3195-8620} \fnm{Alessandro} \sur{Foradori}}\email{alessandro.foradori@unitn.it}

\author[1]{\orcidlink{0009-0000-9519-0879} \fnm{Ilya} \sur{Auslender}}\email{ilya.auslender@unitn.it}

\author[1]{\orcidlink{0000-0002-3361-133X} \fnm{Stefano} \sur{Biasi}}\email{stefano.biasi@unitn.it}

\author[1]{\orcidlink{0009-0009-4747-1057} \fnm{Stefano} \sur{Gretter}}\email{stefano.gretter-1@unitn.it}

\author[1]{\orcidlink{0000-0002-6587-2614} \fnm{Alessio} \sur{Lugnan}}\email{alessio.lugnan.1@unitn.it}

\author[1]{\orcidlink{0000-0002-4167-9158} \fnm{Emiliano} \sur{Staffoli}}\email{emiliano.staffoli@unitn.it}

\author[1]{\orcidlink{0000-0001-7316-6034} \fnm{Lorenzo} \sur{Pavesi}}\email{lorenzo.pavesi@unitn.it}

\newcommand{\orgdiv}[1]{#1}
\newcommand{\orgname}[1]{#1}
\newcommand{\orgaddress}[1]{#1}
\newcommand{\street}[1]{#1}
\newcommand{\city}[1]{#1}
\newcommand{\postcode}[1]{#1}
\newcommand{\country}[1]{#1}

\affil[1]{\orgdiv{Nanoscience Laboratory, Department of Physics}, \orgname{University of Trento}, \orgaddress{\street{Via Sommarive 14}, \city{Trento}, \postcode{38123}, \country{Italy}}}

\affil[2]{\orgdiv{Photonics Research Group, Department of Information Technology}, \orgname{Ghent University-imec}, \orgaddress{\street{Technologiepark Zwijnaarde 15}, \city{Ghent}, \postcode{9052}, \country{Belgium}}}

\renewcommand{\headeright}{}
\renewcommand{\undertitle}{}

\begin{document}

\maketitle


\begin{abstract}
The rapid scaling of artificial neural networks has exposed fundamental limitations of conventional von Neumann computing architectures. In these systems, the physical separation between memory and processing creates a bottleneck, as computational capabilities outpace the ability of memory and interconnects to supply and retrieve data. In contrast, biological neural systems inherently co-localize computation and memory through distributed, dynamical processes. Neuromorphic computing seeks to emulate this paradigm by leveraging physical substrates whose intrinsic dynamics simultaneously encode and process information. Among emerging platforms, silicon photonics offers a compelling approach due to its high bandwidth, low-loss propagation, and inherent parallelism.

This review examines the role of memory in integrated photonic neuromorphic systems, with emphasis on the physical mechanisms that provide volatile and non-volatile memory mechanisms, while distinguishing their retention properties from their temporal behaviour and computational role as state or parameter memory in silicon-on-insulator and hybrid silicon-on-insulator platforms. Drawing inspiration from digital, biological, and photonic memory architectures, we classify existing approaches based on their underlying physical principles. We cover implementations ranging from delay lines and slow-light structures to multistable dynamics and structural memory based on charge trapping and phase-change materials.  We then discuss how these mechanisms support photonic neural network architectures, including feed-forward, reservoir computing, spiking and hybrid optoelectronic recurrent systems, and assess their relevance for time-dependent signal-processing tasks such as channel equalization in telecommunications. This review aims to establish a unified framework for understanding memory and learning in neuromorphic photonics and outlines key challenges and opportunities for scalable, energy-efficient neuromorphic hardware.
\end{abstract}

\keywords{Neuromorphic Photonics, Optical Memory, Silicon Photonics}

\section{Introduction}

The last decade has witnessed remarkable progress in artificial intelligence (AI), driven largely by advances in deep learning\cite{lecun2015deep}.
Artificial neural networks (ANNs) now underpin state-of-the-art systems for natural language processing, artificial perception, and data analysis \cite{mienye2024comprehensive}.
While the algorithmic foundations of ANNs are well established, the rapid growth in model size and complexity has highlighted fundamental limitations of conventional computing hardware.
Moore’s law and Dennard scaling sustained improvements in computational density for several decades\cite{moore1965cramming,dennard2003design}.
Today, these trends have slowed as devices approach thermal and quantum limits in CMOS (complementary metal-oxide-semiconductor) technology\cite{esmaeilzadeh2011dark}.

A key difficulty is architectural.
In von Neumann systems, memory and processing units are physically separated.
Thus, modern AI models must continuously move large volumes of weights and activations between them.
In advanced CMOS technologies, the energy associated with this data transport often exceeds the energy of the arithmetic operations themselves\cite{horowitz20141}.
Reading from memory typically requires tens to hundreds of femtojoules per access, substantially more than the cost of a multiply–accumulate operation on that data.
As neural networks deepen and incorporate mechanisms such as recurrence or attention\cite{vaswani2017attention}, this memory traffic becomes the primary performance bottleneck.
Recent analyses show that server-class compute throughput has increased by a factor of roughly $3\times$ every two years, whereas DRAM (dynamic random-access memory or dynamic RAM) and interconnect bandwidth have grown only by $\sim 1.6\times$ and $\sim 1.4\times$ over the same interval, respectively\cite{gholami2024ai}.
This widening gap, referred to as the AI \textquote{memory wall}, establishes memory access as the dominant performance limiter of large-scale AI systems, surpassing arithmetic throughput as the primary bottleneck.

Biological neural systems operate in a different regime. Neurons do not fetch variables from a distinct memory unit.
Instead, their electrical and biochemical states integrate synaptic inputs over time.
Synapses store information through long-lasting molecular modifications, producing memory traces that span from milliseconds to hours and longer \cite{kandel2000principles,tonegawa2015memory}.
At the circuit level, memories are believed to reside in ensembles of neurons and synapses known as cell assemblies or engrams, whose selective reactivation underlies recall\cite{tonegawa2015memory}.
Because storage and computation are co-located in the same substrate, network activity simultaneously represents ongoing processing and embedded memory.
This contrasts with the rigid separation characteristic of digital hardware and suggests that novel architectures may be required to overcome the von Neumann bottleneck\cite{mead2002neuromorphic,indiveri2015memory}.

Neuromorphic computing (brain-inspired computation) embraces this principle by employing physical systems whose internal states evolve continuously and store information through intrinsic material responses\cite{kudithipudi2025neuromorphic}.
Among the various candidate platforms—including electronic, spintronic, and memristive devices—integrated photonics has emerged \cite{brunner2025roadmap}.
Optical signals propagate with low loss, interfere linearly at high bandwidth, and couple to materials through nonlinear and dynamical effects.
These properties enable photonic circuits to perform computation and encode temporal information in ways not easily achievable with electronics.
In this review, we examine how integrated photonics can implement neural-network computation with both \emph{non-volatile memory}  and \emph{volatile memory}.
We focus on architectures such as MRR-based excitable networks, integrated photonics reservoir computing systems, and recurrent optical processors, and discuss how their physical dynamics support computation and learning.

By memory we denote a physical state or signal path through which the present response of a system depends on a previous input or configuration after the corresponding writing stimulus has ended and under specified holding conditions. We describe each memory mechanism using four independent descriptors (see \cref{tab:memory}).

\begin{table}[t]
	\centering
	\caption{Independent descriptors used to classify memory mechanisms in integrated photonic neural networks. A mechanism qualifies as memory when a previous input or configuration affects the present response after the writing stimulus has ended, under specified holding conditions. Physical retention, computational role, temporal behavior, and holding-power requirements are treated as separate properties.}
	\label{tab:memory}
\renewcommand{\arraystretch}{1.2}
\tablealtrows
\begin{tabular}{@{}p{0.22\textwidth} p{0.35\textwidth} p{0.35\textwidth}@{}}
		\toprule
		Descriptor            & Question addressed         & Categories or quantities  \\
		\midrule
        Memory criterion	& Does the previous input still affect the system after the writing stimulus has ended?	& Yes/no, under specified holding conditions \\
Physical retention	& Does the state survive removal of its state-holding power or bias,  ideally indefinitely? &	Volatile/non-volatile; measured retention interval\\
Computational role	&What does the stored quantity represent?	&State/parameter or configuration\\
Temporal behaviour	& How does the influence of the stored information evolve?	&Finite-window/ fading/ persistent/ oscillatory/ multistable\\
Holding condition	&What is required to maintain the state?&	Zero holding power/continuous optical pump/electrical bias/thermal control\\
		\botrule
	\end{tabular}
\end{table}

The first is \emph{physical retention}, including whether the stored state survives removal of the optical or electrical conditions required to maintain it; this defines the volatile/non-volatile distinction used in this review. The second is the \emph{computational role}, distinguishing state memory, which participates in the evolution of the system during inference, from parameter or configuration memory, which encodes quantities such as weights, biases, or connectivity. The third is the \emph{temporal behaviour}, which may be finite-window, fading, persistent, oscillatory, or multistable. The fourth is the \emph{holding condition}, specifying whether retention occurs without continuous power or requires a sustained optical pump, electrical bias, or thermal control. These descriptors are not interchangeable. A volatile SRAM cell may store a static neural-network parameter; a thermally tuned phase shifter implements parameter memory while requiring continuous holding power; and a powered bistable photonic state may be volatile but non-fading during operation. Conversely, a non-volatile device may retain its programmed state without holding power while still requiring energy for programming, addressing, readout, and detection.

\subsection{Symbolic vs. Subsymbolic Computation: From Logic to Dynamics}\label{sec:symbolic}

To highlight the requirements placed on neuromorphic hardware, it is useful to compare symbolic and subsymbolic paradigms of computation.

\emph{Symbolic AI} follows the classical model of computation based on explicit symbol manipulation\cite{Newell1976PhysicalSymbol,russell1995modern}.
It operates on discrete variables through deterministic rules. Memory is organized as addressable storage.
Computation proceeds through sequences of prescribed operations acting on these stored symbols.

\emph{Subsymbolic AI}, which includes ANNs, adopts a different strategy. Intelligence emerges from the collective interaction of many simple units.
Feed-forward networks implement sequences of learned transformations, whereas recurrent networks additionally evolve an internal state over time. From a physical perspective, Hopfield networks, attractor networks, and other energy-based systems can be interpreted as a high-dimensional dynamical system whose macroscopic behavior reflects these interactions \cite{hopfield1982neural}. Learning generally adjusts the parameters governing the network transformations or dynamics; in energy-based models, this operation can more specifically be viewed as reshaping the energy landscape so that desired states correspond to stable or metastable minima \cite{Rumelhart1986LearningIR,lecun2015deep}. Instead of executing a fixed sequence of logical operations, inference corresponds to the relaxation of system states in an energy landscape shaped by synaptic weights \cite{hopfield1982neural,Rumelhart1986LearningIR}.
Unlike symbolic architectures, where memory and processing are separated into distinct functional units, subsymbolic systems integrate these roles intrinsically: the same distributed state variables that encode past activity also determine the system’s ongoing dynamics and computation \cite{hopfield1982neural}.

\subsection{Neuromorphic Computing as a Physical Paradigm}

Neuromorphic computing implements these principles into diverse hardware platforms.
Information is represented in voltages, currents, optical fields, or material states, and past activity influences the present through internal variables with finite relaxation times (fading memory) or long-lasting physical effects (non-fading memory)\cite{mead2002neuromorphic,indiveri2015memory}.
This physical viewpoint has direct implications for scalability and efficiency.
A neuromorphic architecture naturally supports parallelism through its distributed state, minimizes data movement by co-locating memory and processing, and performs computation through the intrinsic dynamics of the substrate (the hardware).

Co-location arises because the same physical variables, such as charge states, atomic configurations, or electromagnetic fields, both store information and participate directly in computation.
To function effectively, such systems must provide locally evolvable state variables, continuous-time signal evolution, and memory effects with suitable timescales.
Understanding which physical processes can support these requirements is central to evaluating integrated photonic implementations\cite{shastri2021photonics}.

\subsection{Neuromorphic Computing with Photonics}

Photonics offers several attractive properties for neuromorphic computing. Many neural-network operations, especially matrix–vector multiplications, are linear transformations that can be implemented using passive optical interference in scattering, imaging or propagation through different media\cite{reck1994experimental}.
A matrix–vector product $y = W x$ requires $N^2$ multiply–accumulate operations in electronics, each involving charge movement and transistor switching.
In contrast, optical interferometer meshes can implement the same operation through passive interference, with energy consumed primarily in generating light, detecting outputs, and tuning phase shifters\cite{shen2017deep}.
For example, the Taichi photonic processor reports an on-chip efficiency of 160 TOPS/W \cite{xu2024large} (TOPS: Tera Operation Per Second), whereas the packaged 64 × 64 PACE accelerator reports 4.21 TOPS/W excluding laser power and 2.38 TOPS/W when laser power is included \cite{hua2025integrated}.

Photonics also supports wavelength-division multiplexing (WDM), in which multiple independent data channels propagate on different wavelengths within the same optical fiber\cite{agrawal2012fiber}.
This enables large-scale parallelism that is difficult to achieve with electronic interconnects, where many wires are needed to avoid crosstalk.
Furthermore, nonlinear optical phenomena in materials occur on very short time-scales, enabling  high-throughput processing at hundreds of \unit{GHz}, limited primarily by the performance of driver electronics and packaging \cite{li2025all}.

Finally, applying photonic computing to signals that are natively in the optical domain, such as optical telecommunication data and optical sensors' output, allows bypassing or reducing costly and slow optics-to-electric conversions and digitalization, while enabling direct access and exploitation of optical degrees of freedom, such as optical phase, wavelength, propagation modes and polarization.

\subsection{Integrated Photonics for Neuromorphic Computing}

Free-space optical computing has a long history \cite{peyghambarian1985optical,psaltis1990holography,solli2015analog}, but such systems require precise alignment of bulk components and are difficult to scale.
Integrated photonics addresses these issues by defining optical paths lithographically on a chip, providing intrinsic mechanical stability and enabling dense integration of waveguides, resonators, and nonlinear elements\cite{vivien2016handbook}.
Complex circuits can be realized without manual alignment, and photonic components can be co-integrated with electronic drivers and detectors for compact, high-speed operation.

Integrated photonics is well suited to high-throughput neuromorphic architectures, which often combine linear transformations, nonlinear responses, and feedback.
This platform enables long delay lines, high-Q resonators, and nonlinear elements on a single chip, and further benefits from heterogeneous material integration, high-volume CMOS manufacturing, and support for multiple device technologies.
These characteristics allow relatively compact and densely integrated components with favorable cost scaling, providing a versatile substrate for implementing a range of temporal dynamics and memory mechanisms \cite{brunner2025roadmap}.

At the same time, photonic platforms face several constraints \cite{prucnal2017neuromorphic,shastri2021photonics}.
Optical nonlinearities in common materials are relatively weak, so realizing nonlinear activation functions often requires electro-optic conversion or resonant enhancement.
Photonic components occupy larger footprints than transistors, limiting integration density.
Thermo-optic tuning consumes continuous power and has microsecond-scale response times, while faster electro-optic tuning typically requires higher voltages, large power or specialized materials.
Photodetection and amplification also imply non-negligible energy and area overheads, particularly for large arrays.
These considerations motivate hybrid optoelectronic architectures in which photonics handles high-bandwidth linear transformations and distributed signal routing, while electronics provides nonlinear processing, memory consolidation, and control.

\subsection{Silicon Photonics and Related Platforms}

Silicon photonics, particularly the silicon-on-insulator (SOI) platform, has become a leading technology for integrated photonic circuits due to its compatibility with CMOS fabrication, enabling high-yield and wafer-scale manufacturing \cite{vivien2016handbook}. Silicon offers strong optical confinement (high refractive index contrast), transparency in telecommunications bands, and a sizable thermo-optic coefficient for phase tuning. Although it lacks efficient light emission and requires heterogeneous integration for gain and detection, its fabrication maturity and integration density make it attractive for neuromorphic architectures. Crucially, silicon photonics based on CMOS technology should not just be seen as an alternative to electronics, but as a means of enabling close integration with electronic circuits.

Alternative platforms, including silicon nitride (\ce{SiN}), indium phosphide (\ce{InP}), and thin-film lithium niobate on insulator (\ce{LNOI}), offer complementary properties such as lower propagation loss, monolithic light sources, and strong electro-optic modulation.\cite{smit2014introduction,zhu2021integrated} In neuromorphic computing, these platforms support different types of memory mechanisms: \ce{SiN} for low-loss delay lines, \ce{InP} for active spiking elements and gain dynamics, and \ce{LNOI} for fast modulators and low-power weight tuning. \Cref{tab:platform-comparison} summarizes key characteristics of these platforms that are relevant for neuromorphic photonics. \Cref{tab:platform-comparison} separates intrinsic physical response times from representative device bandwidths. A fast material response is not by itself a memory mechanism: memory requires a relaxing or persistent state, resonant storage, propagation delay, or feedback path.

\begin{table}[t]
    \caption{Comparison of the main integrated photonic platforms for
    neuromorphic computing at wavelengths around \qty{1.5}{\micro\meter}.
    Intrinsic response or relaxation time refers to the material response or
    memory-bearing state, whereas device bandwidth also depends on geometry,
    loaded quality factor, electrodes, parasitics, and packaging. The
    memory-bearing-state column identifies the physical variable or path that
    retains information; a Kerr or Pockels response alone does not constitute
    memory. Values are indicative and drawn from representative literature
    \cite{shastri2021photonics,smit2014introduction,zhu2021integrated,soref2006past,reed2010silicon,wang2018integrated,valdez2022110,ren2025photorefractive}.}
    \label{tab:platform-comparison}

    \renewcommand{\arraystretch}{1.2}
    {\footnotesize
    \setlength{\tabcolsep}{2.5pt}
    \tolerance=9999
    \hyphenpenalty=0
    \exhyphenpenalty=0
    \emergencystretch=1em
    \tablealtrows

    \begin{tabular}{@{}
        >{\centering\arraybackslash}m{.055\linewidth}
        >{\centering\arraybackslash}m{.100\linewidth}
        >{\raggedright\arraybackslash}m{.130\linewidth}
        >{\raggedright\arraybackslash}m{.145\linewidth}
        >{\raggedright\arraybackslash}m{.140\linewidth}
        >{\raggedright\arraybackslash}m{.235\linewidth}
        >{\raggedright\arraybackslash}m{.105\linewidth}
        @{}}
        \toprule

        & Index contrast / loss
        (\unit{\deci\bel\per\centi\meter})
        & Relevant physical response
        & Intrinsic response or relaxation time
        & Representative device bandwidth
        & Memory-bearing state or feedback path
        & Electronic integration maturity
        \\

        \midrule

        Si
        &
        High / $<1$
        &
        Kerr, two-photon absorption, free-carrier and thermo-optic effects
        &
        Kerr/TPA: instantaneous; carriers:
        \unit{\nano\second}; thermal:
        \unit{\nano\second}--\unit{\micro\second}
        &
        Thermal: \unit{\kilo\hertz}--\unit{\mega\hertz};
        carrier-based: multi-\unit{\giga\hertz} to $>50$~\unit{\giga\hertz};
        resonators: loaded-$Q$-factor limited
        &
        Delay or feedback; photon lifetime; carrier population; temperature.
        Kerr/TPA alone do not provide memory
        &
        Mature CMOS platform; high electronic-integration maturity
        \\

        SiN
        &
        Medium / $<0.1$
        &
        Kerr and thermo-optic effects
        &
        Kerr: instantaneous; thermal:
        \unit{\micro\second};
        delay and photon lifetime are device-defined
        &
        Broadband waveguides; loaded $Q$-factor limited resonators, typically
        \unit{\mega\hertz}--\unit{\giga\hertz}
        &
        Propagation delay; photon lifetime; filtering and feedback.
        Kerr alone does not provide memory
        &
        Mature passive platform; active functions generally require
        hybrid integration
        \\

        InP
        &
        Medium / moderate
        &
        Carrier-induced gain, absorption and index changes
        &
        Carrier and gain recovery:
        \unit{\pico\second}--\unit{\nano\second}
        &
        Typically tens of \unit{\giga\hertz} for active devices;
        resonators and feedback are architecture-dependent
        &
        Carrier or gain population; mode competition and bistability;
        delay or feedback. Injection selects or writes the state
        &
        Mature active PIC platform; electronics generally hybrid
        \\

        LNOI
        &
        Medium / low
        &
        Pockels, Kerr, photorefractive and pyroelectric effects
        &
        Pockels/Kerr: instantaneous; internal-field relaxation:
        condition-dependent
        &
        $>100$~\unit{\giga\hertz} electro-optic modulation demonstrated;
        resonators: loaded-$Q$-factor limited
        &
        Delay or feedback; photon lifetime; long-lived photorefractive,
        pyroelectric or ferroelectric states. Pockels modulation alone
        does not provide memory
        &
        Rapidly improving; electronics and active devices generally
        hybrid
        \\

        \botrule
    \end{tabular}}
\end{table}

For neuromorphic applications, silicon and related platforms (in short silicon photonics) provide a useful compromise between optical performance, integration density, and manufacturing maturity. While the thermo-optic effect provides a mature mechanism for configuring quasi-static weight matrices, its inherent thermal diffusion limits and microsecond-scale response times are often insufficient for high-frequency dynamics. To address these limitations, other modulation methods are employed such as plasma dispersion or electro-optic effects via heterogeneous integration of other materials like Lithium Niobate, Barium Titanate or organic materials \cite{vivien2016handbook}. The fundamental limitation that silicon cannot provide efficient light emission or optical gain is mitigated in hybrid-SOI platforms through the integration with III–V sources and amplifiers, a technology now well established for telecommunications and for complex photonic circuits\cite{shekhar2024roadmapping,kaur2021hybrid}.

The high refractive index contrast of silicon photonics enables the dense integration of Mach-Zehnder Interferometer (MZI) meshes or microring resonator (MRR) arrays required for high-dimensional tensor operations. Consequently, the selection of a specific modulation physics is governed by the required balance between update volatility, energy-per-bit, and the desired timescale of the network's evolution.

\subsection{The Architecture of Memory: Addresses vs. Engrams}
\label{sec:memory-architecture}

Memory is a fundamental aspect of computation, but different physical and computational systems organize it in profoundly different ways.
In conventional digital architectures, information is stored in explicitly addressable locations, physically separated from the processing unit.
Biological systems distribute memory across networks of interacting elements (cell assemblies or engram) \cite{hebb2005organization}, where synaptic efficacies encode long-term information and internal state variables encode short-lived traces of recent activity.
Artificial neural networks adopt a similar distributed paradigm: training determines weights, biases, and other parameters governing the network transformation or dynamics, while inference propagates activations and, in recurrent architectures, evolves internal working states. For attractor and energy-based models, this process admits the more specific interpretation of evolution within a learned energy landscape \cite{lecun2015deep,hopfield1982neural}.

Understanding these contrasting memory organizations, namely localized and addressable versus distributed and dynamical, is essential for analyzing how memory can be implemented in integrated photonics. Photonic devices do not naturally support static storage of the information carriers themselves, since photons cannot be held at rest, but they do support a variety of material and resonant states that act as physical memory variables. These state variables differ in volatility, programmability, and characteristic decay time, and together they define the functional memory architecture available to a photonic neural network (PNN).

\begin{figure}[!htb]
	\centering
	\includegraphics[width=\linewidth]{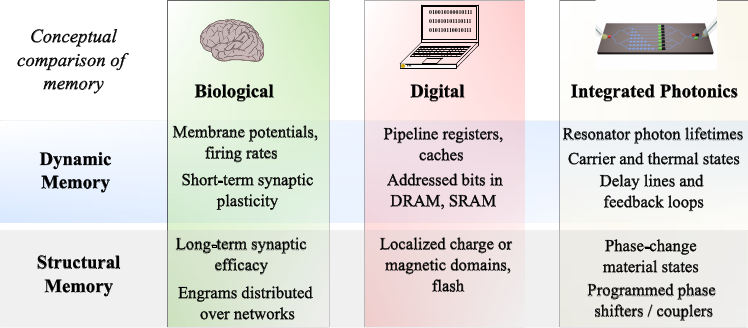}
	\caption{Conceptual comparison of structural  and dynamic  memory across digital, biological, and integrated photonic systems. Structural memory corresponds to long-lived parameters (e.g. weights, device states), while dynamic memory is encoded in transient state variables that evolve during computation.These computational roles are independent of physical volatility, retention time, temporal behaviour, and holding-power requirements. The correspondence between columns indicates a functional analogy and does not imply physical equivalence between the listed mechanisms.}
	\label{fig:memory-taxonomy}
\end{figure}

The distinction between volatile and non-volatile memories parallels the contrast between address-based memory in digital systems, distributed engram-like representations in biological circuits, and the parameter–state decomposition characteristic of artificial neural networks \cite{indiveri2015memory,tonegawa2015memory}.
\Cref{fig:memory-taxonomy} summarizes this comparison across digital, biological, and photonic substrates.

\subsection{The Physics of Memory and Learning}

Learning in any adaptive system requires that internal parameters change in response to correlations between past states and present outcomes.
In digital implementations of neural networks, this temporal dependency is reconstructed algorithmically by storing intermediate activations during a forward pass and propagating error signals backward through time \cite{mienye2024comprehensive}.
Training a photonic neural network can involve different degrees of physical participation. In hybrid systems, intermediate activations may be stored electronically and parameter updates calculated using digital backpropagation or hardware-in-the-loop optimization \cite{pai2023experimentally}. Alternatively, gradients may be measured in situ through physical forward and adjoint propagation \cite{ashtiani2026integrated}. At the other end of this spectrum, correlation-based local learning rules update a parameter using only quantities available at or near the corresponding physical element. Local learning can reduce activation storage and data movement and may therefore improve scalability, but it is not the only physically realizable training approach.

Learnability in physical dynamical systems, such as a photonic integrated circuit (PIC), requires that parameter updates correlate with the direction of improvement in the underlying cost function, even if exact gradients cannot be computed. In integrated photonics, such updates must be driven by physically available signals—typically optical intensities, voltages, or local interference terms—combined with structural mechanisms such as refractive-index tuning, phase-change reconfiguration, or slow thermal shifts.\cite{prucnal2017neuromorphic,feldmann2019all}

The dynamical regime in which a neuromorphic photonic system operates therefore determines not only its memory characteristics but also the effectiveness of any training algorithm applied to it. Learning requires that local parameter updates remain sensitive to the temporal structure of internal state trajectories without being overwhelmed by noise or instability. In some reservoir and recurrent systems, operation close to a stability boundary can enhance sensitivity, memory capacity, or nonlinear feature generation. Operating near the edge of chaos \cite{langton1990computation}, i.e. a region between stable and chaotic trajectories, provides a balance between sensitivity and stability.  This enables local physical processes—such as carrier accumulation, thermal relaxation, or refractive-index changes—to encode meaningful correlations between pre- and post-synaptic signals. For example, these correlations form the physical basis of plasticity mechanisms such as those discussed in \cref{sec:bio}, i. e. Hebbian updates or spike-timing-dependent plasticity (STDP), which rely on the interaction between transient state variables and external teaching or reward signals \cite{van2000stable,lajoie2014structured}. However, the optimum condition is task-, architecture-, and noise-dependent: approaching or entering a chaotic regime can reduce consistency, input reproducibility, and noise tolerance. The edge of chaos should therefore be regarded as a useful operating regime to investigate, rather than as a generally optimal condition for learning \cite{bertschinger2004edge,legenstein2007edge,boedecker2012information}.

When gradients and intermediate activations are not explicitly computed or stored by an external controller, local physical learning rules provide an attractive alternative. In this case,  the update of a given parameter depends only on quantities available at that location, such as local intensities, carrier densities, temperatures, or co-propagating signals \cite{indiveri2015memory}. In integrated photonics, several material-level mechanisms provide the basis for such local plasticity. Phase-change materials (PCMs) can undergo long-lived refractive-index changes when triggered by optical or electrical pulses, enabling non-volatile updates to structural memory \cite{rios2015integrated,feldmann2019all}. Carrier-based and thermal nonlinearities can transiently modulate effective refractive indices or absorption, providing short-lived state dependencies that can interact with external supervisory signals \cite{biasi2024exploring}. Properly organized, these mechanisms can approximate the temporal correlations required for learning \cite{lugnan2025emergent}.

In this view, the different learning paradigms correspond primarily to differences in how local plasticity is gated or modulated by global or semi-global signals. Supervised, unsupervised, and reinforcement learning can all be understood as specific ways of combining local state-dependent updates with externally provided error, statistical, or reward signals acting on appropriate timescales.

\subsection{Scope and Organization of This Review}

Neuromorphic photonics has expanded significantly in recent years \cite{shastri2021photonics,lei2025progress,han2025integrated,wu2025intelligent,wang2026algorithms}. Many contributions focus on feed-forward architectures such as optical matrix multipliers and static classifiers, which primarily exploit structural memory. In contrast, the temporal dimension, that is the physical realization of dynamic memory and plasticity, has received comparatively less systematic attention. As emphasized in \cref{fig:memory-taxonomy}, neuromorphic computation requires both long-lived structural parameters and short-lived dynamical state variables. Integrated photonics provides several mechanisms to support these roles. The purpose of this review is not to identify a single optimal photonic memory mechanism. Each implementation introduces a different balance among retention, bandwidth, insertion loss, holding power, write and read energy, observability, stability, drift, programmability, and integration complexity. The relevant design question is therefore how to match the timescale and temporal behavior of state memory to the task and how to combine useful transient dynamics with controllable parameter or configuration memory. We also distinguish the maturity of the proposed systems, separating simulated concepts, component-level demonstrations, hybrid optoelectronic experiments, and fully integrated implementations.

In this review, we focus on PNN, with emphasis on the transient physical memory mechanisms available in current platforms. \Cref{sec:bio} reviews only those aspects of biological memory that are directly relevant to the physical implementation of memory in photonic systems: the coexistence of short- and long-timescale state variables, and the locality of plasticity rules. \Cref{sec:memory} surveys integrated optical memory mechanisms in silicon photonics, providing a classification based on the physical mechanism, including resonant storage, delay lines, carrier and thermal dynamics, and PCMs. It also analyzes how these processes can serve as state variables with different characteristic decay times. \Cref{sec:theo} presents the theoretical and experimental framework for characterizing computational memory in PNNs. It introduces neural architectures, memory metrics, and benchmark tasks, classifying and comparing systems based on how they process time and implement memory.
\Cref{sec:PNNs} examines architectural PIC implementations such as integrated reservoir computing systems based on delay memory or nonlinear dynamics, recurrent processors, networks with long-term physical plasticity and delay-based system for high speed telecom applications. Throughout, we link the behavior of each architecture to the memory concepts introduced earlier. Finally \Cref{sec:conclusion} ends the review by discussing limitations and perspectives of PNN.

\section{Biological Background: Hebb and Synaptic Plasticity}\label{sec:bio}

\subsection{Biological Memory Mechanisms}
Memory in biological organisms is the fundamental capacity to store and retrieve information across time scales ranging from seconds to years \cite{tulving1983elements,squire2003memory}.
It is widely regarded as a core component of learning, enabling organisms to adapt their behavior based on past sensory experiences to increase their chances of survival and reproduction \cite{gallistel1990organization}.
In animals with nervous systems, memory is generally viewed as an emergent property of brain activity, arising largely from biochemical and structural modifications at neuronal synapses, together with changes in intrinsic excitability and network dynamics \cite{hebb2005organization,bliss1973long,kandel2001molecular,abbott2004synaptic}. The study of biological memory, fundamental to learning and behavior \cite{squire2003memory,gallistel1990organization,baddeley2012working}, is increasingly driving AI research \cite{hassabis2017neuroscience,marblestone2016toward} toward the development of robust, energy-efficient architectures that emulate biological learning mechanisms \cite{bengio2015towards,lillicrap2020backpropagation,bi1998synaptic,indiveri2015memory}.
This section reviews the primary mechanisms of biological memory to establish a comparative foundation for the photonic memories and PNN explored in the following sections.

\subsubsection{General Description of Biological Memory}

There is still no universally accepted definition of \textquote{memory} in biological systems, partly because memory spans multiple levels of organization, from molecular signaling to network-level dynamics and behavior \cite{dudai2004neurobiology,mayford2012synapses,lisman2018memory}.
A general consensus exists that memory comprises three fundamental processes: \emph{encoding}, in which new information is initially represented; \emph{consolidation}, during which the encoded representation is stabilized and integrated into existing neural circuits; and \emph{retrieval} or recall, the reactivation of stored representations to guide behavior or cognition \cite{tulving1983elements,kandel2014molecular,squire2015memory}.

\begin{figure}[!htb]
	\centering
	\includegraphics[width=1\linewidth]{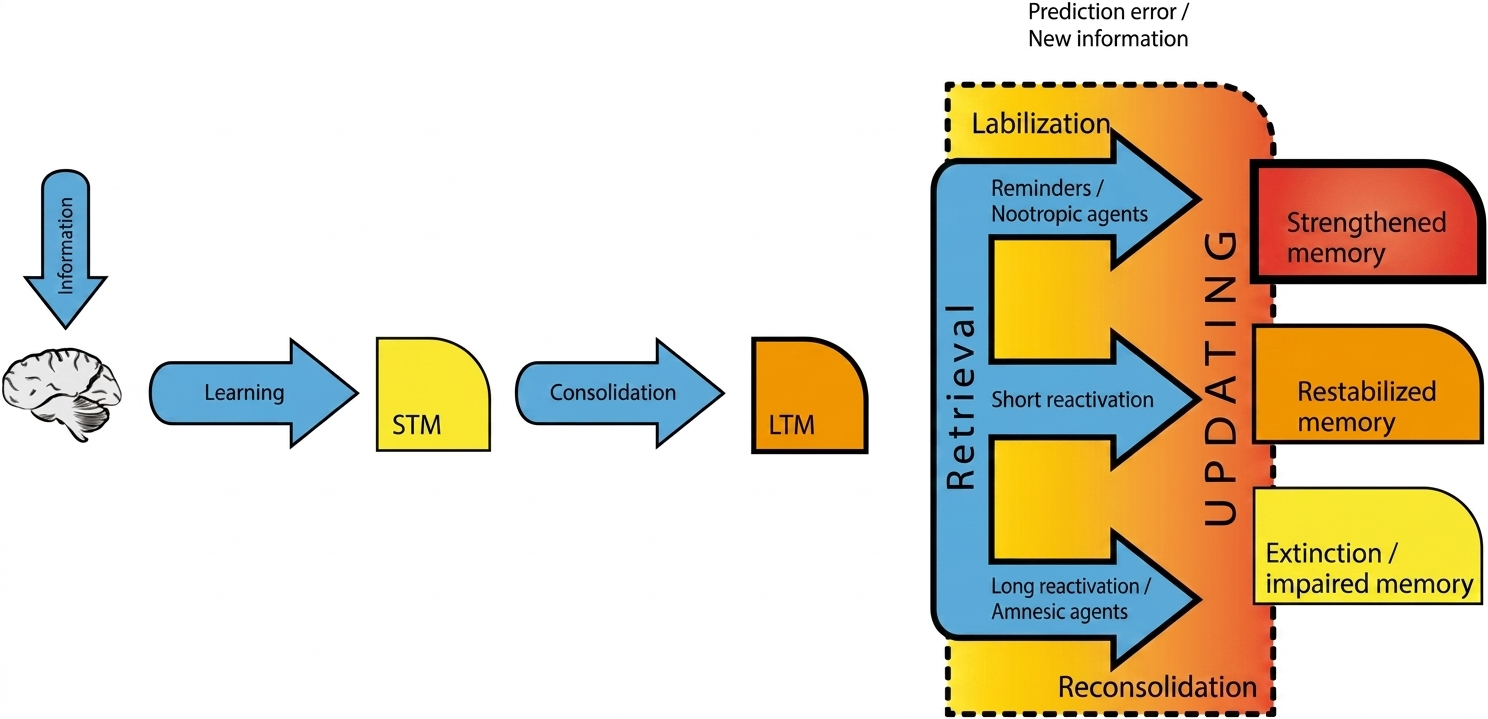}
	\caption{%
		The general description of biological memory from short-term (STM) to long-term memory (LTM), through encoding, learning and retrieval of information.
		Reproduced from \cite{osorio2023transforming}, licensed under \href{https://creativecommons.org/licenses/by/4.0/}{CC BY 4.0.}
	}
	\label{fig:stm-and-ltm}
\end{figure}

\paragraph{Short-term and working memory}
A key distinction is usually made between \emph{Short-Term Memory} (STM) and \emph{Long-Term Memory} (LTM), which differ in their time scales, capacity, and underlying biological mechanisms (\cref{fig:stm-and-ltm}) \cite{atkinson1968human,baddeley1992working}.
STM supports the temporary maintenance of information over seconds to minutes, relying primarily on sustained neural activity and short-lived biochemical processes \cite{fuster1971neuron,compte2000synaptic}.
A related but distinct construct is \emph{working memory}, which not only stores information briefly but also manipulates it to support ongoing tasks such as reasoning and decision-making \cite{baddeley2007working}.
Both STM and working memory are capacity-limited and strongly dependent on active neural circuitry, particularly within the prefrontal cortex \cite{goldman1995cellular,miller2001integrative,funahashi2017working}.

\paragraph{Long-term memory}
In contrast, LTM encompasses memories that persist from hours to years and is generally associated with more durable forms of plasticity, including synaptic modifications, structural changes, and systems-level consolidation across brain regions \cite{kandel2014molecular,bliss1973long,bailey1993structural,frankland2005organization}. LTM is commonly divided into two broad categories: \emph{non-declarative} and \emph{declarative} memory.
Non-declarative memory refers to learning processes that occur gradually through repetition or conditioning without conscious awareness, such as motor skills, habits, and sensorimotor reflexes \cite{squire2004memory,knowlton1996neostriatal,dickinson1985actions}. Declarative memory, on the other hand, involves the conscious recall of facts and personal experiences, typically further classified into \emph{semantic} (knowledge about the world) and \emph{episodic} (memory of specific events) components \cite{tulving1972episodic,cabeza2000imaging}.

\subsubsection{Biological Foundations of Memory: From Synaptic Communication to Engram Consolidation}


The formation and stabilization of memories emerge from a hierarchy of cellular, synaptic, and systems-level processes that enable the brain to encode, store, and retrieve information \cite{kandel2014molecular,squire1995retrograde,dudai2004neurobiology}.

\begin{figure}
    \centering
    \includegraphics[width=0.75\linewidth]{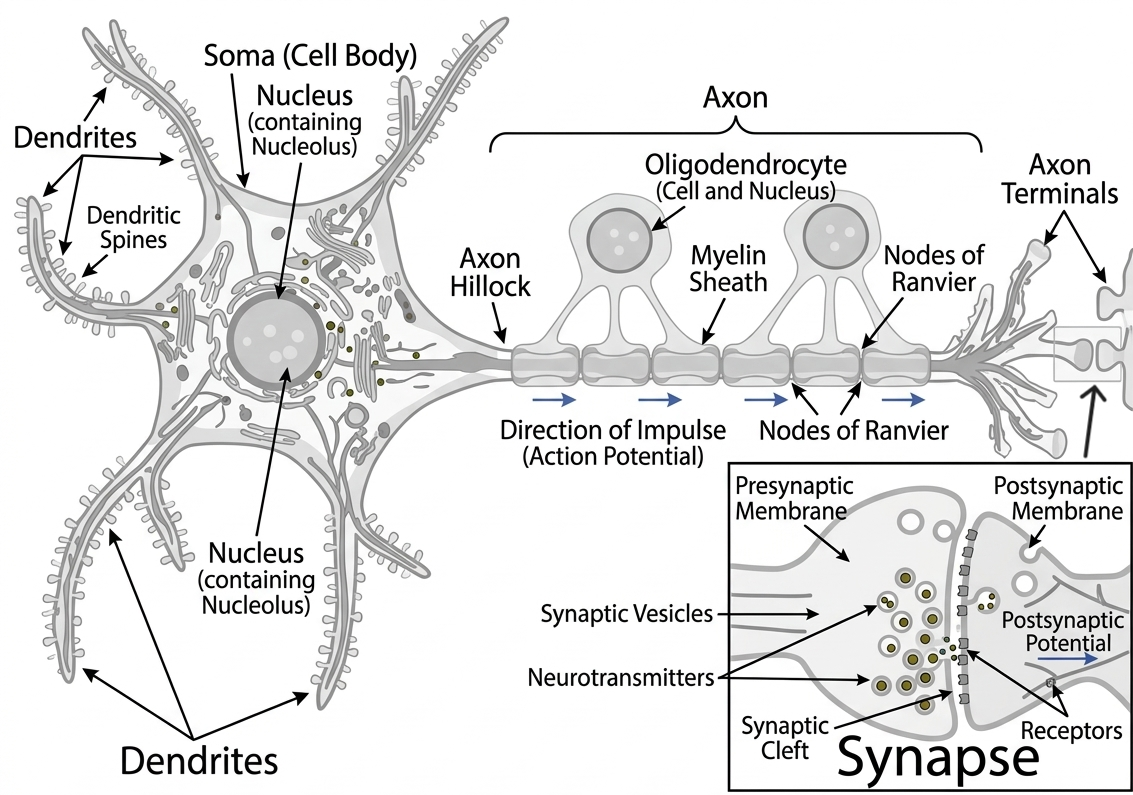}
    \caption{A general scheme of a biological neuron and a synapse with their main components. The terminal of a presynaptic neuron comes into close contact with a postsynaptic cell at the synapse. Adapted from SwissBioPics, licensed under \href{https://creativecommons.org/licenses/by/4.0/}{CC BY 4.0} (indications of cell components are added)}
    \label{fig:neuron}
\end{figure}

At the most fundamental level, neurons communicate through action potentials (APs) and chemical synapses (\cref{fig:neuron}) \cite{kandel2000principles}. APs are rapid, transient changes in the membrane potential that propagate along the axon, enabling electrical signals to travel over long distances within the neuron. When an action potential reaches the presynaptic terminal, it triggers the release of neurotransmitters into the synaptic cleft. These chemical messengers bind to receptors on the postsynaptic membrane, modulating the membrane potential of the receiving neuron and thereby influencing its probability of generating subsequent action potentials. Through this electrochemical signaling mechanism, information is transmitted and processed across neuronal networks.
The arrival of an AP at the presynaptic terminal opens voltage-gated \ce{Ca^2+} channels, initiating neurotransmitter release and thereby modulating postsynaptic excitability through ligand-gated ion channels and metabotropic receptors \cite{katz1969release,sudhof2012presynaptic}.
This basic mode of communication provides the substrate on which experience-dependent changes in synaptic strength, collectively termed \emph{synaptic plasticity}, are built \cite{martin2000synaptic}.

\paragraph{Short-term Synaptic Plasticity}
Short-term synaptic plasticity (STP) provides a mechanism through which synaptic efficacy dynamically depends on recent neuronal activity, primarily through short-term facilitation and depression \cite{zucker1989short}. Facilitation typically arises from residual presynaptic \ce{Ca^2+}, which transiently increases neurotransmitter release probability during closely spaced spikes, whereas depression can result from depletion of the readily releasable synaptic vesicle pool during sustained activity. By continuously adapting synaptic strength according to recent spike history, these mechanisms enable neural circuits to encode temporal patterns and transiently retain information about previous activity. STP can therefore contribute to short-term memory and working-memory processes, providing a dynamic form of memory that operates on timescales from milliseconds to seconds without requiring persistent changes in long-term synaptic weights \cite{mongillo2008synaptic}.

Related memory processes operate at the neuronal and circuit levels \cite{benda2021neural}. The membrane potential provides a form of leaky integration, in which synaptic inputs are accumulated while their influence progressively decays toward baseline \cite{gerstner2014neuronal}. Neuronal adaptation introduces slower activity-dependent changes in excitability, causing the present neuronal response to depend on its recent firing history \cite{benda2003universal,prescott2008spike}. At the circuit level, recurrent excitation can maintain a population state after the initiating input has disappeared, thereby supporting working memory under continuous metabolic activity \cite{compte2000synaptic}. Leaky integration and neuronal adaptation therefore provide fading state memory, whereas recurrent circuits can support persistent but actively maintained states. Their photonic counterparts include cavity-field decay, carrier and thermal relaxation, gain recovery, and recurrent optical or optoelectronic feedback.

\begin{figure}[h!]
    \centering
    \includegraphics[width=0.5\linewidth]{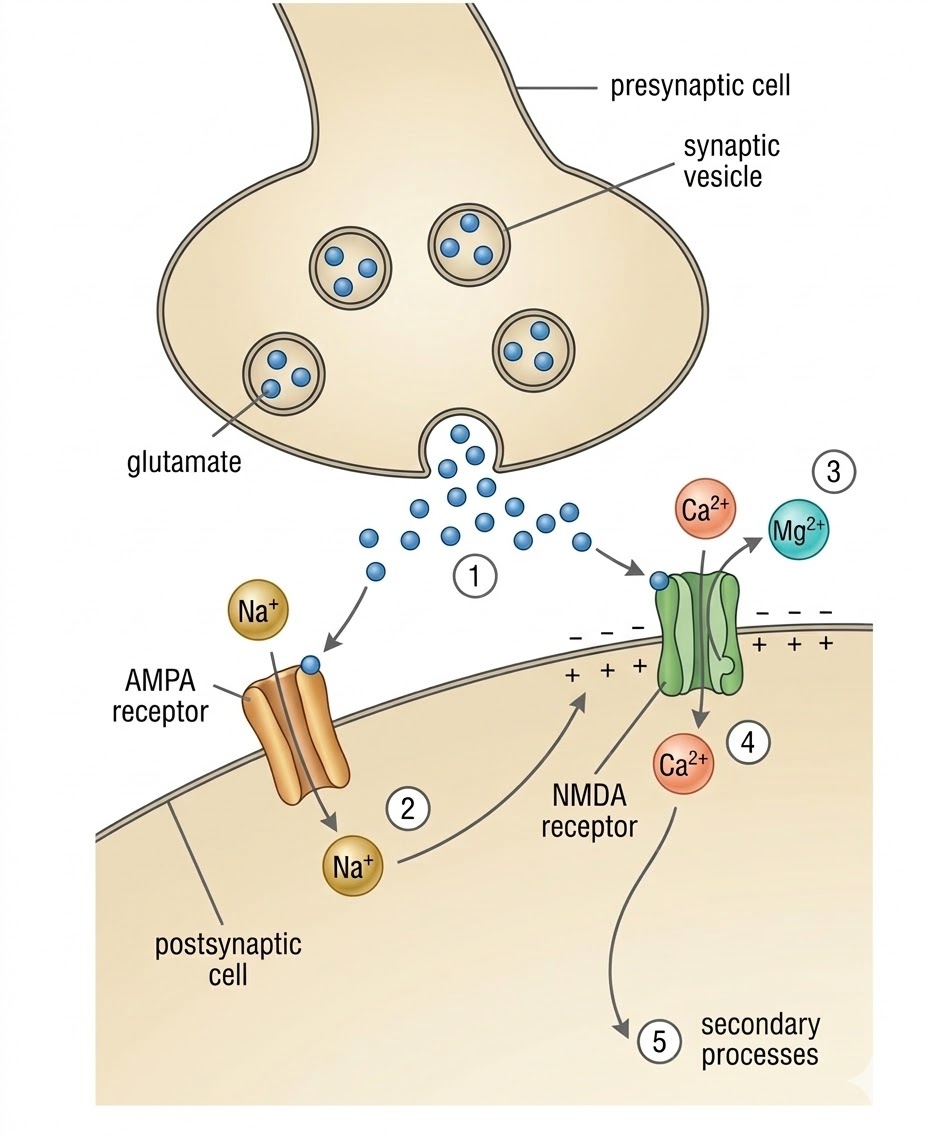}
    \caption{Synaptic communication and LTP. An action potential on the pre- synaptic cell stimulates the release of the neurotransmitters into the synaptic cleft by the synaptic vesicles. In the case of long term potentiation, the involved neurotransmitter is glutamate (1). It can bind to AMPA and NMDA receptors on the postsynaptic membrane. (2) The binding with glutamate induces AMPA receptor opening and a flux of \ce{Na+} occurs. The following depolarization of the cell led to the opening of the NMDA receptor with the release of the \ce{Mg^2+} (3) and the influx of \ce{Ca^2+} (4). The increase of calcium concentration in the cell activates some secondary processes that lead to changes in the postsynaptic cell. Adapted from \cite{zaccaria2022light} with permission.}
    \label{fig:LTP}
\end{figure}

\paragraph{Long-term Potentiation}
Long-term potentiation (LTP) is one of the most extensively investigated mechanisms for memory encoding \cite{bliss1973long,morris1986selective}.
Canonical LTP induction requires coincident presynaptic glutamate release and postsynaptic depolarization sufficient to relieve the voltage-dependent block of N-methyl-D-aspartate (NMDA) receptors (\cref{fig:LTP}) \cite{collingridge1983excitatory}.
The subsequent \ce{Ca^2+} influx activates intracellular signaling pathways. Activation of these pathways promotes the phosphorylation of AMPA receptors ($\alpha$-amino-3-hydroxy-5-methyl-4-isoxazolepropionic acid receptors), the principal ionotropic glutamate receptors mediating fast excitatory synaptic transmission. Phosphorylation facilitates the trafficking and membrane insertion of additional AMPA receptors into the postsynaptic membrane, thereby strengthening synaptic transmission and increasing synaptic efficacy \cite{malinow2002ampa,lisman2002molecular,huganir2013ampars}.

LTP is typically divided into two temporal phases: an early, protein-synthesis–independent phase (E-LTP) lasting minutes to hours, and a late, protein-synthesis–dependent phase (L-LTP) that can persist for days and requires transcriptional regulators such as CREB (cAMP -cyclic adenosine monophosphate- response element–binding protein) \cite{frey1997synaptic,nguyen1994requirement,silva1998creb}.

\paragraph{Long-term Depression}
Complementary to LTP is long-term depression (LTD), a sustained weakening of synaptic transmission driven by lower levels or distinct temporal patterns of \ce{Ca^2+} signaling \cite{bear1994synaptic, dudek1992homosynaptic}.
The balance between LTP and LTD permits both the stabilization and refinement of neural representations \cite{abbott2000synaptic}.

\begin{figure}[!htb]
	\centering
	\includegraphics[width=0.5\linewidth]{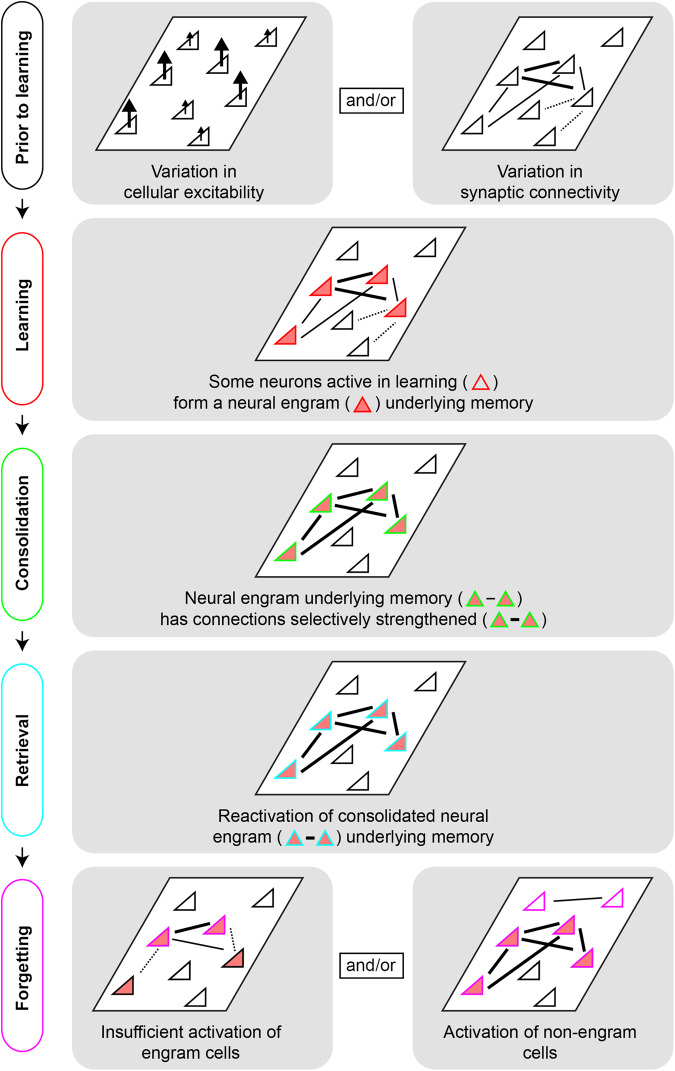}

    \caption{%
    Cellular and synaptic organization across memory stages, from encoding to the formation of engram cells. The process begins with an increase in intrinsic neuronal excitability during encoding, which biases certain neurons toward allocation into the engram. During learning, activity in these cells promotes synaptic consolidation, leading to strengthened connections that can be further reinforced upon reactivation during the consolidation phase. Once an engram has been consolidated and stored, its activation enables memory retrieval through the reactivation of neurons recruited during learning. Forgetting is associated with ongoing synaptic remodeling. Reproduced from \cite{guskjolen2023engram}, licensed under \href{https://creativecommons.org/licenses/by/4.0/}{CC BY 4.0}.
}
	\label{fig:engram}
\end{figure}

\paragraph{Engram Cells}
Beyond synaptic potentiation \cite{zaccaria2026investigation}, memory formation engages broader cellular mechanisms. Neurons allocated to a given memory, often referred to as \emph{engram cells}, exhibit transient increases in intrinsic excitability, due to elevated CREB levels, biasing them toward recruitment during learning and create competitive interactions among neurons (\cref{fig:engram}) \cite{josselyn2015finding,han2007neuronal,yiu2014neurons,rogerson2014synaptic,zaccaria2026investigation}.

At the structural level, long-term plasticity is accompanied by the enlargement and stabilization of dendritic spines, cytoskeletal remodeling, and the formation or elimination of synaptic contacts, providing a physical substrate for durable memory traces \cite{bailey1993structural,holtmaat2009experience,yang2009stably}.

\paragraph{Memory Consolidation}
Memory consolidation operates on multiple timescales.
On the cellular scale, sustained kinase activity, local protein synthesis, and synaptic remodeling stabilize LTP \cite{frey1997synaptic,sajikumar2004late}.
On the systems level, hippocampal-cortical interactions gradually reorganize and redistribute memory representations \cite{frankland2005organization,squire2015memory,teyler1986hippocampal}.
During sleep, hippocampal sharp-wave ripples reactivate recently encoded neuronal ensembles, promoting synaptic plasticity in neocortical circuits and supporting long-term storage \cite{wilson1994reactivation,buzsaki2015hippocampal,diekelmann2010memory}.
This \emph{systems consolidation} process mitigates interference and allows memories to become increasingly independent of the hippocampus over time \cite{squire2015memory,teyler1986hippocampal}.

\begin{figure}[!htb]
	\centering
	\includegraphics[width=0.75\linewidth]{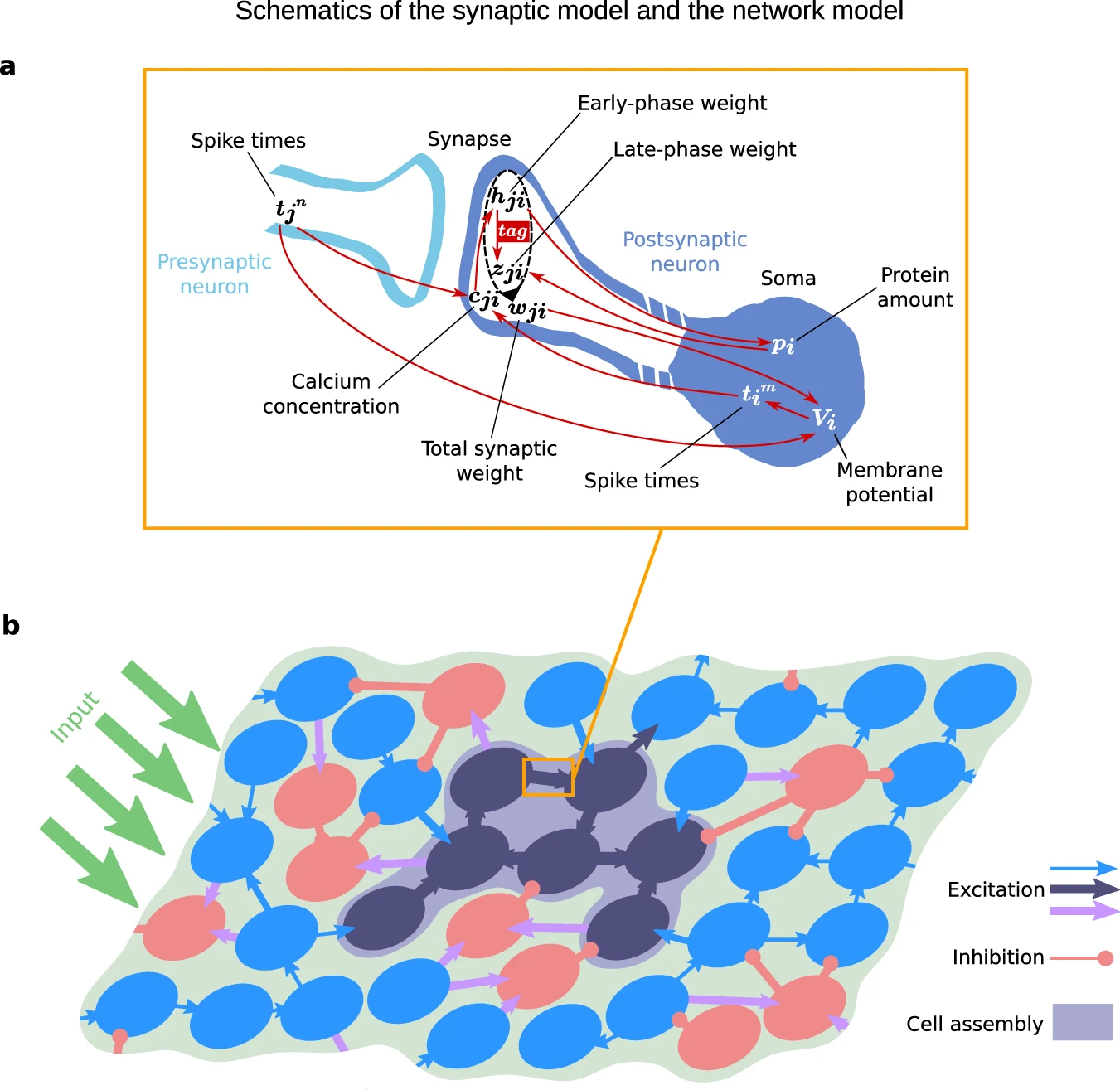}

	\begin{subcaptiongroup}
		\phantomcaption\label{fig:syn-tag-a}
		\phantomcaption\label{fig:syn-tag-b}
	\end{subcaptiongroup}

	\caption{%
		\subref{fig:syn-tag-a} The synaptic model integrates the interplay between various mechanisms of calcium-dependent synaptic plasticity and the STC hypothesis.
		\subref{fig:syn-tag-b} Schematic of a part of the neural network that consists of excitatory (blue and dark blue circles) and inhibitory neurons (red circles), and receives external input from other brain areas (green arrows).
		Only synapses between excitatory neurons (blue and dark blue arrows) undergo plastic changes by the processes shown in \subref{fig:syn-tag-a}.
		A Hebbian cell assembly represents a memory and consists of a group of strongly interconnected neurons (dark blue).
		Reproduced from \cite{luboeinski2021memory}, licensed under \href{https://creativecommons.org/licenses/by/4.0/}{CC BY 4.0}.
	}
	\label{fig:syn-tag}
\end{figure}

Additional regulatory mechanisms ensure that synaptic plasticity remains precise and context-dependent.
Neuromodulators such as dopamine, acetylcholine, and norepinephrine gate the induction of LTP and LTD, linking plasticity to behavioral relevance, attention, novelty, and reward \cite{lisman2005hippocampal,hasselmo2006role,sara2009locus,schultz2016dopamine}.
Synaptic tagging and capture (STC) further explain how weakly activated synapses can achieve long-lasting changes by capturing plasticity-related proteins synthesized in response to strong events (\cref{fig:syn-tag}) \cite{redondo2011making,rogerson2014synaptic,luboeinski2021memory}.

Finally, \emph{metaplasticity}, the plasticity of synaptic plasticity, modifies the threshold and magnitude of subsequent LTP or LTD according to prior activity, thereby helping to maintain stable but adaptable synaptic dynamics  \cite{abraham1996metaplasticity,cooper2012bcm}.
Together, these intertwined processes transform transient patterns of neuronal activity into persistent changes in synaptic and circuit organization.
Through the orchestration of electrical signaling, biochemical cascades, structural remodeling, and large-scale network interactions, the brain constructs and consolidates the memory engrams that underlie learning and experience \cite{squire1995retrograde,dudai2004neurobiology,tonegawa2015memory}.

\paragraph{Hebbian learning}
Hebbian learning refers to a class of synaptic plasticity mechanisms in which the strength of a connection between two units is modified according to the correlation of their activity. The principle, first proposed by Donald Hebb, states that if a presynaptic neuron repeatedly contributes to the activation of a postsynaptic neuron, the synaptic connection between them is strengthened—often summarized as “cells that fire together wire together” \cite{hebb2005organization}. In formal terms, Hebbian plasticity encompasses a family of local, activity-dependent learning rules. A canonical phenomenological formulation updates synaptic weights proportionally to the product of pre- and post-synaptic activities, thereby capturing the correlation-based nature of Hebbian learning. This mechanism provides a theoretical explanation for how memory traces can emerge in neural networks: repeated co-activation of groups of neurons leads to the strengthening of their mutual connections \cite{zaccaria2022light}, forming stable assemblies whose activation patterns can later be re-evoked by partial or related inputs. In this sense, memory is often interpreted as being encoded in the distributed pattern of synaptic weights shaped by past activity.

Several mathematically formulated variants of Hebbian learning have been proposed to address stability and normalization issues, including Oja’s rule \cite{oja1982simplified} and spike-timing-dependent plasticity (STDP) models that incorporate temporal correlations between spikes \cite{markram1997regulation,gerstner2014neuronal}. Beyond these two-factor formulations, eligibility-trace and three-factor learning rules introduce an additional modulatory signal, such as a reward, error, or global teaching signal, that interacts with locally accumulated pre- and post-synaptic activity traces to determine the eventual synaptic update \cite{fremaux2016neuromodulated}. This framework provides a useful compromise between local synaptic dynamics and global task-dependent modulation, since the eligibility trace can be computed from local activity while the third factor supplies information that is not locally available at the synapse. Such formulations are particularly relevant to hybrid photonic-electronic learning architectures, where local pre- and post-synaptic processing can be implemented in the physical substrate while a global error or reward signal can be supplied and processed electronically. Consequently, Hebbian learning need not be interpreted as requiring strictly two-factor, fully local weight updates; rather, local plasticity mechanisms can be combined with global modulatory signals to support task-dependent learning and improved stability.

\begin{table}[t]

\caption{Comparison between biological and integrated photonic memory: physical substrate, state variables, and memory mechanisms.}
\label{tab:bio_photonic_memory_physical}
\renewcommand{\arraystretch}{1.2}
\tablealtrows
\begin{tabular}{@{}p{0.22\textwidth} p{0.35\textwidth} p{0.35\textwidth}@{}}
\toprule
Aspect & Biological memory & Integrated photonic memory \\
\midrule
Physical substrate & Neurons, synapses, ion channels, intracellular biochemical pathways, and network connectivity & Optical fields, waveguides, resonators, free carriers, thermal populations, gain media, and non-volatile materials such as phase-change, ferroelectric, ionic, or charge-trapping elements \\

Basic information carrier & Membrane voltage, spike trains, neurotransmitter release, and molecular state & Optical amplitude, phase, wavelength, intensity, carrier density, temperature, and material phase/polarization/charge state \\

Short-term memory mechanism & Persistent neural activity, membrane time constants, short-term synaptic plasticity, and recurrent network dynamics & Delay lines, cavity photon lifetime, free-carrier relaxation, thermo-optic relaxation, gain dynamics, and transient feedback states \\

Long-term memory mechanism & Long-term potentiation/depression, structural synaptic remodeling, engram formation, and protein-synthesis-dependent consolidation & Phase-change materials, ferroelectric domain states, ionic resistive switching, charge-trapping/floating-gate states, and programmed interferometric weights \\

Where memory resides & Distributed across synapses, neurons, and network ensembles & Distributed across device states and circuit dynamics; may also reside in local non-volatile elements \\

Retention / holding condition & Dynamic states require continuing activity and metabolic support; synaptic and structural changes may persist after activity ceases & Passive transient states, power-sustained dynamical states, and zero-holding-power material states coexist \\

Typical timescales & Milliseconds to years & Picoseconds to hours or years, depending on the mechanism \\

Fading memory & Arises from relaxation of membrane and synaptic states; Leaky-integration; recent inputs gradually lose influence & Resonator decay; carrier, thermal and gain relaxation; and filtered or multi-path propagation. An ideal non-dispersive delay provides finite-window rather than fading memory. \\

Persistent memory & Recurrent working-memory states; Stabilized by synaptic efficacy changes and systems consolidation & Stabilized by structural/material hysteresis or trapped charges; can persist without continuous optical drive \\
\botrule
\end{tabular}
\end{table}

\section{Integrated Photonic Memory}\label{sec:memory}

It is useful to compare biological and integrated photonic memory at the level of physical substrate, state variables, and retention mechanisms. \Cref{tab:bio_photonic_memory_physical} summarizes this correspondence and highlights both the analogies and the important physical differences that motivate the classification developed in this section.

\begin{figure}[!htb]
	\centering
	\includegraphics[width=\linewidth]{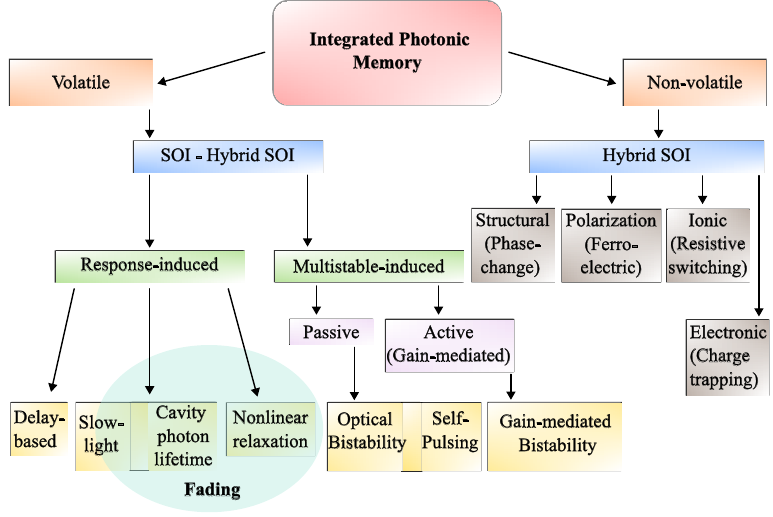}
	\caption{Schematic overview of the proposed classification of integrated photonic memory in silicon photonics. Memory mechanisms are categorized as volatile or non-volatile, with volatile memory further subdivided into response-induced and multistable-induced mechanisms according to the physical process responsible for memory formation. Non-volatile memory is further classified, based on the same principle, into structural, polarization, ionic and charge-trapping memory. Hybrid SOI platforms incorporate additional materials to enable optical functionalities unavailable in silicon. These include phase change materials, bonded III-V semiconductors, optofluids, or nonlinear polymers to enable tunable, active, or reconfigurable photonic responses and to expand the range of mechanisms for implementing memory. The shaded ellipse groups the mechanisms whose memory is fading, i.e. those in which the influence of a past input decays with a characteristic time rather than disappearing abruptly after a fixed time window; delay-based memory lies outside it, while slow-light lies only partly inside. Slow-light memory is associated with the fading-memory regime  in resonator-based implementations, whereas grating-based slow light provides finite-window delay memory.
}
	\label{fig:MemoryClassification}
\end{figure}

In integrated photonics, memory can be classified along at least three complementary axes: physical origin, such as propagation delay, resonant storage, carrier or thermal relaxation, gain dynamics, phase transition, ferroelectric switching, ionic motion, or charge trapping; phenomenology, such as fading, multistable, or persistent behavior; and computational role, such as transient state storage, recurrent-state evolution, or non-volatile parameter storage. In what follows, we use the volatile/non-volatile distinction as the primary organizational principle, while noting that some devices can occupy more than one category depending on operating regime and signal bandwidth. Integrated photonic memory in silicon and hybrid silicon platforms therefore provides both transient and persistent information storage through physical mechanisms that differ in retention time and underlying dynamics. Photonic literature commonly refers to these two regimes as volatile and nonvolatile memory. This taxonomy is adopted in the remainder of this section and is summarized in \cref{fig:MemoryClassification}.

\subsection{Volatile memory}

Volatile photonic memory refers to memory effects that arise from the intrinsic temporal response or dynamical states of the photonic system and naturally decay once the driving field is removed.
In silicon photonics, such memory is enabled by the finite response times of optical, electronic, and thermal degrees of freedom, as well as by the existence of stable or metastable operating states.
Depending on the memory-inducing mechanism, volatile memory can be further divided into response-induced memory and multistable-induced memory.

\subsubsection{Response-induced memory} \label{subsubsec:response_mem}

Photonic response-induced memory arises from the finite temporal response of a system to an optical stimulus.
This includes delay lines, slow-light effects (either material-related or structure-induced dispersion), finite photon lifetimes in optical cavities, and nonlinear relaxation dynamics such as free-carrier or thermo-optic processes. In particular, in time-delay-based memory and slow-light effect systems, the memory is fixed by the group delay $\tau_g$, which is generally defined as:
\begin{equation}\label{eq:groupDelay}
	\tau_g(\omega) = -\dv{\omega} \arg \left[ H_\mathrm{}(\omega) \right],
\end{equation}
where $H_\mathrm{}(\omega)$ denotes the transfer function of the system, while $\omega$ is the angular frequency of the input signal.
The group delay has a direct physical interpretation as a temporal delay only when the phase of the transfer function can be approximated as linear over the signal bandwidth and the amplitude response is approximately flat.

In the case of cavity photon lifetimes or nonlinear relaxation processes, memory manifests as a dependence of the system output on the recent history of the input signal. This often results in dynamical memory effects whose strength is governed by the characteristic relaxation times (e.g. the free carrier lifetime or the temperature relaxation time) or by a characteristic time which is determined by the interplay of different effects and the operating sample rate.

A memory is said to be \emph{fading} when the influence of past inputs decreases with time, not necessarily monotonically, becoming negligible beyond a characteristic time. While related, volatile and fading memory are not equivalent: the former refers to the absence of persistent storage, whereas the latter describes a progressive decay of past input influence. Notably, volatile systems may also exhibit multistable or self-pulsing regimes, where memory does not necessarily fade \cite{lugnan2025reservoir}.

In a dynamical system, fading memory arises from the relaxation of internal state variables toward equilibrium: past inputs leave decaying traces in the field amplitude, the carrier population or temperature, and these traces modulate the current output in a history-dependent manner. While cavity lifetimes lead to linear filtering, nonlinear relaxation processes can induce nonlinear reshaping.

It is useful at this point to explicitly distinguish between delay and fading memories, as the two are often conflated in the reservoir-computing literature. As shown in \cref{fig:MemoryClassification}, delay-based memory lies outside the fading memory regime. An ideal delay element, $y(t)=x(t-\tau)$, provides exact access to a past input sample over a finite time window of duration $\tau$. Once this window has elapsed, the sample leaves the system without being progressively attenuated or mixed with neighboring samples. This \textit{finite-window}, non-fading memory is the regime realized by the non-resonant delay lines described by eqs. (2)–(6): memory capacity as a function of lag is limited by loss, crosstalk, and readout noise, rather than by an intrinsic decay of the delayed signal.

If the delay path introduces significant dispersion, frequency-dependent attenuation, or other filtering, for example, when the higher-order terms $\beta_i(\omega-\omega_0)^{i-1}$, $i>1$, in Eq. (3) become non-negligible, successive samples are reshaped and mixed during propagation. The resulting memory can therefore be regarded as \textit{dispersive} (or \textit{filtered}) delay memory: it is not fading in the strict sense of an exponentially decaying kernel, but it no longer provides a clean sample-and-hold operation. By contrast, \textit{resonant} fading memory results from the finite photon lifetime of a resonant structure, which causes the response to an input to decay while interacting with subsequent inputs. \textit{Material relaxation} memory, instead, originates from the intrinsic relaxation of physical degrees of freedom in the medium, such as free-carrier recombination or thermo-optic relaxation in silicon photonics, independently of resonant enhancement (3.1.1).

These four regimes differ in their memory–bandwidth trade-offs and typical applications: finite-window delay provides a broadband, low-distortion memory buffer; dispersive delay trades bandwidth for a longer effective memory; resonant fading memory enables compact, high-Q memory at the cost of limited bandwidth; and material relaxation provides memory timescales determined by carrier or thermal dynamics, largely independent of device length, but with more nonlinear and less easily invertible dynamics.

Within this framework, the equivalent of a biological working memory can be a volatile memory in integrated photonics, as they both provide temporary storage that is lost after the required holding drive is removed as well as support ongoing computation.

\paragraph{Delay-based memory}
Delay-based memory relies on the propagation delay of light along waveguide paths, i.e. on the time-of-flight of photons.
In this case, the delay is equivalent to the group delay (\cref{eq:groupDelay}). Indeed, the transfer function of a lossless waveguide can be modeled as:

\begin{equation}
    H(\omega)= \exp{-\imath \beta(\omega) L},
\end{equation}where $\beta(\omega)$ is the propagation constant and $L$ is the waveguide length. $\beta(\omega)$ can be expanded as
\begin{equation}
    \beta(\omega)= \beta_0 + \beta_1 (\omega-\omega_0)+
    \beta_2 (\omega-\omega_0)^2+...,
\end{equation}

where $\beta_i$ are the coefficients of the Taylor expansion of $\beta(\omega)$ evaluated at $\omega_0$. Therefore,
it follows that
\begin{equation} \label{eq:T}
	\tau_g = -\dv{\omega}(-\beta(\omega) L) \simeq \beta_1 L = \frac{L}{v_g}.
\end{equation}
\Cref{eq:T} shows that $\tau_g$ is equal to the propagation delay and is proportional to  $L$ and the group velocity $v_g$, provided that $\beta_i(\omega-\omega_0)^{i-1}$ for $i>1$ are negligible compared to $\beta_1$. The delays are often implemented on the PIC as long spirals or compact geometries. In this case, memory arises because the optical signal physically traverses the waveguide; it is not due to resonant photon storage.

\begin{figure}[!htb]
	\centering
	\includegraphics[width=1\linewidth]{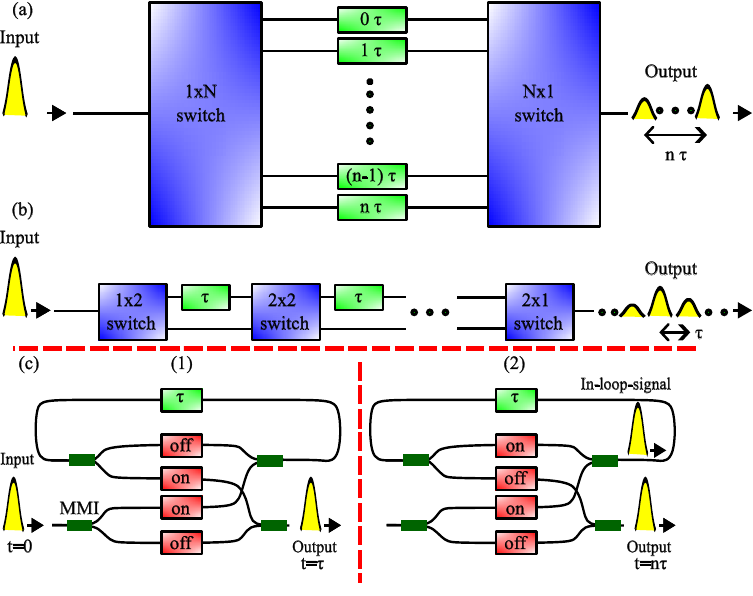}

	\begin{subcaptiongroup}
		\phantomcaption\label{fig:par_ser-a}
		\phantomcaption\label{fig:par_ser-b}
		\phantomcaption\label{fig:par_ser-c}
	\end{subcaptiongroup}

	\caption{%
		Schematic representation of three used architectures for implementing time delays in PICs. \subref{fig:par_ser-a} Parallel, \subref{fig:par_ser-b} serial and \subref{fig:par_ser-c} recirculating-loop optical delay line scheme in (1) bar state and (2) cross state.
		Optical switching in an integrated photonic circuit is typically realized using an interferometric configuration incorporating MMI couplers and phase shifters, enabling active control of the different optical paths.	}
	\label{fig:par_ser}
\end{figure}

\emph{Parallel}, \emph{serial} and \emph{recirculating-loop} optical delay lines \cite{zhou2018integrated} represent three widely used architectures for implementing discrete and reconfigurable time delays in PICs (\cref{fig:par_ser}).
In all schemes, optical switches are employed.
These switches are typically implemented using cascaded MZIs with integrated phase shifters.
More generally, they can be realized through interferometric architectures incorporating multimode interference (MMI) couplers and phase shifters, in which the optical path is actively controlled.

In parallel delay-line architectures (\cref{fig:par_ser-a}), the input optical signal is routed into one of several waveguide paths with different physical lengths, each providing a fixed delay.
As a result, the number of distinct output delay states (or output pulses in response to an impulsive input) is equal to the number of parallel channels. Let us define the achievable delay granularity as the temporal spacing between output pulses for an impulsive input. The delay granularity is here determined by the incremental length difference between adjacent paths \cite{aboketaf2010optical}.

For a system comprising $n$ parallel delay lines each with an incremental delay (see \cref{fig:par_ser-a}), the output field amplitude $\text{y}(t)$ is related to the input signal $\text{x}(t)$ by
\begin{equation}
	\text{y}(t) \propto \sum_{r=0}^{n-1} \text{x}(t - r \tau) \exp(-\alpha  r L) \exp\left( \imath \frac{2\pi n_{\text{eff}} r L}{\lambda} \right),\label{eq:prop}
\end{equation}
where $r L$ and $r \tau$ represent the physical length and the corresponding propagation delay of the spiral indexed by $r$. Typical parameters for silicon waveguides include a loss constant $\alpha$ of approximately \qty{1}{dB \per\cm}, an effective refractive index $n_{\text{eff}}$ of \num{2.5}, and an operating wavelength $\lambda$ of \qty{1550}{\nm}.
When the different signals are summed up in the output port (\cref{eq:prop}), the larger $r L$, the lower the contribution to the output of the $r$th channel, due to propagation loss.

In contrast, serial (or cascaded) delay-line architectures (\cref{fig:par_ser-b}) consist of multiple delay stages connected sequentially, typically controlled by optical switches.
In this case, the delay granularity can increase significantly with only a modest increase in the number of cascaded stages characterized by an appropriate delay time.
These granularity characteristics are particularly attractive for high-resolution and large-delay-range implementations. Granular systems are advantageous with respect to nongranular systems with the same memory extent because they can provide a more tunable delay.

For a configuration of $n$ units in series, the output amplitude relates to the input signal as
\begin{equation}
	\text{y}(t) \propto \sum_{r=0}^{n}
	\binom{r}{n}_{\exp(\imath \phi_i)}
	\text{x}(t - r \tau) \exp(-\alpha r L) \exp\left( \imath \frac{2\pi n_{\text{eff}} r L}{\lambda} \right),
\end{equation}
where the coefficient $\binom{r}{n}_{\exp(\imath \phi_i)}$ represents the collective interference of the $\binom{n}{r}$ possible paths that result in the same total delay. It is defined as a sum of complex phasors, each accounting for the specific phase shifts $(\phi_i)$ introduced by phase shifters or random waveguide imperfections \cite{secondini2003adaptive}.

The recirculating loop delay lines (\cref{fig:par_ser-c}) can be viewed as a loop-based memory element controlled by a tunable coupler \cite{zhou2018integrated}.
The optical signal is injected into the loop waveguide by setting the coupler to the cross state.
Once the coupler is switched to the bar state, the signal is confined and repeatedly propagates within the loop.
To extract the stored data back into the access waveguide, the coupler is returned to the cross state.
Variable time delays are obtained by adjusting the number of round trips performed by the optical signal inside the loop, with the delay granularity defined by the loop round-trip time.
The maximum attainable delay is limited by the accumulated round-trip losses.
It is worth noting that, to ensure proper storage, the temporal duration of the optical packet must be shorter than the loop’s round-trip time minus the switching time from the bar to the cross state.
This may impose a significant constraint on the minimum loop length and implies that delaying an arbitrarily long data stream is not possible.

A notable example of parallel delay-line architecture is provided in \cite{aboketaf2010optical}, where a fully integrated optical time-division multiplexer is realized on a SOI (silicon-on-insulator) platform.
An input pulse train is split into multiple waveguide paths with linearly increasing delays and then recombined to achieve repetition-rate multiplication.
Using this architecture, \qty{20}Gbps and \qty{40}Gbps  signals are generated from a \qty{5}Gbps  input with a compact footprint (\qty{\sim 1}{\mm\squared} ).
The approach is inherently broadband, as the delays are implemented using non-resonant silicon waveguides.

Another example of a parallel delay line is provided by \cite{moralis2017chip}, where three spiral waveguides of different lengths (equivalent to \qtylist{6.5;11.3;17.2}{\ns} delays) are used to implement a variable optical delay bank.
The PIC implements a parallel delay line architecture, where multiple spiral waveguides of different lengths are present.
Each optical packet is routed to a specific spiral based on its wavelength, so while the lines operate in parallel, each packet travels through only one path at a time.
The delay lines are combined with semiconductor optical amplifier coupled MZI (SOA-MZI) wavelength converters to demonstrate error-free optical buffering of \qty{10}Gbps Non-Return-to-Zero (NRZ) optical packets, achieving a high on-chip delay efficiency of \qty{2.6}{\ns\per\mm\squared}.

As an example of serial implementations of delay lines, \cite{horst2013cascaded} demonstrates a serial architecture based on cascaded MZI lattice filters.
The device exploits multiple precisely engineered waveguide delay stages to achieve flat pass-bands, high extinction ratios, and low insertion loss.
Cascaded MZI filters can be considered as a serial delay-line structure, although in \cite{horst2013cascaded}, they are only used as wavelength filters and not for time multiplexing.

One of the earliest demonstrations of an integrated recirculating loop delay line on an SOI platform is reported in \cite{park2008integrated}. The device realizes an optical buffer that combines a silicon delay line with a gate matrix switch, operates error free at \qty{40}Gbps, and provides a packet delay of approximately \qty{1.1}{\ns} within a compact footprint.

Beyond single-stage implementation, parallel, serial, and recirculating architectures can be combined to realize a more complex and recursive form of memory through optical recurrences and on-chip delay lines.
In \cite{van2025real}, an example of such a hybrid delay-line memory is demonstrated within a reservoir computing framework for signal equalization.

An alternative to conventional integrated delay-line networks is the use of programmable waveguide meshes to implement reconfigurable optical delays. In \cite{perez2018programmable}, the optical signal is routed through configurable paths within a hexagonal mesh of tunable basic units, enabling precise control of both the delay magnitude and the optical path geometry. This approach can realize parallel, serial, or hybrid architectures, offering flexible delay-line configurations.

\paragraph{Slow-light-induced memory}
Slow-light-induced memory uses engineered dispersion to reduce the group velocity ($v_g$) of light in waveguides or resonant structures \cite{baba2009dispersion}. This produces a time delay (\cref{eq:T}). Let us define the group index $n_g$:
\begin{equation}
	n_g = c \dv{k}{\omega} = \dv{(n_\text{eff}\omega )}{\omega} = n_\text{eff} + \omega \dv{n_\text{eff}}{\omega},
\end{equation}
where $k=n_\text{eff} \omega /c$ is the wave number. Then
\begin{equation}
	v_g = \frac{1}{\beta_1} = \frac{c}{n_g}.
\end{equation}

The slow-light effect reduces $v_g$ by increasing $n_g$. The quantity $n_g$ is primarily enhanced through the increase of the first-order dispersion ($\dv*{n_\text{eff}}{\omega}$).
Dispersion is engineered via tailoring the material properties to yield, e.g., narrowband spectral gain or loss. In fact, through the Kramers–Kronig relations \cite{hutchings1992kramers,lucarini2005kramers,de1926theory,kramers1928diffusion}, sharp spectral features in gain or loss induce steep variations in the real part of the refractive index.
Such narrow spectral features can also arise in optical fibers from nonlinear effects such as stimulated Brillouin scattering (SBS) \cite{thevenaz2008slow}.
The acoustic wave generated by SBS creates a dynamic Bragg grating that diffracts light from the higher-frequency pump to the lower-frequency Stokes wave.
Using a 2-meter-long fiber, $n_g$ can be increased to approximately \num{4.26} \cite{gonzalez2005optically}, corresponding to a nearly threefold slowdown of light.

Otherwise, narrow spectral features can be caused by interference phenomena like electromagnetically induced transparency (EIT) \cite{harris1997electromagnetically}.
EIT arises when narrow interfering excitation channels suppress absorption of a broadband resonance, creating a transparency window in the spectrum.
This phenomenon has been observed in atomic vapors, cold atomic gases, as well as in photonic systems such as MRRs coupled to external cavities \cite{zhang2016electromagnetically}, coupled photonic-crystals cavities \cite{yang2009all}, or Side-Coupled Integrated Spaced Sequence Of Resonators (SCISSORs) \cite{xu2006experimental}. SCISSORs consist of a series of $n$ MRR coupled to side waveguides \cite{mancinelli2011optical}.

In atomic vapor \cite{boller1991observation}, EIT occurs when a control laser couples a metastable level to an excited state (the interfering excitation channel), which renders the medium transparent to a probe beam operating at the broader resonance.
In \cite{hau1999light}, the speed of light in an ultracold atomic gas reaches values as low as \qty{17}{\m\per\s} by exploiting EIT.

\begin{figure}[!htb]
	\centering
	\includegraphics[width=1\linewidth]{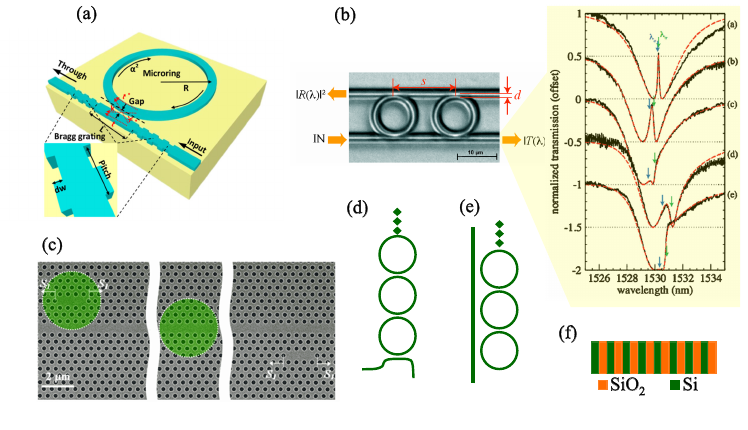}

	\begin{subcaptiongroup}
		\phantomcaption\label{fig:slow-light_examples-a}
		\phantomcaption\label{fig:slow-light_examples-b}
		\phantomcaption\label{fig:slow-light_examples-c}
		\phantomcaption\label{fig:slow-light_examples-d}
		\phantomcaption\label{fig:slow-light_examples-e}
		\phantomcaption\label{fig:slow-light_examples-f}
	\end{subcaptiongroup}

	\caption{%
		Structures achieving slow light by means of
		\subref{fig:slow-light_examples-a} a MRR coupled to a Bragg grating cavity \cite{zhang2016electromagnetically},
		\subref{fig:slow-light_examples-b} two coupled MRRs in a Side-Coupled Integrated Spaced Sequence of Resonators (SCISSOR) configuration where the yellow box shows the narrow transparency peak which appears within the broad coupled MRR resonance \cite{xu2006experimental},
		\subref{fig:slow-light_examples-c} two coupled photonic crystal cavities \cite{yang2009all},
		\subref{fig:slow-light_examples-d} a series of coupled-resonator optical waveguide (CROW) configurations, \subref{fig:slow-light_examples-e} a series of coupled MRRs in an all-pass filter (APF), and \subref{fig:slow-light_examples-f} a grating waveguide delay line on SOI. \subref{fig:slow-light_examples-a} is adapted with permission from  \cite{zhang2016electromagnetically} $\copyright$ Optical Society of America. \subref{fig:slow-light_examples-b} is reprinted with permission from \cite{xu2006experimental} $\copyright$ (2006) by the American Physical Society. \subref{fig:slow-light_examples-c} is reprinted with permission from \cite{yang2009all} $\copyright$ (2009) by the American Physical Society.}
	\label{fig:slow-light_examples}
\end{figure}

In a single MRR coupled to a Bragg grating (\cite{zhang2016electromagnetically}, \cref{fig:slow-light_examples-a}), EIT arises from destructive interference between the resonant mode of the ring and the diffracted wave in the Bragg grating. This interference causes the formation of a narrow transparency window in the transmission spectrum.
In a SCISSOR formed by two MRR (\cref{fig:slow-light_examples-b}), interference occurs between different resonant pathways: the single ring resonance and the waveguide-ring-waveguide-ring resonance. Indeed EIT in SCISSOR is properly named coupled resonator induce transparency (CRIT) and is observed also in longer SCISSORs (eg. an eight-MRR SCISSOR in \cite{mancinelli2011optical}). CRIT results in a slow-light transparency window \cite{xu2006experimental}.
EIT can also be found in coupled photonic crystal cavities (\cite{yang2009all}, \cref{fig:slow-light_examples-c}), where the standard excitation channel is the direct coupling of light from the waveguide into a cavity and back into the waveguide.
The interfering pathway is the indirect coherent coupling through the other cavity (or cavities), which interferes destructively with the standard path in the waveguide.

Slow light can also be engineered using optical structures, without the need for EIT.
These structures include coupled MRRs \cite{xia2007ultracompact}, photonic crystals \cite{johnson2000linear,jamois2003silicon, vlasov2005active, krauss2007slow} and grating delay lines \cite{jean2019slow}.
Sketches of different kinds of systems that achieve slow-light can be found in \cref{fig:slow-light_examples}.
The rapid variation of the optical phase near resonant features or band edges leads to a large derivative $\dv*{\phi}{\omega}$.
Since the accumulated phase satisfies $\phi = kL$ (where $L$ denotes the physical optical path length), the group velocity is given by
\begin{equation}
	v_g = \dv{\omega}{k} = \frac{L}{\dv*{\phi}{\omega}}.
\end{equation}
Therefore, steep phase--frequency dependence, corresponding to a large value of $\dv*{\phi}{\omega}$, results in an enhanced group index and slow-light effects.

More generally, in photonic periodic structures, slow light arises from the strong modification of the dispersion relation induced by periodicity of the refractive index, which leads to band gaps, band edges, or mode anti-crossings where the phase varies rapidly with frequency.
However, the enhancement of the group index in slow-light waveguides is generally accompanied by increased optical absorption and scattering losses, since disorder-induced scattering and material absorption are effectively amplified as the optical energy density and interaction time increase \cite{o2010loss}.

An illustrative example of a coupled-resonator architecture used to engineer slow-light propagation is reported in \cite{xia2007ultracompact}.
The authors investigate structures composed of single-MRRs in a CROW architecture (\cref{fig:slow-light_examples-d}) or arranged in an all-pass filter (APF) configuration (\cref{fig:slow-light_examples-e}), with each ring side-coupled to a common bus waveguide. In the latter architecture, a single ring exhibits low insertion loss, a high quality factor, and a substantial group delay at resonance, corresponding to approximately 55 roundtrips of light circulating inside.
In their study, 56 rings are combined, leading to a broadened overall transmission spectrum primarily due to small random variations in the rings' perimeters or waveguide widths, rather than variations in coupling.
An on-resonance group delay of \qty{510(10)}{\ps} is measured, using the APF response \qty{2}{\nm} off-resonance as a reference.

Building on this work, \cite{xie2014continuously} introduces a significant improvement by implementing a reflective-type optical delay line based on an APF structure of 13 silicon MRRs terminated with a Sagnac loop reflector.
Unlike the previous APF configuration, in this design light passes through each MRR twice, effectively doubling the delay while reducing the number of resonators required for a given buffering capacity.
Moreover, each MRR integrates a lateral p-i-p junction acting as a resistive micro-heater. This enables continuous tuning of the resonance wavelengths via the thermo-optic effect, which mitigates the uncertainty in the resonance positions present in the previous APF.
The fabricated device achieves a maximum group delay of \qty{110}{\ps} with a bandwidth of approximately \qty{168}{\GHz}, while the delay can be continuously tuned from \qty{10}{\ps} to \qty{110}{\ps} with a power efficiency of \qty{0.34}{\ps\per\mW}.
Experimental verification using \qty{20}Gbps PRBS signals demonstrates preserved signal fidelity after the applied delays, highlighting the potential of this compact, tunable on-chip optical buffer and the clear advancement over the original APF approach.
Note that the same structure may exhibit pure cavity photon-lifetime memory, which explains its overlap with slow-light behavior in the classification scheme of \cref{fig:MemoryClassification}. This occurs when the input bandwidth is sufficiently large so that the phase response becomes nonlinear with respect to $\omega$, while the amplitude response is no longer uniform across the spectrum.
In other words, if the input signal changes too rapidly in time, the cavity cannot respond instantaneously, causing successive samples to influence each other rather than being rigidly shifted back in the time domain \cite{biasi2019time}.
This fading memory scales with the typical photon lifetime and is therefore related to the energy stored in the cavity, which in turn depends on the quality factor. In silicon MRRs with a loaded quality factor $Q_L$ \num{1e4} at \qty{1550} nm, the photon lifetime (energy-decay time) is $\tau_E=Q_L/\omega_0\simeq$\qty{8.2}{\ps}. It is worth noting that this fading memory corresponds to the resonant fading regime introduced above: unlike the finite-window or purely dispersive delay of a non-resonant waveguide, the system output at a given time is here a weighted superposition of many past inputs, with exponentially decreasing weights determined by the cavity's energy decay time.

For photonic crystal implementations of slow-light, a notable example is presented in \cite{vlasov2005active}.
Here, a photonic crystal waveguide, formed by a periodic triangular lattice of holes in a suspended silicon membrane, enables an extremely compact MZI footprint of \qty{0.04}{\mm\squared}.
Near the photonic band-edge, the group index increases by more than two orders of magnitude, reaching values above \num{300}, and active thermo-optic tuning allows practical control of the group index from about \num{20} to \num{60} with only \qty{2}{\mW} of electrical power.
Similarly, the microgear photonic crystal ring in \ce{Si3N4} \cite{lu2022high}, a one-dimensional photonic crystal implemented via periodic modulation of a MRR's inner boundary, achieves slow-light modes with record-high Q values exceeding \num{1e6}, strongly localizing the optical field.
In both cases, the combination of slow light and structural design leads to enhanced light–matter interactions in ultra-compact footprints, concentrating optical power where it is most needed for efficient nonlinear optical effects.

Another photonic structure that achieves slow-light is the grating delay line, a structure in which the optical properties, such as the refractive index, vary periodically across its surface.
In silicon photonics, gratings are created by periodically modulating the effective refractive index of a waveguide. Several grating-based structures have been demonstrated on silicon photonic platforms for use in optical delay lines \cite{wang2012narrow,jean2019slow}.
An example is reported in \cite{jean2019slow}, where it is demonstrated that structural slow light in subwavelength grating waveguides on SOI can drastically reduce the group velocity, achieving group indices up to about \num{47.7} and a significant group index over a broad optical bandwidth.
However, this slow-light regime is accompanied by increased propagation losses.

A propagation loss of \qty{\sim 12}{\dB\per\mm} was measured, due to enhanced interaction with imperfections and scattering, almost an order of magnitude larger than the long-wavelength reported loss of \qty{\sim 2.5}{\dB\per\mm}.
By varying the grating period, the position of the band‑edge and thus the slow‑light operating region can be shifted over a wide wavelength range, providing tunability of the slow‑light response in the devices.

Besides conventional resonance-based slow light, it is possible to achieve slow light topologically. In \cite{he2025valley}, by placing two dielectric photonic crystals with a mirror-symmetric interface, valley-Hall edge states appear, confined along the interface because the bulk prohibits propagation at their frequencies.
These states can reach very high group indices (\num{1000}, \num{1378}) at two frequency bands and are robust and broadband, unlike narrowband resonant or band-edge schemes.

\paragraph{Nonlinear relaxation–induced memory}
Nonlinear relaxation–induced memory arises from light–matter interaction inside resonant cavities, such as MRRs.
The optical energy stored in a cavity is governed by its quality factor Q, which determines both photon lifetime and field enhancement.
High-Q cavities enable strong intracavity field buildup even for low input powers, thereby enhancing nonlinear effects \cite{borghi2017nonlinear}.
Consequently, even at relatively low optical power, the polarization vector $\mathbf{P}$ is no longer proportional to the optical field $\mathbf{E}$, but is instead expressed as \cite{boyd2008nonlinear}:
\begin{equation}
	\mathbf{P} = \varepsilon_0 \left( \chi^{(1)} \mathbf{E} + \chi^{(2)} \mathbf{E}^2 + \chi^{(3)} \mathbf{E}^3 + \dots \right),
\end{equation}
where $\varepsilon_0$ is the vacuum permittivity and $\chi^{(i)}$ denotes the $i$-th order susceptibility.
In silicon photonics, due to the centrosymmetry of silicon, native second-order nonlinearities are absent ($\chi^{(2)}=0$) \cite{leuthold2010nonlinear}.
As a result, nonlinear effects are governed by the third-order susceptibility, while higher-order contributions are relevant only at very high optical intensities.

\begin{figure}[!htb]
	\centering
	\includegraphics[width=1\linewidth]{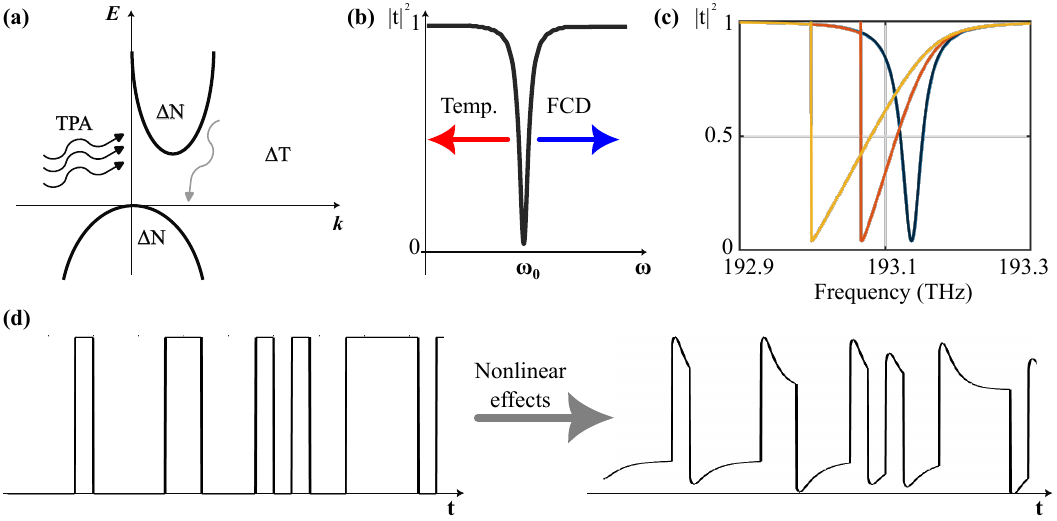}

	\begin{subcaptiongroup}
		\phantomcaption\label{fig:Non_ser-a}
		\phantomcaption\label{fig:Non_ser-b}
		\phantomcaption\label{fig:Non_ser-c}
        \phantomcaption\label{fig:Non_ser-d}

	\end{subcaptiongroup}

	\caption{%
		    Nonlinear effects and fading memory in a silicon MRR. \subref{fig:Non_ser-a} Two-photon absorption (TPA) of the cavity modes leads to the generation of free carriers ($\Delta N$), as illustrated by the energy–momentum diagram for indirect band-gap silicon. The relaxation of these carriers results in an increase in the effective cavity temperature ($\Delta T$). Both $\Delta N$ and $\Delta T$ feed back into the system, modifying the cavity modes. \subref{fig:Non_ser-b} Shift of the cavity resonance frequency ($\omega_0 (2\pi)$) toward shorter wavelengths due to free-carrier effects and toward longer wavelengths due to thermal effects. $|t|^2$ is the MRR transmission intensity. \subref{fig:Non_ser-c} Transmittance as a function of the input frequency sweep (from higher to lower $\omega$). The nonlinearity induces a power-dependent resonance shift, resulting in a triangular shape for the transmission. \subref{fig:Non_ser-d} Fading-memory effect induced by the cavities: the system is driven by a sequence of input samples (rectangular pulses of different durations) at a rate matching the nonlinear timescales (left). This produces an output  that retains a memory of past inputs (right). }
	\label{fig:Non_ser}
\end{figure}

The third-order susceptibility is complex, $\chi^{(3)} = \chi_R^{(3)} + \imath \chi_I^{(3)}$, giving rise to a variety of nonlinear phenomena.
The real part, $\chi_R^{(3)}$, produces the optical Kerr effect, resulting in an intensity-dependent refractive index change, while the imaginary part, $\chi_I^{(3)}$, governs nonlinear absorption processes, primarily two-photon absorption (TPA) \cite{leuthold2010nonlinear, nikitin2021carrier}.
TPA occurs when two photons are absorbed simultaneously, with their combined energy sufficient to excite an electron from the valence band to the conduction band (see \cref{fig:Non_ser-a}).
This generates free carriers, which in turn lead to free-carrier absorption (FCA) and free-carrier dispersion (FCD), modifying the imaginary and real parts of the refractive index, respectively.
Absorption and carrier thermalization heat the waveguide, causing an additional refractive index change through the thermo-optic (TO) effect \cite{borghi2017nonlinear}. Overall, $n_\mathrm{eff}$ in a MRR becomes complex and intensity-dependent:
\begin{equation}
	n_\mathrm{eff}(I) = n_\mathrm{eff}(0) + n_2 I + \delta n_\mathrm{FCD}(\Delta N) + \delta n_\mathrm{TO}(\Delta T)
	- \imath  \frac{\lambda}{4 \pi} \left( \alpha + \beta_\mathrm{TPA} I + \alpha_\mathrm{FCA}(\Delta N) \right),
\end{equation}
where $n_2 I$ represents the Kerr-induced intensity-dependent index change, $\delta n_\mathrm{FCD}(\Delta N)$ accounts for FCD, and $\delta n_\mathrm{TO}(\Delta T)$ captures the thermo-optic effect. $\Delta N$ and $\delta T$ are the increase in free carrier concentrations and in the temparature, respectively. The imaginary part includes the linear propagation loss $\alpha$, the TPA-induced absorption $\beta_\mathrm{TPA} I $ associated with $\chi_I^{(3)}$, and the FCA loss $\alpha_\mathrm{FCA}(\Delta N)$ due to carriers generated by TPA.
The interplay of Kerr, TPA, free-carrier, and thermo-optic effects gives rise to a nonlinear fading memory characteristic of these resonant systems \cite{biasi2024photonic}.

Along with the quasi-instantaneous Kerr effect, FCD and TO nonlinearities determine a resonance shift of the cavity by $n_\mathrm{eff}$ modifications.
They contribute differently to the cavity’s resonant frequency shift: FCD shifts the resonance toward the blue, while TO shifts it toward the red (see \cref{fig:Non_ser-b}).
Notably, these nonlinearities are characterized by different relaxation times: the thermal relaxation time $\tau_\mathrm{th}$ and the free-carrier lifetime $\tau_\mathrm{fc}$.
Moreover, they exhibit distinct dependencies on the intracavity field amplitude. Specifically, $\Delta N$ scales with the square of the field intensity, as it is generated via TPA, whereas $\Delta T$ depends linearly on the absorbed optical power and is thus directly proportional to the field intensity \cite{borghi2017nonlinear}.
It is worth noting that the Kerr response in silicon is maximal for input wavelength around \qtyrange{1.8}{1.9}{\um}, while TPA is significantly reduced for wavelengths exceeding \qty{2}{\um}, beyond the half-bandgap threshold \cite{leuthold2010nonlinear}.

At \qty{1.5}{\um} where silicon photonics is technologically mature, TPA remains pronounced, giving rise to FCA and FCD that limit purely Kerr-based nonlinear performance.
Here, at high optical powers, the transmission spectrum of a cavity deviates from a simple Lorentzian shape.
The nonlinearities induce a power-dependent resonance shift, resulting in a distorted, typically triangular, line shape in transmission (see \cref{fig:Non_ser-c}) \cite{priem2005optical, chen2012bistability}.
A sudden jump in transmission immediately after the resonance is a hallmark of optical bistability (see\cref{subsubsec:multistable_mem}) \cite{almeida2004optical}.

In silicon MRRs, $\tau_\mathrm{th} \simeq \qtyrange{60}{150}{\ns}$, while $\tau_\mathrm{fc} \simeq \qtyrange{1}{10}{\ns}$ \cite{van2012simplified}.
However, variations in the fabrication process or different fabrication runs can lead to different values (e.g., $\tau_\mathrm{th} \simeq \qty{280}{\ns}$ and $\tau_\mathrm{fc} \simeq \qty{45}{\ns}$ \cite{borghi2021modeling}).
The fading-memory effect induced by MRRs becomes evident when the system is excited with a time sequence of rectangular pulses (samples) at a rate that matches the characteristic timescales of the nonlinear dynamics.
Depending on the spacing between consecutive pulses and the typical timescales of the free-carrier and thermal responses, a given sample can be influenced by different numbers of preceding samples in the past (see \cref{fig:Non_ser-d}).
Thus, these nonlinearities allow the fast optical dynamics to be extended from picoseconds (linear response) up to microseconds, enabling interactions on timescales compatible with a broader range of physical phenomena.
As we will discuss in detail in \cref{sec:PNNs}, this type of memory has been leveraged to implement PNNs \cite{denis2018all, borghi2021reservoir, giron2024effects, donati2024time}. It is important to note that not only is memory generated, but a nonlinearity is also applied to the raw signal. As a result, these two roles become intertwined, making it difficult to disentangle their respective contributions within a network architecture. This is further complicated by the fact that the readout is often based on optical intensity, which inherently introduces an additional nonlinear transformation (square-law detection) \cite{bazzanella2022microring}.

In hybrid platforms, the nonlinearities of silicon, such as $\chi^3$ and carrier/thermal dynamics, are complemented by high-$\chi^2$ materials (e.g., electro-optic polymers). In this case, ultrafast $n_\mathrm{eff}$ modulation are possible via the Pockels effect, with speeds up to \qty{100}{\GHz} \cite{alloatti2014100}.
However, to the best of our knowledge, there are no reports that exploit $\chi^2$ to achieve fading memory in silicon photonics.
Similarly, even if heterogeneous integration of III–V materials with silicon is possible, there are currently no published experimental demonstrations showing that such hybrid integrations provide fading memory in PNN.

\subsubsection{Multistability as a memory mechanism} \label{subsubsec:multistable_mem}

Multistable-induced memory relies on the existence of two or more stable or metastable states in a dynamical system. In this case, information is stored in the operating point of the system rather than in its transient dynamics.
In silicon photonics, we can have passive and active systems, encompassing nonlinear optical bistability as well as memory schemes based on injection locking and master–slave configurations, enabled by heterogeneous integration of silicon with III–V gain materials.

In multistable systems, stationary operating points correspond to fixed points in phase space, an abstract mathematical space in which the state of a dynamical system is fully described by a set of variables \cite{gibbs1902elementary, nolte2010tangled}.
Each point in phase space represents a unique state of the system, defined by its dynamical degrees of freedom.
\emph{Stable} fixed points act as attractors of trajectories, toward which the system evolves under small perturbations, while \emph{unstable} fixed points behave as repellers, usually separating distinct basins of attraction.
There are also \emph{saddles}, which are unstable points with at least a stable direction.
In addition to fixed points, the system may exhibit \emph{limit cycles}, another kind of attractor, corresponding to stable periodic orbits in phase space and associated with sustained oscillatory behavior.
To show non-fading memory, a system needs to have more than one stable equilibrium point or, more generally, multiple attractors, such as limit cycles and stable equilibrium points.

\begin{figure}
	\centering

	\begin{subcaptiongroup}
		\phantomcaption\label{fig:2_level_system-a}
		\phantomcaption\label{fig:2_level_system-b}
		\phantomcaption\label{fig:2_level_system-c}
		\phantomcaption\label{fig:2_level_system-d}
		\phantomcaption\label{fig:2_level_system-e}
	\end{subcaptiongroup}

	\includegraphics[width=1\linewidth]{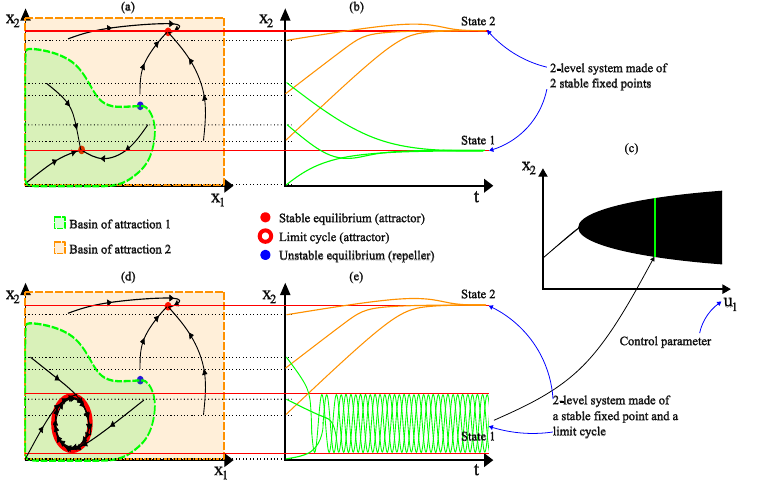}
	\caption{Phase-space dynamics, basins of attraction, and bifurcation-induced transitions in two-state nonlinear systems. The system is described by two variables $x_1$ and $x_2$.  \subref{fig:2_level_system-a}, \subref{fig:2_level_system-d} Bi-dimensional phase space containing attractors (red), repellers (blue), and trajectories (black). Basins of attraction are shown in green and orange. Different trajectories are shown which start at different initial conditions \subref{fig:2_level_system-b}, \subref{fig:2_level_system-e} Temporal evolution of the variable $x_2$ for different initial conditions. \subref{fig:2_level_system-c} Topological change of an attractor from a stable fixed point to a limit cycle by slowly varying a control parameter ($u_1$).}
	\label{fig:2_level_system}
\end{figure}

\Cref{fig:2_level_system} shows an example of a bistable system composed of states 1 and 2 and characterized by two variables $x_1$ and $x_2$. \Cref{fig:2_level_system-a,fig:2_level_system-d} report the phase space, while \cref{fig:2_level_system-b,fig:2_level_system-e} show the corresponding temporal evolutions of the system output variable. Finally, \cref{fig:2_level_system-c} shows an example of bifurcation where the attractor turns from a stable fixed point to a limit cycle.
The state of the system at a given time depends on its initial conditions, or more precisely, on the history of the system.

As an illustrative example, let us consider a system with two stable fixed points, which we label as fixed point 1 and fixed point 2 (\cref{fig:2_level_system-a,fig:2_level_system-b}). If the initial condition is set sufficiently close to the fixed point 1, the system trajectory (green lines in \cref{fig:2_level_system-b}) will evolve around this equilibrium, and the system state can be identified as state 1. Conversely, if the initial condition is chosen near the fixed point 2, the dynamics will remain in the neighborhood of that equilibrium (orange lines in \cref{fig:2_level_system-b}), and the system state will be identified as state 2.

Different equilibrium states are generally characterized by different values of the system variables, i.e., the phase-space coordinates, which evolve according to the system equations.
One of these variables may represent, for instance, the energy stored inside a laser cavity.
As a consequence, distinct equilibria typically correspond to different values of measurable output quantities, such as the output optical power of a laser.
In practice, it is highly unlikely for two distinct equilibria to yield exactly the same measurable output.
Among the different dynamical regimes, a peculiar example is shown in \cref{fig:2_level_system-d,fig:2_level_system-e}, characterized by a phase space featuring a stable fixed point and a limit cycle.
Also in this case, depending on the initial conditions or on the system history, the dynamics may converge either to the stable state 2 (see orange curves in \cref{fig:2_level_system-e}) or be captured by the limit cycle, enabling self-sustained oscillations in state 1 (see green curves in \cref{fig:2_level_system-e}).
Furthermore, by plotting the system output variable versus a control parameter (\cref{fig:2_level_system-c}), one can see an example of how a stable fixed point after a threshold value of $u_1$ transforms into a limit cycle, where the system repeatedly visits a range of values in the $x_2$ dimension (vertical black lines for each value of $u_1$ after threshold).

For the system to function as a memory, however, a mechanism must exist to induce transitions between different equilibria, thereby allowing the system state to be switched.
A commonly adopted approach consists of applying a transient perturbation (or impulse) to one of the control parameters, such as the input power of the laser.
This perturbation temporarily alters the system’s possible trajectories and, consequently, shifts the position of the equilibria in the phase space or destroys one or more of the equilibria in the phase space. As a result, the system trajectory starts evolving toward the new equilibrium configuration.

When the perturbation is removed, the control parameters return to their original values, and so do the corresponding equilibrium positions.
At that moment, however, the system state does not necessarily coincide with any of the original equilibria; instead, it lies within the basin of attraction of one of them.
If the applied impulse has been such that the system is driven into the basin of attraction of the second equilibrium, the system will relax toward that, thus completing a state transition and effectively switching the memory state.

This mechanism can be extended for a system with multiple attractors, and it is more general than a simple hysteresis cycle.
Nevertheless, for specific classes of systems, it may give rise to the standard hysteretic behavior, since two equilibrium states can coexist for the same values of the control parameters.

\paragraph{Dynamical systems}
In this paragraph, we briefly introduce the mathematical framework that stands behind and generalizes the example shown in \cref{fig:2_level_system}.
Within the theory of dynamical systems \cite{strogatz2024nonlinear}, it is typical to consider a system described by
\begin{equation}
	\dot{\vb{x}} = \vb{f}(\vb{x}, \vb{u}),
\end{equation}
where $\vb{x}$, usually a vector, denotes the system state, $\vb{u}$ is a set of control parameters and $\vb{f}$ is a set of $n$ functions.

Due to the presence of nonlinearities, the equation $\dot{\vb{x}}=0$ may admit multiple stationary solutions for the same value of $\vb{u}$.
Note that $\dot{\vb{x}}$ is a set of $n$ derivatives, and when they are all set to zero, they give rise to $n$ equations that can be represented as $n-1$ dimensional manifolds in the phase space.
These manifolds are called nullclines, and the intersections of all nullclines in the phase space are the equilibrium points.
Each stationary solution corresponds to a possible equilibrium state of the system.

The stability of each equilibrium can be systematically analyzed by linearizing the system around that equilibrium
$\vb{x}_0$ and computing the Jacobian matrix:
\begin{equation}
	J_{ij}(\vb{x}_0) = \eval{\pdv{f_i}{x_j}}_{\vb{x} = \vb{x}_0}.
\end{equation}
The complex eigenvalues of $\vb{J}(\vb{x}_0)$ determine the local behavior of trajectories near the equilibrium: if all eigenvalues have negative real parts, the equilibrium is stable; if any eigenvalue has a positive real part, it is unstable.
Moreover, the way these eigenvalues change as the control parameters $\vb{u}$ are varied determines the type of bifurcation the system undergoes, signaling the creation, destruction, or qualitative change of equilibria.

As the control parameters $\vb{u}$ are varied, the number and nature of the stable equilibria may change.
In particular, parameter variations can induce the appearance or disappearance of stable states, enabling transitions between different dynamical regimes.
This parameter-dependent multiplicity of stable states provides a dynamical framework that describes multistability-based memory operation.

\subsubsection{Passive and active implementations} \label{subsubsec:multistable}

Optical multistabilities in integrated photonics can be broadly classified into \emph{passive} and \emph{active}. A passive multistability arises intrinsically from the nonlinear response of the system, allowing at least two stable steady states under identical external conditions. In contrast, active multistability relies on external gain or control signals to sustain stability and regulate transitions between these states.

\paragraph{Passive optical multistability}

\begin{figure}[!htb]
	\centering
	\includegraphics[width=1\linewidth]{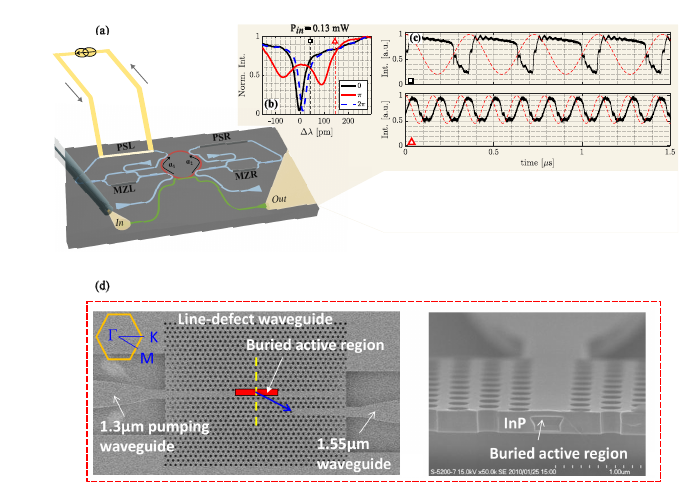}
	\begin{subcaptiongroup}
		\phantomcaption\label{fig:nonlinear_bistability-a}
		\phantomcaption\label{fig:nonlinear_bistability-b}
		\phantomcaption\label{fig:nonlinear_bistability-c}
        \phantomcaption\label{fig:nonlinear_bistability-d}
	\end{subcaptiongroup}

	\caption{
		Passive and active optical bistability implementations.
        \subref{fig:nonlinear_bistability-a} Scheme of a Dynamically Reconfigurable Unified Microring Resonator (DRUM).
		Labels refer to phase shifter left/right (PSL/PSR), MZI left/right (MZL/MZR), and cavity energy field propagating clockwise/counterclockwise ($\alpha_\text{1}$/$\alpha_\text{2}$). \subref{fig:nonlinear_bistability-b} Low-power linear transmission spectra for a phase imprinted by PSL of $0,\pi,$ and $2\pi$ as a function of the wavelength detuning $\Delta\lambda$ with respect to the cold resonance. The input power is $P_{in}=\qty{0.13}{\mW}$. \subref{fig:nonlinear_bistability-c} In black, examples of different self-pulsing states. In red, the first harmonic of the self-pulsing trace. \subref{fig:nonlinear_bistability-a}, \subref{fig:nonlinear_bistability-b}, \subref{fig:nonlinear_bistability-c} are adapted from \cite{biasi2025phase} \href{https://creativecommons.org/licenses/by/4.0/}{CC BY 4.0}.
		\subref{fig:nonlinear_bistability-d} SEM images of the buried heterostructure photonic crystal (BH-PhC) laser, showing the top view on the right and the cross-sectional view along the yellow dashed line on the left \cite{chen2011all}. \subref{fig:nonlinear_bistability-d} is adapted with permission from \cite{chen2011all} $\copyright$ Optical Society of America.}
	\label{fig:nonlinear_bistability}
\end{figure}

In MRRs, passive optical multistability typically originates from the interplay between cavity feedback and nonlinear refractive index changes, resulting in hysteretic input–output characteristics, without the need for active elements.
Optical bistability is experimentally demonstrated in simple silicon MRRs \cite{tanabe2005fast,xu2006carrier,nikitin2022optical}, making them potential candidates for photonic memory elements.
The bistable response is generally controlled by varying the input laser power or detuning with respect to the cold MRR resonance. This provides limited flexibility and robustness for practical memory operation. Recently, a novel MRR design, called Dynamically Reconfigurable Unified Microresonator (DRUM), has been introduced (\cref{fig:nonlinear_bistability-a}) \cite{aslan2025coherent,biasi2025phase}.
The DRUM  enables controlled coupling of the clockwise (CW) mode and of the counter-clockwise (CCW) mode which propagates in a MRR. The complex coupling coefficient is controlled via a sequence of a MZI and phase shifters driven by injected currents. By using the current as a control parameter, hysteresis can be induced by annihilating the stable fixed point at which the system resides with a saddle fixed point. This leads to the disappearance of the stable state and triggers a transition to a limit cycle at a different position in the phase space \cite{biasi2025phase}. \Cref{fig:nonlinear_bistability-b} shows the low-power linear transmission spectra for phase shifts of $0$, $\pi$, and $2\pi$ applied by the left phase shifter.

Another intriguing phenomenon arises from the interplay of Kerr, TPA, FCD, and TO effects and is commonly referred to as self-pulsing oscillation \cite{johnson2006self, pavesi2021thirty}.
As shown in \cref{fig:nonlinear_bistability-c}, in this regime, a continuous-wave input signal with a frequency near the MRR's resonance is converted into a periodically oscillating output.
Whether self-pulsing constitutes a memory mechanism, and whether that memory fades, depends on which dynamical regime is realized. Reusing the phase-space language introduced above (section 3.1.2), at least four situations should be distinguished: (i) a single stable limit cycle, which by itself provides no multistable memory, since there is only one accessible attractor, although the transient relaxation towards the cycle still carries fading memory of recent inputs; (ii) coexistence of a stable fixed point and a limit cycle (\cref{fig:2_level_system-d,fig:2_level_system-e}), where the system can be switched between a static and an oscillatory state; (iii) coexistence of multiple stable limit cycles, where information is stored in the discrete choice of attractor basin and is, in this sense, genuinely non-fading, limited mainly by noise-induced escape over the basin boundary; and (iv) a single limit cycle along which information is encoded as a retained phase offset. Case (iv) is qualitatively different from cases (ii)-(iii): because phase along a limit cycle is a neutrally stable (non-restoring) direction, a retained phase offset is not protected by an attractor boundary, and its lifetime is set by the accumulation of phase diffusion and noise rather than by a basin-escape process. It is therefore not equivalent to storage in separate attractor basins, and should not automatically be classified as non-fading memory.

In the self-pulsing regime, the temporal dynamics of the output are directly linked to the multiple parameters involved, such as the temperature, the input
signal phase, and the free-carrier population.
When one or more of these parameters are perturbed, the dynamics can experience slight variations or change dramatically, depending on the type of evolution taking place.
As we will discuss in detail in \cref{sec:PNNs}, self-pulsing oscillations can be used to realize persistent volatile memory maintained by a continuous optical pump in SCISSOR architectures \cite{biasi2024exploring} or in a 8 $\times$ 8 array of coupled MRRs \cite{lugnan2025reservoir}.
The persistent memory observed in \cite{biasi2024exploring} can be interpreted in terms of the system’s dynamics being trapped in different attractors.
In particular, excitation pulses of different types, superimposed on a constant input, drive the system state toward one of the several limit cycles in the phase space.
After the excitation ends, the system can relax into a new limit cycle or into the same limit cycle but with a modified phase, depending on the characteristics of the excitation.
The former corresponds to case (iii) above and is the regime typically accessed in systems of multiple coupled MRRs, as in \cite{biasi2024exploring}; the latter corresponds to case (iv), and its retention is accordingly expected to be limited by phase diffusion rather than by attractor stability. The large coupled-MRR network of \cite{lugnan2025reservoir} similarly relies on multiple accessible dynamical states of the network. However, the complete attractor landscape was not experimentally reconstructed, and the precise distinction between cases (ii) and (iii) remains open.
A systematic analysis of the number of distinct limit cycles accessible in the phase space of such photonic architectures could therefore provide an estimate of the number of stable states that the system can support.

In silicon MRRs, the self-pulsing effect leads to temporal oscillations of the signal ranging from a few megahertz to several tens of megahertz, as it originates from the competition between thermal and free-carrier effects.
The oscillation frequency can be increased or decreased by using p–i–n junctions integrated across the MRR, enabling free carrier injection or depletion. This can increase the self-pulsing frequencies up to hundreds of MHz which, however, remain far from the GHz regime \cite{shetewy2024demonstration}.

The use of cavities based on photonic crystals enables light–matter interactions with markedly different self-pulsing temporal dynamics.
In \cite{cazier2013high}, a silicon photonic crystal nanocavity is reported in which self-pulsing occurs at gigahertz frequencies, driven by the interplay between nonlinear responses and the photon lifetime of the cavity.
Silicon photonic crystal nanocavities also exhibit passive bistability due to TPA \cite{notomi2008chip}.
TPA triggers a nonlinear feedback between the intracavity field and the carrier and thermal dynamics which gives rise to bistability.
By using set and reset pulses with energies below \qty{100}{\fJ}, switching between two ouptuput power states was demonstrated, with a holding bias power of about \qty{0.4}{\mW}. Since the cavity is small (on the micron scale), carrier diffusion dominates and $\tau_{fc}$, which sets the length of the set and reset pulse, is reduced to \qtyrange{50}{100}{\ps}.

Passive nonlinear bistability is also provided by optofluidic memories \cite{gao2023optofluidic}.
In these systems, a thin liquid film covers the surface of a photonic waveguide, and a small patch of gold is deposited below.
When light propagates through the waveguide, part of its power is dissipated as heat in the gold layer, causing a local temperature increase.
This temperature rise alters the thickness of the liquid film, which in turn modifies the effective refractive index of the waveguide.
As a result, the light experiences a phase shift in the vicinity of the heated gold patch. This phase shift shows the bistability cycle as a function of the input power.

\paragraph{Active optical multistability}
Active optical multistability relies on an external gain source or control signal to maintain stability and control transitions between at least two stable states. Since retention always originates in the competition and saturation of an optical gain medium, we name this class \emph{gain-mediated bistability} in \Cref{fig:MemoryClassification}.
One important example in this context is injection locking, a mechanism in which an oscillator synchronizes its frequency and phase to an external driving signal.
Injection locking can exhibit multistability.
In the case of the Adler model of locking \cite{adler2006study,paciorek2005injection}, there is just an attractor, which gives rise to fading memory when excited.
More generally, a device can be regarded as implementing multistable memory only if it exhibits a reproducible bistable or multistable response at fixed holding/control parameters, together with independently triggerable write (state-selection) and reset operations, and retention of the selected state after the writing stimulus is removed.
In bistable injection locking-based systems, a self-sustained laser oscillator exhibits bistability between a free-running and a locked emission state when driven by an external coherent field.

This phenomenon can be harnessed for all-optical memory applications: the injected field acts as a transient trigger that selects between the oscillator’s two internally self-sustained states, while state retention after the field is removed is ensured by the underlying gain-mediated bistability. Switching to the opposite state requires crossing a distinct threshold, thereby giving rise to hysteresis.
A clear demonstration of this principle is found in the microdisk laser flip-flop integrated on silicon photonics with a III–V (\ce{InP}/\ce{InGaAsP}) active region \cite{liu2010ultra}. In this case, short injected optical pulses act as external fields that
selectively excite one of the two counter-propagating whispering-gallery modes.
When the injected power exceeds a threshold, the laser is forced into a unidirectional emission state determined by the propagation direction of the injected field.
The gain competition, together with nonlinear saturation, ensures that this state remains stable after the pulse ends.
Beyond microdisk lasers, injection locking can take many forms.
In \cite{chen2011all}, a photonic crystal laser (\cref{fig:nonlinear_bistability-d}) is frequency-locked to an external master laser, illustrating how an external coherent field can control a laser’s dynamics.

A different scenario is presented by self-injection locking, in which a laser is continuously stabilized by light backscattered into its own cavity from a high-Q resonator, rather than by a separate master laser. In \cite{lihachev2022platicon}, a distributed feedback laser is frequency-stabilized through backscattering in a high-Q \ce{Si3N4} MRR, with the scattered light locking the laser frequency to the MRR resonance. Unlike the bistable devices discussed above, this configuration maintains a single self-consistent locked condition for as long as the laser and resonator are powered, and is used for linewidth narrowing and frequency stabilization rather than for switching between distinct states; we therefore do not classify it here as a memory demonstration. Together, the microdisk-laser flip-flop of \cite{liu2010ultra} and the photonic-crystal laser of \cite{chen2011all} illustrate how a transient injected field can trigger gain-mediated bistability and thereby enable memory-like operation, whereas continuous self-injection locking, as in \cite{lihachev2022platicon}, provides frequency control without state retention.

Another operational principle in \emph{gain-mediated bistable}  systems is the master–slave configuration, in which one optical signal, generated by the master, controls the dynamics of another system (the slave).
In master–slave wavelength-based all-optical flip-flops, a class of devices that can overlap with injection-locked systems, bistable operation is achieved through the competition of two optical signals at distinct wavelengths circulating within a common cavity or coupled active structure.
At any given time, only one wavelength can dominate the system, while the competing wavelength remains suppressed due to gain saturation effects.
The logical state of the memory element is therefore encoded in the wavelength of the dominant optical signal.
The active component emitting the dominant wavelength acts as the master, as it suppresses the gain available to the competing device.
Conversely, the suppressed component operates as the slave, remaining inactive while its gain is clamped by the master wavelength.
State switching is performed by injecting an external optical signal at an appropriate wavelength and power level into the master device.
This perturbation suppresses the master operation and allows the slave component to recover its gain and reach its equilibrium emission state.
Once the slave wavelength becomes dominant, it propagates back into the previously dominant component, enforcing suppression even after the external injection is removed.
This feedback mechanism enables stable state retention and underpins the set–reset flip-flop functionality employed in optical static RAM cells.
This operating principle has been experimentally demonstrated in SOA-MZI-based optical static RAM cells \cite{pleros2008optical} and later refined through wavelength-diverse input schemes enabling reduced component count and improved scalability \cite{fitsios2012dual}.

\subsection{Non-volatile memory} \label{subsec:non-volatile}
Non-volatile photonic memory relies on permanent or quasi-permanent modifications of material properties, enabling information retention in the absence of optical excitation and without the need for continuous electrical power to sustain the memory state (e.g., heaters or biasing).
In silicon photonics, non-volatile memory is typically achieved by using materials whose optical properties can be reversibly switched between distinct states with long retention times.
These functional materials exhibit hysteresis or bistability. The underlying
 mechanisms include structural, polarization, trap states and ionic effects, each relying on a different microscopic degree of freedom.
Based on this, the non-volatile memory can be grouped into four main categories: structural memory, polarization-based memory, ionic memory, and charge-trapping memory.

\begin{figure}[!htb]
	\centering
	\includegraphics[width=1\linewidth]{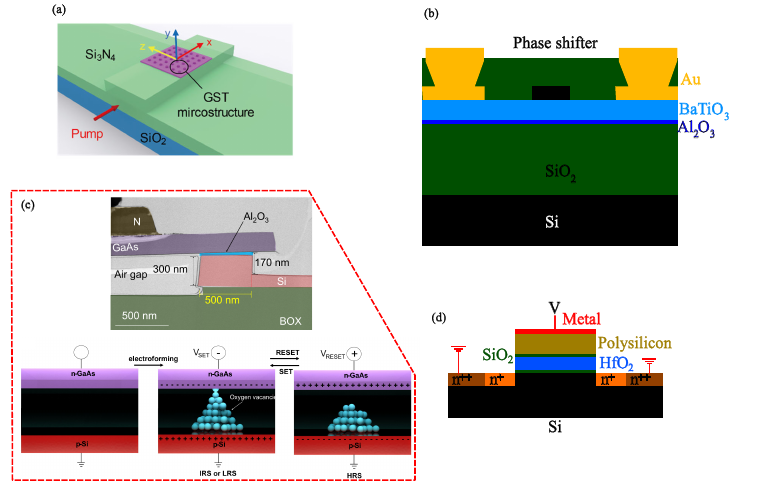}
	\begin{subcaptiongroup}
		\phantomcaption\label{fig:nonvolatile-a}
		\phantomcaption\label{fig:nonvolatile-b}
		\phantomcaption\label{fig:nonvolatile-c}
		\phantomcaption\label{fig:nonvolatile-d}
	\end{subcaptiongroup}

	\caption{%
		Implementation of non-volatile memories.
        \subref{fig:nonvolatile-a} Micro-cylinders GST pattern on a \ce{Si3N4} waveguide atop an SOI substrate. It is reprinted from \cite{gao2025structure} \href{https://creativecommons.org/licenses/by/4.0/}{CC BY 4.0}.
		\subref{fig:nonvolatile-b} the layout of a photonic phase shifter based on ferroelectric \ce{BaTiO3} \cite{geler2022ferroelectric}.
		\subref{fig:nonvolatile-c} Cross-section of a memristor based on a MRR with an illustration of the device's working mechanism, adapted from \cite{tossoun2024high} \href{https://creativecommons.org/licenses/by/4.0/}{CC BY 4.0}. \subref{fig:nonvolatile-d} Cross-section of the charge-trapping structure in a MRR \cite{olivares2021non} , where stored charges in \ce{HfO2} change the effective refractive index via the plasma dispersion effect.}
	\label{fig:nonvolatile}
\end{figure}

\paragraph{Structural memory}
Structural (phase-change) memories store information in the atomic configuration of the material \cite{feldmann2019all}.
This involves a transition between amorphous and crystalline phases, which correspond to different atomic arrangements.
The transition, which can be triggered by an external optical or electrical field exceeding a threshold, produces large and persistent changes in its electrical conductivity, its effective refractive index, and its absorption.
Non-volatility originates from the significant energy barrier separating these structural phases, which stabilizes the stored state at room temperature.

In \cite{wuttig2005towards}, the authors report a non-volatile memory based on PCMs (PCM), in which electrical pulses reversibly switch the material between amorphous and crystalline states with distinct resistances, enabling fast read/write operations at low power.
However, limitations in switching speed, power consumption, and crystallization stability remain challenges compared to more mature technologies such as flash or MRAM (magnetoresistive RAM).

Extending this concept to the optical domain, \cite{wei2023electrically} integrates a PCM (\ce{Sb2Se3}) on a p-i-n junction embedded in a silicon photonics MRR. The PCM  exhibits distinct optical transmission depending on its structural phase. Crystallization and amorphization are electrically induced using voltage pulses, with typical values of \qtyrange{3}{3.3}{\volt} for crystallization and \qtyrange{4.4}{8.2}{\volt} for amorphization, with pulse durations from hundreds of nanoseconds to tens of microseconds.
These phase transitions modify the refractive index of the PCMs, shifting the MRR resonance and thus changing the transmitted optical power, enabling both nonvolatile storage and nanosecond-scale volatile modulation. Refs.
\cite{rios2015integrated,cheng2018device,li2018fast,gao2025structure} employ an all-optical programming scheme, where ultrashort optical pulses are used to directly trigger the phase transition in the PCM via the absorption of light, rather than relying on electrical heating mechanisms.
\cite{gao2025structure} demonstrates that patterning GST (\ce{Ge2Sb2Te5}) into micro-cylinder arrays on top of the PCM film (\cref{fig:nonvolatile-a}) significantly enhances device performance.
These improve thermal dissipation and distribute stress during phase transitions, enabling over \num{2e6} reversible switching cycles, to be compared with the \num{6e3} in the case of a GST film.
Additionally, the arrays maintain high optical transmission contrast: the micro-cylinder geometry concentrates the optical field within the PCMs regions, amplifying the difference in absorption between amorphous and crystalline states and thus enhancing transmission contrast.
Simultaneously, the presence of gaps between the cylinders allows light to propagate through low-loss regions, minimizing overall absorption and lowering insertion loss compared to a continuous PCM film.
A complementary route to improving PCM device performance relies on plasmonic field enhancement rather than array-based geometric optimization. By embedding the PCM within or adjacent to a metallic nanostructure, the optical near-field is locally concentrated, so that only a very small PCM volume needs to switch phase to produce a strong optical response. \cite{sun2022all} proposes, via FDTD simulation, a metal double-ring structure embedding GST that confines the switching region to a footprint of approximately 50 × 90 nm², with simulated insertion losses of 2.3 dB (amorphous) and 1.0 dB (crystalline) and a transmission contrast of $15.8\%$ at 1550 nm, while keeping the temperature variation within the PCM cell below 2 °C during crystallization to avoid inhomogeneous switching. This design builds on earlier proposals for plasmonically-enhanced all-optical phase-change memory \cite{gemo2019plasmonically}, and, compared to the array-based approach of \cite{gao2025structure}, trades a potentially much smaller footprint and lower switching energy against the additional optical (ohmic) loss introduced by the metallic structure.
It is also possible to induce a molecular configuration change by sending set ($\lambda$=\qty{275}{\nm}) and reset ($\lambda$=\qty{635}{\nm}) pulses onto a MRR covered with a thin (\qty{165}{\nm}) layer of photochromic material \cite{bilodeau2024all}.
This layer slightly changes its thickness, and thus its refractive index, by \qty{2.3}{\percent}, which shifts the MRR resonance up to 5 different wavelengths limited by the free spectral range.

\paragraph{Polarization-based memory}
Polarization-based (ferroelectric) memories rely on the bistable orientation of electric dipoles in non-centrosymmetric materials. Once an external electric field aligns the polarization, the remanent dipole moment persists after the field is removed. Information is therefore encoded in the direction or magnitude of the polarization state.

\cite{he2022optical} demonstrates a multidimensional non-volatile memory based on Polyvinylidene Fluoride, where optical absorption and ferroelectric polarization independently encode multiple logic states within the same physical cell. In this work, the hysteresis of the ferroelectric material is exploited to achieve multilevel phase transitions and 8-state storage with high density and long retention. While this represents a materials-level proof of concept, it is not directly compatible with silicon photonics due to the use of polymeric ferroelectrics and atomic force microscopy-based writing mechanisms.

Building on this concept, \cite{zhang2024thin} demonstrates a non-volatile photonic-electronic memory using an \ce{Al}-doped \ce{HfO2} (HAO) ferroelectric thin film integrated on a silicon MRR.
As in \cite{he2022optical}, this approach is fully compatible with silicon photonics, supports electrical and optical programming/erasing, achieves a high extinction ratio of \qty{6.6}{\dB}, and an endurance of \num{4e4} cycles at \qty{5}{\volt}. The different memory states are encoded in the absorption of the material, i.e., in the imaginary part of the refractive index, allowing readout through the amplitude of the optical output.

A related technique is presented in \cite{geler2022ferroelectric}, which demonstrates a multistate, non-volatile photonic phase shifter based on ferroelectric \ce{BaTiO3} integrated with silicon waveguides (\cref{fig:nonvolatile-b}).
In contrast to \cite{zhang2024thin}, here the different memory states are encoded in the real part of the refractive index, so that the amplitude remains essentially unchanged while the optical phase is modified.
By applying short electrical pulses, ferroelectric domains are switched via nucleation and domain wall propagation, changing the effective Pockels coefficient and thus the refractive index.
Each distinct domain configuration corresponds to a different refractive index, enabling stable, low-loss, and reprogrammable optical states suitable for programmable photonic circuits and hybrid electronic-photonic systems.

\paragraph{Ionic memory}
Ionic or resistive-switching memories exploit the field-driven migration of ions or vacancies within a solid electrolyte or oxide.
The formation and rupture of conductive filaments, or more generally the redistribution of ionic species, leads to stable changes in conductivity or optical response.
The slow relaxation dynamics of ionic motion ensure long retention times.

\cite{kumar2023double} presents a CMOS-compatible double-slot nanophotonic resistive switch with optical readout capabilities.
The device operates by forming and breaking \ce{Ag} conductive filaments in \ce{SiO2} under an applied voltage, modulating the optical transmission and achieving a high extinction ratio of \qty{35}{\dB}.
However, this device primarily functions as a binary switch (ON/OFF).

In contrast, the memresonator in \cite{tossoun2024high} integrates a memristor within a MRR (\cref{fig:nonvolatile-c}), where conductive filaments formed by oxygen vacancies can create multiple intermediate resistance states.
These states alter the carrier distribution and the $n_\mathrm{eff}$ in the waveguide, shifting the MRR resonance. This enables multiple optical levels, providing true multilevel non-volatile memory operation.
Thus, while \cite{kumar2023double} demonstrates a high-extinction-ratio optical switch, \cite{tossoun2024high} represents an advancement toward multistate, optically addressable memory.

\paragraph{Charge-trapping memory}
Charge-trapping (electronic) memories store information in the electrostatic potential created by charges confined in isolated regions, such as floating gates or trap layers.
The surrounding insulating barriers suppress charge leakage, allowing the stored state to persist for extended periods.
Here, non-volatility originates from charge confinement rather than from material hysteresis.

This mechanism can often be implemented using materials and process steps standard in CMOS technology, such as polysilicon or high-k dielectrics like HfO$_2$ and Al$_2$O$_3$, rather than the material systems required by the other non-volatile mechanisms discussed in this section, such as chalcogenide phase-change compounds (e.g., GST) or ferroelectric materials (e.g., BaTiO$_3$, Al-doped HfO$_2$, or ferroelectric polymers); we therefore describe such implementations as CMOS-compatible silicon photonics.

Refs. \cite{song2016integrated} and \cite{olivares2021non} exploit the plasma dispersion effect in silicon to achieve non-volatile photonic memory functionality. Ref. \cite{song2016integrated} demonstrates a SOI- and CMOS-compatible, electrically programmable, non-volatile multi-level photonic memory cell integrated in a silicon photonic circuit.
The device is based on a polycrystalline-silicon floating gate in a waveguide structure, here injected electrons modify the effective refractive index, allowing an optical readout via a MRR with distinct ON/OFF states and multiple intermediate levels controlled by the programming pulse width.

Ref. \cite{olivares2021non} reports a 2-D TCAD simulation of tunneling and trap-assisted charge transport in a non-volatile photonic memory based on a charge-trapping configuration using HfO$_2$ and Al$_2$O$_3$ (\cref{fig:nonvolatile-d}). Similar to \cite{song2016integrated}, the stored charges are predicted to modulate the effective refractive index of silicon via the plasma dispersion effect, enabling electrically written and optically read memory states. Embedded in a MRR, the simulated device is predicted to achieve high extinction ratios, low insertion losses, and a retention time exceeding 10 years; the authors note that this result does not account for extrinsic tunnel-oxide defects, and that experimental validation of the structure remains future work. Compared to the floating-gate approach of \cite{song2016integrated} (which was experimentally demonstrated but not tested for long-term retention) the simulated SAHAS (silicon-aluminum oxide-hafnium aluminum oxide) configuration is predicted to allow significantly faster write and erase times (in the microsecond range) and operational speed.
This speed-up is due to the optimization of the design of the trapping structure, designed to optimize the TE polarized optical mode overlap with the carrier accumulation layer.

\paragraph{}

A related example, based on the same storage principle but on a different readout mechanism,  is given in \cite{ren2025photorefractive}, where light photoexcites electrons from impurity levels into the conduction band of thin-film lithium niobate. The resulting charge redistribution and retrapping builds up a long-lived internal (photorefractive) space-charge field, reinforced by pyroelectric fields arising from local optically-induced heating. The information is therefore still stored in trapped charge; however, unlike  the charge-trapping devices discussed above, the refractive-index shift is not produced directly by the trapped-carrier population itself, but is mediated by the linear electro-optic (Pockels) response of lithium niobate to this internal field.  The relaxation back to the pristine state occurs over several hours.

\subsection{Comparative summary}

To summarize, \cref{tab:volatile} and \cref{tab:nonvolatile} collect the quantitative characteristics of the volatile and non-volatile mechanisms discussed above, using the entries of \Cref{fig:MemoryClassification} as column headers so that the classification and the numbers can be read side by side. They report retention/relaxation time, write/erase time, switching energy, bias power, insertion loss, optical contrast, endurance, number of addressable levels, footprint, and experimental or simulated status. Values are those given in the cited works: entries marked \emph{NA} are not applicable to that mechanism.  Read across the retention row, the two families overlap in duration but not in kind: volatile mechanisms range from finite propagation delays of a few picoseconds to attractor-based states that persist as long as the holding drive is maintained. After removal of the drive, the information either leaves the circuit after a finite propagation time or decays according to the relevant relaxation dynamics. By contrast, non-volatile states are stabilized by an energy barrier and persist without holding power, although energy is still required for writing, erasing and readout.

\newcommand{\NA}{\textit{\textcolor{gray}{NA}}}
\begin{sidewaystable}[t]
	\caption{Quantitative comparison of volatile photonic memory mechanisms. Classification entries (\cref{fig:MemoryClassification}) are given as column headers. The specific quantities reported here have been selected from representative literature. Optical contrast is reported using the metric adopted in each original publication, including extinction ratio, transmission variation, resonance shift or refractive-index change. Values expressed using different metrics are not directly comparable. \NA{} = Not Applicable.}
	\label{tab:volatile}
    \footnotesize
	\renewcommand{\arraystretch}{1.2}
	\setlength{\tabcolsep}{2pt}
	\tablealtrows
	\centering
	\begin{tabular}{@{}>{\raggedright\arraybackslash}p{0.08\textwidth} *{7}{>{\raggedright\arraybackslash}p{0.123\textwidth}}@{}}
		\toprule
		\textbf{}
		& \textbf{Delay-based}
		& \textbf{Slow-light}
		& \textbf{Cavity photon lifetime}
		& \textbf{Nonlinear relaxation}
		& \textbf{Optical Bistability}
		& \textbf{Self-Pulsing}
		& \textbf{Gain-mediated Bistability}
		\\

		\midrule

		{Retention / relaxation}
		& Buffered packet delay $\approx$\qty{1.1}{\nano\second} \cite{park2008integrated}; delay bank up to \qty{17.2}{\nano\second} (3 lines: \qty{6.5}{\nano\second}\slash\qty{11.3}{\nano\second}\slash\qty{17.2}{\nano\second}) \cite{moralis2017chip}; \qty{\sim 150}{\pico\second} \cite{aboketaf2010optical}
		& Group delay \qty{510}{\pico\second}, 56-ring APF \cite{xia2007ultracompact}; delay set by $n_g$=47.74 \cite{jean2019slow};
		& Usually $\approx$\qty{8}{\pico\second} for $Q=10^4$ at \qty{1550}{\nano\meter}; $\tau_p\approx$\qty{0.6}{\nano\second} for $Q_0\approx7.5\times10^5$ \cite{lu2022high}
		& $\tau_{th}\approx$\qty{65}{\nano\second}, $\tau_{fc}\approx$\qty{5.3}{\nano\second} \cite{van2012simplified}; $\tau_{th}\approx$\qty{280}{\nano\second}, $\tau_{fc}\approx$\qty{45}{\nano\second}, different fab.\ run \cite{borghi2021modeling}
		& Stable while driven/biased, no fixed duration reported \cite{aslan2025coherent,biasi2025phase,notomi2008chip}
		& $\geq$\qty{10}{\micro\second} \cite{biasi2024exploring}; $\geq$\qty{75}{\micro\second} after perturbation \cite{lugnan2025reservoir}
		& Verified $\sim$\qty{1}{\nano\second} \cite{liu2010ultra}; \qty{1}{\micro\second} demonstrated \cite{chen2011all}; demonstrated $\sim$\qty{1}{\nano\second} \cite{pleros2008optical}; verified $\sim$\qty{1}{\micro\second} \cite{fitsios2012dual}
		\\

		{ Write / erase time}
		& \NA{}
		& \NA{}
		& \NA{}
		& \NA{}
		& \qtyrange{50}{300}{\pico\second} \cite{notomi2008chip}
		& 20 pulses of \qty{50}{\nano\second}, variable separation \cite{biasi2024exploring}; \qtyrange{60}{200}{\nano\second} pulse width \cite{lugnan2025reservoir}
		& \qty{60}{\pico\second} \cite{liu2010ultra,chen2011all}; \qty{200}{\pico\second} \cite{pleros2008optical}; \qty{125}{\nano\second} \cite{fitsios2012dual}
		\\

		{Switching energy}
		& \NA{}
		& \NA{}
		& \NA{}
		& \NA{}
		& $\sim$\qty{100}{\fJ} \cite{notomi2008chip}
		& /
		& \qty{1.8}{\fJ} \cite{liu2010ultra}
		\\

		{Bias / holding power }
		& CW bias $\approx$\qty{2.14}{\milli\watt} \cite{moralis2017chip}
		& /
		& \qty{290}{\milli\watt} for \qty{100}{\pico\second} delay \cite{xie2014continuously}
		& \NA{}
		& \qty{0.4}{\milli\watt} \cite{notomi2008chip}; $>$\qty{8}{\milli\watt} \cite{gao2023optofluidic};
		\qty{20}{\milli\watt} \cite{aslan2025coherent,biasi2025phase}
		& \qtyrange{0.75}{5.05}{\milli\watt} \cite{biasi2024exploring}; \qtyrange{5.5}{16.4}{\milli\watt} \cite{lugnan2025reservoir}
		& Electrical pump of a few\,mW \cite{liu2010ultra}; \qty{25}{\micro\watt} pump \cite{chen2011all}; CW optical drive of \qtyrange{1.2}{3.55}{\milli\watt} \cite{pleros2008optical}
		\\

		{Optical loss}
		& $\approx$\qty{3}{\decibel\per\centi\meter} waveguide \cite{aboketaf2010optical}; \qty{20}{\decibel}\slash\qty{25}{\decibel}\slash\qty{30}{\decibel} total, \qty{0.12}{\decibel\per\centi\meter} waveguide \cite{moralis2017chip}; $<$\qty{0.1}{\decibel} per basic unit \cite{perez2018programmable}
		& $\approx$\qty{22}{\decibel} on-resonance, 56-ring APF \cite{xia2007ultracompact}; \qty{12.5}{\decibel\per\milli\meter} max (band-edge) / \qty{1.4}{\decibel\per\milli\meter} min \cite{jean2019slow};
		& $\approx$\qty{16}{\decibel} fiber-to-fiber off-resonance, $+\approx$\qty{5}{\decibel} on-resonance, \qtyrange{3}{4}{\decibel\per\centi\meter} waveguide \cite{xie2014continuously}
		& /
		& /
		& /
		& /
		\\

		{Contrast / ER}
		& \qty{6.6}{\decibel}\slash\qty{6.4}{\decibel}\slash\qty{6.2}{\decibel} (3 lines) \cite{moralis2017chip}
		& /
		& /
		& /
		& /
		& /
		& \qty{11}{\decibel} \cite{liu2010ultra}; \qty{6.75}{\decibel} \cite{chen2011all}; \qty{10}{\decibel} \cite{fitsios2012dual}
		\\

		Endurance
		& \NA{}
		& \NA{}
		& \NA{}
		& \NA{}
		& \NA{}
		& \NA{}
		& \NA{}
		\\

		Addr.\ levels
		& 4 parallel channels, \qty{20}{\Gbps} \cite{aboketaf2010optical}; 3 discrete delay lines \cite{moralis2017chip}; reconfigurable continuous mesh \cite{perez2018programmable}
		& \NA{}
		& \NA{}
		& \NA{}
		& 2 \cite{aslan2025coherent,biasi2025phase,notomi2008chip,gao2023optofluidic}
		& Multiple limit cycles, $n$ not quantified \cite{biasi2024exploring,lugnan2025reservoir}
		& 2 \cite{liu2010ultra,chen2011all,pleros2008optical,fitsios2012dual}
		\\

		Footprint
		& $\approx$\qty{1}{\milli\meter\squared} \cite{aboketaf2010optical}; \qty{4.4}{\milli\meter}\allowbreak$\times$\allowbreak\qty{1.5}{\milli\meter} (\qty{17.2}{\nano\second} line) \cite{moralis2017chip}; coupler \qty{1235}{\micro\meter} per basic unit \cite{perez2018programmable}
		& \qty{0.09}{\milli\meter\squared}, 56-ring APF \cite{xia2007ultracompact}
		& 13 MRRs (radius \qty{10}{\micro\meter}) \cite{xie2014continuously}
		& /
		& \qty{560}{\micro\meter}\allowbreak$\times$\allowbreak\qty{160}{\micro\meter} \cite{aslan2025coherent,biasi2025phase}; mode volume \qty{0.1}{\micro\meter\cubed} \cite{notomi2008chip}; liquid cell \qty{50}{\micro\meter}\allowbreak$\times$\allowbreak\qty{50}{\micro\meter}\allowbreak$\times$\allowbreak\qty{3}{\micro\meter} \cite{gao2023optofluidic}
		& \qty{20}{\micro\meter}\allowbreak$\times$\allowbreak\qty{50}{\micro\meter} \cite{biasi2024exploring}; \qty{0.15}{\milli\meter\squared} \cite{lugnan2025reservoir}
		& \qty{7.5}{\micro\meter}\allowbreak$\times$\allowbreak\qty{7.5}{\micro\meter} \cite{liu2010ultra}; \qty{4}{\micro\meter}\allowbreak$\times$\allowbreak\qty{0.3}{\micro\meter}\allowbreak$\times$\allowbreak\qty{0.16}{\micro\meter} \cite{chen2011all}; 3\allowbreak$\times$\allowbreak SOA \qty{1.5}{\milli\meter} each, bulk/fibre \cite{fitsios2012dual}
		\\

		Status
		& Experiment \cite{aboketaf2010optical,moralis2017chip,park2008integrated,perez2018programmable}
		& Experiment \cite{xia2007ultracompact,jean2019slow};
		& Experiment \cite{xie2014continuously,lu2022high}
		& Experiment \cite{borghi2021modeling}; Simulation \cite{van2012simplified}
		& Experiment \cite{aslan2025coherent,biasi2025phase,notomi2008chip,gao2023optofluidic}
		& Experiment \cite{biasi2024exploring,lugnan2025reservoir}
		& Experiment \cite{liu2010ultra,chen2011all,pleros2008optical,fitsios2012dual}
		\\

		\botrule
	\end{tabular}
\end{sidewaystable}
\begin{sidewaystable}[t]
	\caption{Quantitative comparison of non-volatile photonic memory mechanisms. Classification entries (\cref{fig:MemoryClassification}) are given as column headers. The specific quantities reported here have been selected from representative literature. Optical contrast is reported using the metric adopted in each original publication, including extinction ratio, transmission variation, resonance shift or refractive-index change. Values expressed using different metrics are not directly comparable. \NA{} = Not Applicable.}
	\label{tab:nonvolatile}
    \footnotesize
	\renewcommand{\arraystretch}{1.2}
	\setlength{\tabcolsep}{1.5pt}
	\tablealtrows
	\centering
	\begin{tabular}{@{}>{\raggedright\arraybackslash}p{0.15\textwidth} *{4}{>{\raggedright\arraybackslash}p{0.205\textwidth}}@{}}
		\toprule
		\textbf{Category}
		& \textbf{Structural (Phase-change)}
		& \textbf{Polarization (Ferro-electric)}
		& \textbf{Ionic (Resistive switching)}
		& \textbf{Electronic (Charge trapping)}
		\\

		\midrule

		{Retention / relaxation}
		& $>$\qty{1e5}{\second} ($\approx$\qty{28}{\hour}, measured, single-cell) \cite{gao2025structure}; $\sim$days \cite{bilodeau2024all}
		& $\approx$3 months, room temp.; degrades $>$\qtyrange{80}{85}{\celsius} in \qtyrange{5}{20}{\minute} \cite{he2022optical}; $\geq$\qty{1000}{\second}, no noticeable change \cite{zhang2024thin}; $\geq$\qty{10}{\hour}, stable \cite{geler2022ferroelectric}
		& /
		& Bi-exponential relaxation, $\tau_1\approx$\qty{1.7}{\hour} / $\tau_2\approx$\qty{10}{\hour}, not restored after \qty{20}{\hour} \cite{ren2025photorefractive}; \qty{20}{\hour} monitored \cite{song2016integrated}; $>$\qty{10}{\yr} simulated, not measured \cite{olivares2021non}
		\\

		Write / erase time
		& \qty{2}{\micro\second} \cite{wei2023electrically}
		& \qtyrange{50}{400}{\nano\second} pulses \cite{geler2022ferroelectric}
		& \qtyrange{300}{900}{\pico\second} pulses \cite{tossoun2024high}
		& $\approx$\qtyrange{350}{600}{\milli\second} program / \qtyrange{75}{225}{\milli\second} erase \cite{song2016integrated}; \qty{100}{\micro\second} \cite{olivares2021non}
		\\

		Switching energy
		& \qty{0.96}{\pico\joule}\slash operation, single-cell \cite{gao2025structure}; \qty{2}{\nano\joule} (SET) / \qty{11}{\nano\joule} (RESET) \cite{bilodeau2024all}
		& \qty{1}{\micro\joule}\slash spot, AFM-IR \cite{he2022optical}; \qtyrange{4.6}{26.7}{\pico\joule} \cite{geler2022ferroelectric}
		& \qtyrange{0.15}{0.36}{\pico\joule} \cite{tossoun2024high}
		& \qty{17.2}{\pico\joule}\slash\qty{11.4}{\pico\joule} \cite{song2016integrated}; \qty{1.25}{\pico\joule}\slash\qty{5.7}{\pico\joule} \cite{olivares2021non}
		\\

		{Bias voltage}
		& \qty{3}{\volt} \cite{wei2023electrically}, set
		& \qty{-15}{\volt}, writing example \cite{he2022optical}; \qty{1}{\volt} read,  \cite{zhang2024thin}; \qty{2}{\volt}, read \cite{geler2022ferroelectric}
		& $\approx$\qty{5}{\volt} \cite{kumar2023double}, set
		& \qty{20}{\volt} (program) / \qty{6}{\volt} (erase) \cite{song2016integrated}; \qty{21.5}{\volt}, write / \qty{-30}{\volt}, erase \cite{olivares2021non}
		\\

		Insertion loss
		& \qty{2.3}{\decibel} / \qty{1.0}{\decibel} \cite{sun2022all}
		& \qty{0.07}{\decibel} \cite{geler2022ferroelectric}
		& $\approx$\qty{0.5}{\decibel\per\micro\meter} ($\approx$\qty{10}{\decibel} total, \qty{20}{\micro\meter} device) \cite{kumar2023double}; $\approx$\qty{0.05}{\decibel} \cite{tossoun2024high}
		& $\approx$\qty{8.6}{\decibel} total \cite{song2016integrated}; $<$\qty{1}{\decibel} \cite{olivares2021non}
		\\

		Contrast / ER
		& \qty{14}{\decibel} \cite{wei2023electrically}; \qty{8.69}{\decibel} improvement, array vs.\ film-free array \cite{gao2025structure}; $\Delta n$=\qty{2.3}{\percent} \cite{bilodeau2024all}; \qty{15.8}{\percent} \cite{sun2022all}
		& \qty{6.6}{\decibel} \cite{zhang2024thin}
		& \qty{35}{\decibel} \cite{kumar2023double}
		& \qty{325}{\pico\meter} $\lambda$-shift \cite{song2016integrated}; \qty{15}{\decibel} \cite{olivares2021non}
		\\

		Endurance
		& 6000 cycles (single-cell), $>2\times10^6$ (array) \cite{gao2025structure}
		& $10^6$ cycles at \qty{4}{\volt} \cite{zhang2024thin}
		& 100 cycles measured; up to 1000 ML-forecasted, NOT measured \cite{kumar2023double}; 1000 cycles, measured \cite{tossoun2024high}
		& 30 cycles tested \cite{song2016integrated}
		\\

		Addr.\ levels
		& 34 states \cite{wei2023electrically}; 209 distinguishable states, single-cell \cite{gao2025structure}
		& 8 \cite{he2022optical,geler2022ferroelectric}; multilevel
		& 2, binary \cite{kumar2023double}
		& $\sim$50 theoretical / 4-bit demonstrated \cite{song2016integrated}
		\\

		Footprint
		& $\approx$\qty{50}{\nano\meter}\allowbreak$\times$\allowbreak\qty{90}{\nano\meter} \cite{sun2022all}
		& \qty{150}{\micro\meter}\allowbreak$\times$\allowbreak\qty{200}{\micro\meter}, $Q=\num{40000}$ \cite{geler2022ferroelectric}
		& \qty{20}{\micro\meter} device length \cite{kumar2023double}
		& \qty{150}{\micro\meter}\allowbreak$\times$\allowbreak\qty{150}{\micro\meter} \cite{ren2025photorefractive}; MRR $r\sim$\qty{12}{\micro\meter} \cite{song2016integrated}
		\\

		Status
		& Experiment \cite{wei2023electrically,gao2025structure,bilodeau2024all}; Simulation (FDTD) \cite{sun2022all}
		& Experiment, not Si-compatible \cite{he2022optical}; Experiment \cite{zhang2024thin,geler2022ferroelectric}
		& Experiment \cite{kumar2023double,tossoun2024high}
		& Experiment \cite{ren2025photorefractive,song2016integrated}; Simulation \cite{olivares2021non}
		\\

		\botrule
	\end{tabular}
\end{sidewaystable}

\section{Characterizing Dynamic Memory of Neural Networks}
\label{sec:theo}

Having discussed how to realize memories in PIC, let us now discuss the theoretical and experimental framework to characterize the computational memory of PNNs used in machine learning (ML). %
This section is organized into four parts.

First, \cref{sec:nn_architectures} introduces the principal neural network architectures from a theoretical standpoint, classifying them according to the mechanism by which they handle the time dimension.

Second, \cref{sec:memory_char} %
presents the mathematical tools and experimental protocols used to characterize fading memory in dynamical systems.
Because these metrics are agnostic to the underlying physical substrate, they apply equally to digital simulations and to photonic hardware, providing a common language for comparison across platforms.

Third, \cref{sec:persistent_state_metrics} introduces the complementary metrics used to characterize the write-retain-read-reset cycle of persistent-state systems.

Fourth ,  \cref{sec:benchmarks} surveys the standard benchmark tasks %
commonly reported  in the photonic literature.
Together, these benchmarks probe complementary aspects of their computational capability and serve as the primary figures of merit to compare PNNs.

\subsection{Neural Network Architectures}
\label{sec:nn_architectures}

\begin{figure}[!htp]
    \centering
    \includegraphics[width=.9\linewidth]{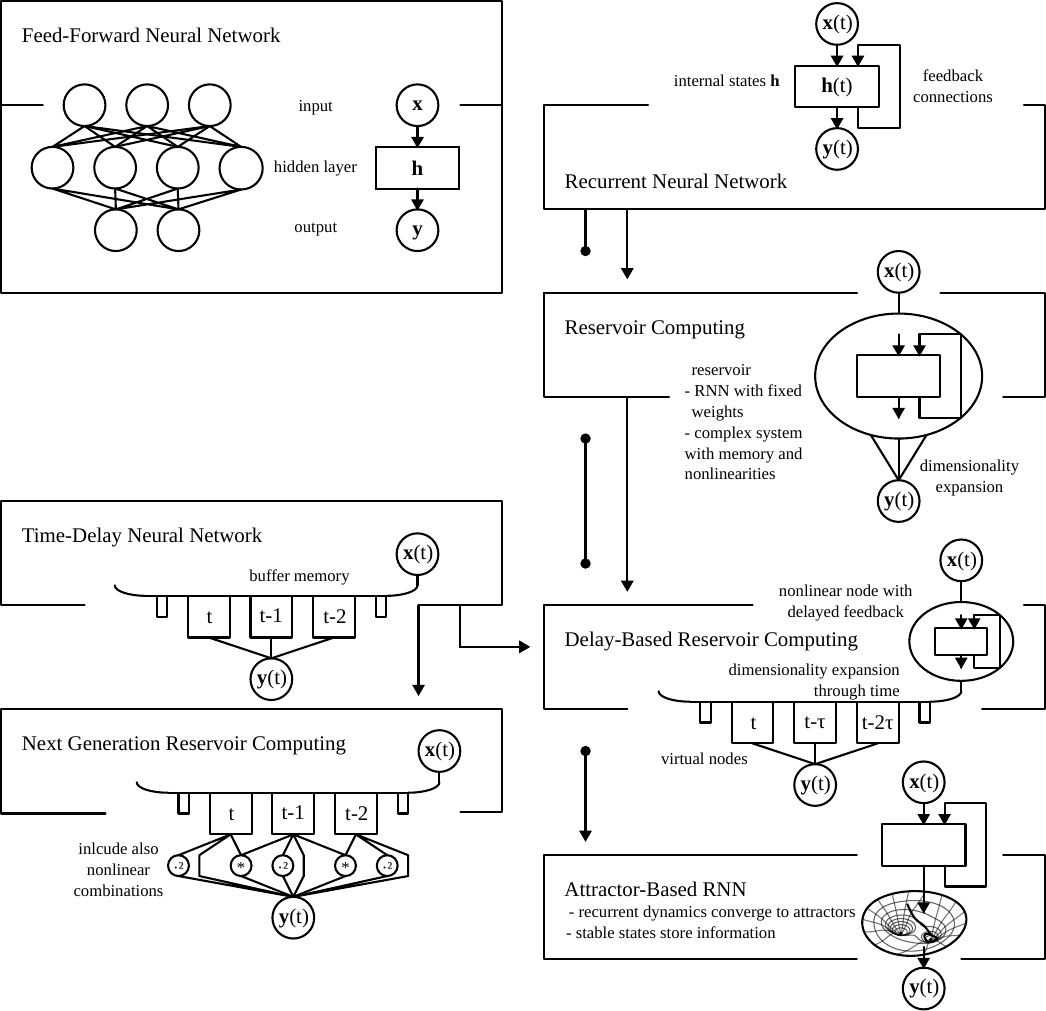}
    \caption{%
Overview of neural network architectures relevant to temporal memory processing and neuromorphic photonic implementations.
A \emph{feed-forward neural network} maps an input vector $\vb{x}$ through one (or more) hidden layers $\vb{h}$ to produce an output $\vb{y}$, where
information propagates only in the forward direction.
A \emph{recurrent neural network} includes feedback connections, so the internal state $\vb{h}(t)$ depends on the current input $\vb{x}(t)$ and previous hidden states, enabling temporal sequence processing.
In \emph{reservoir computing}, a fixed, randomly connected recurrent neural network acts as a high-dimensional nonlinear dynamical system, called the reservoir.
Only the linear readout weights mapping reservoir states to the output $\vb{y}(t)$ are trained.
A \emph{time-delay neural network} introduces explicit buffer memory by providing the input at multiple delayed time steps, allowing the use of temporal context without recurrent connections.
\emph{Delay-based reservoir computing} combines delay-based buffering with reservoir dynamics using a single nonlinear node with delayed feedback.
Its continuous output is sampled at discrete times to form \textquote{virtual nodes}, whose states together define the reservoir state vector for the readout.
\emph{Next generation reservoir computing} replaces the random reservoir with structured nonlinear combinations of time-delayed input features, keeping the key memory and nonlinearity of conventional reservoirs while removing the need for a recurrent network.
An \emph{attractor-based recurrent neural network} encodes memory through attractors shaped by the recurrent dynamics. A write input brings the network state into the basin of the corresponding attractor, toward which the recurrent dynamics subsequently converge. The resulting stable or persistent state stores the encoded information after the write signal is removed, under the specified holding condition. 
    }
    \label{fig:summary_nn}
\end{figure}

A schematic summary of the key architectural families and of their conceptual relationships is presented in \cref{fig:summary_nn}.

\paragraph{Feed-Forward Neural Networks}

The perceptron represents the simplest network architecture.
It operates by computing a linear weighted combination of input features and eventually applying a nonlinear activation function to the result \cite{rosenblatt1958perceptron, mcculloch1943logical, minsky2017perceptrons}. The training of the perceptron (i.e. learning) is performed by adjusting the weights in the linear combination to achieve the desired output.
The perceptron is a simple form of a Feed-Forward Neural Network (FFNN) because the output relies exclusively on the current input without temporal dependencies.
In particular, a FFNN is a network of computational units (network nodes or neurons) in which information propagates strictly
from input to output, with no feedback connections.
To address more complex tasks, additional layers can be introduced.
A \emph{layer} is a collection of neurons that receive the same input signal and whose outputs are passed collectively to the next layer of the network. It can be realized by connecting more perceptrons in a complex geometry, e.g. in a multi-layer perceptron network.
By stacking multiple layers, a Deep Neural Network (DNN) is realized. DNN learns increasingly abstract representations of the input data. Individual layers may take various forms, yet they all adhere to the feed-forward paradigm as long as signal
propagation remains strictly unidirectional \cite{lecun2015deep, Rumelhart1986LearningIR, schmidhuber2015deep}. Training is performed by backpropagation, which is a gradient based technique where the error at the output are minimized by adjusting the weights and biases of the neurons backward through the layers \cite{rumelhart1986learning,hughes2018training}.

The most fundamental layer type is the \emph{dense} layer, also referred to as a fully-connected layer, in which every neuron receives all outputs of the preceding layer and computes a weighted sum of its inputs eventually followed by a nonlinear activation function. A \emph{convolutional} layer, by contrast, exploits the spatial structure of the input by sliding a set of learnable filters across it, each producing a feature
map. It is robust to local translations in the input, making them particularly effective for image processing tasks.

FFNN are fundamentally stateless architectures.
These systems possess no intrinsic memory mechanism to retain information from prior time steps. Consequently, the output is determined exclusively by the instantaneous input and remains independent of the temporal history of the input signal.

\paragraph{Recurrent Neural Networks}

On the other hand, Recurrent Neural Networks (RNNs) are designed  to process sequential and time-varying data, by storing temporal information across different time steps. To this end, RNNs incorporate internal hidden states $h$ that evolve through
feedback connections, encoding information from the history of the input signal.
For digital networks, time is discretized:
\begin{equation}
	y(t) = f(h(t)) \text{ and } h(t) = g(x(t), h(t-1))
\end{equation}
where $x(t)$ is the input at time step $t$, $h(t)$ is the hidden state encoding the history of the signal up to time $t$, $h(t-1)$ is the hidden state from the previous time step, $y(t)$ is the network output, and $f$ and $g$ are functions with trainable parameters.
More generally, if time is not discretized, the state evolution is governed by a differential equation
\begin{equation}
	\dv{h}{t} = G(h(t), x(t))
\end{equation}
where $h(t)$ is the continuous-time hidden state, $x(t)$ is the input signal, and $G$ is a function governing the dynamics of the system.
Consequently, the internal state acts as a memory trace of the input history. Thus, the network exhibits intrinsic and dynamic memory \cite{elman1990finding, funahashi1993approximation}.

\paragraph{Reservoir Computing}

A robust method for analyzing the statistical properties of RNNs is to treat them as \emph{reservoirs}. This paradigm is known as Reservoir Computing (RC). In the reservoir, a large number of neurons are randomly connected in a sparse and static recurrent network. The input is thus mapped into a high-dimensional state space (a process of dimensional expansion) before reaching the output layer.
The internal parameters of the recurrent network remain fixed, while only the readout layer is trained \cite{jaeger2004harnessing, maass2002real, lukovsevivcius2009reservoir}. In this way, time dependent data can be processed without incurring the difficulties and computational cost of training RNNs.
In fact, RC circumvents the difficulties associated with the training of RNN, namely numerical instabilities due to backpropagation through time, and reduces the computational cost of training to a simple linear regression on the reservoir states.

A central mechanism underlying the computational power of reservoirs is the dimensionality expansion of the input signal.
The reservoir maps low-dimensional inputs into a high-dimensional space, where linearly inseparable patterns become separable in the expanded representation.
This separation is achieved because the reservoir combines recurrent connectivity, which provides fading memory and temporal integration, and nonlinear activations, which generate diverse high-dimensional responses from simple inputs.
Consequently, any physical or computational system that naturally exhibits recurrent dynamics and nonlinearity is, in principle, a candidate reservoir.

The primary limitation, however, is that the fixed internal dynamics constrain the expressive capacity of the network.

Moreover, this technique enables the systematic quantification of intrinsic properties, such as memory capacity, fading timescales, correlation structure, and generalization ability, without the confounding effects of training recurrent weights.

\paragraph{Time-Delay Neural Networks}
Feed-forward architectures can process temporal information by spatializing time, as seen in Time-Delay Neural Networks (TDNNs), where a sliding window of inputs $\vb{\tilde{x}}(t)$ is presented simultaneously to the network \cite{waibel2013phoneme, wan2019multivariate}.
\begin{equation}
	\vb{\tilde{x}}(t) = [x(t), x(t-1), \dots, x(t-k)]
\end{equation}
While the network topology is feed-forward (no cycles), the system as a whole possesses a finite memory depth $k$ determined by the input buffer \cite{appeltant2011information}.


\paragraph{Next Generation Reservoir Computing}
Time-Delay Neural Networks can be generalized to include nonlinear transformations and products of the input history.
\begin{equation}
	\vb{\tilde{x}}(t) = [x(t), x(t-1), \dots, x(t-k), x^2(t), x(t)x(t-1), \dots , x^2(t-k) \dots]
\end{equation}
This paradigm is termed Next Generation Reservoir Computing (NGRC). It has been demonstrated that a linear readout applied to a vector of delayed inputs and their nonlinear combinations exhibits functional behavior equivalent to that of traditional recurrent reservoirs \cite{gauthier2021next}.
\paragraph{Delay-Based Reservoirs}
A widely adopted architecture in photonics is delay-based RC, which utilizes a physical system typically described as a nonlinear node with delayed feedback and a TDNN\cite{larger2012photonic, paquot2012optoelectronic}.
This configuration transforms the input signal by mixing current values with past states through intrinsic nonlinear dynamics and a feedback loop.
To emulate a network of neurons using this single component, the continuous temporal output is sampled at discrete intervals to define \textquote{virtual nodes} \cite{appeltant2011information, duport2012all, soriano2014delay}.
The virtual nodes
collectively form the reservoir state vector, which is fed to the readout layer.

\paragraph{%
Attractor-Based Recurrent Neural Networks }

RNNs can implement both fading-memory and persistent-state memory. In a fading-memory RNN, the influence of the initial state and remote inputs progressively decays, so that the current state is determined primarily by the recent input history. In an attractor-based RNN, by contrast, the recurrent dynamics support multiple stable attractors. Initialization or a write operation places the network within the basin of a particular attractor, whose identity represents the stored information after the write input is removed \cite{khona2022attractor,hopfield1982neural}. Long-lived metastable states provide an analogous form of retention over a finite interval \cite{brinkman2022metastable,stern2023reservoir}. 

The discrete-time evolution of an RNN can be written as $h(t) = g_{\theta}(x(t),h(t-1))$, where $\theta$ denotes the parameters of the recurrent state-update function. In an attractor-based RNN, each memory state $\mu$ is represented by an attractor $\mathcal{A}_{\mu}$. Its basin of attraction $\mathcal{B}_{\mu}$ contains the network states that evolve toward this attractor when the write input is removed \cite{ye2023controlling}. Let $t_w$ denote the time at which the write operation ends. The operation stores $\mu$ by placing the network state $h(t_w)$ inside $\mathcal{B}_{\mu}$. The subsequent convergence can be summarized as
\begin{equation}
	h(t_w) \in \mathcal{B}_{\mu}
	\quad \Longrightarrow \quad
	\lim_{t \to \infty} \operatorname{dist}\left(h(t),\mathcal{A}_{\mu}\right) = 0,
	\qquad x(t) = x_{\text{hold}} \text{ for } t > t_w,\label{Eq:18}
\end{equation}
where $x_{\text{hold}}$ denotes the operating condition maintained after the write signal is removed. It may be zero for a self-retaining material state or nonzero when continuous optical pumping, electrical bias or feedback is required to maintain the attractor. In \cref{Eq:18}, $\mathcal{A}_{\mu}$ is the set of network states $h$ that belong to the attractor encoding memory $\mu$. The quantity $\operatorname{dist}(h(t),\mathcal{A}_{\mu})$ therefore measures how far the current network state is from this set, and its convergence to zero indicates that the network has reached the attractor representing the written memory. For a fixed-point memory, the set contains only one stable state $h_{\mu}^{*}$ and is written as $\mathcal{A}_{\mu} = \{h_{\mu}^{*}\}$. In the absence of an input, this state satisfies $h_{\mu}^{*} = g_{\theta}(0,h_{\mu}^{*})$, meaning that the recurrent update leaves it unchanged. For a limit cycle, $\mathcal{A}_{\mu}$ instead contains all the states along a repeating trajectory. A continuous attractor consists of a continuous family of states, with the position within this set representing the stored information \cite{wimmer2014bump,seeholzer2019stability}. By contrast, a metastable state retains the stored information for a finite interval but can eventually transition away \cite{brinkman2022metastable}.

Training such a network means adjusting $\theta$, which may include recurrent weights, biases, and input or write couplings, so that the desired memory states become stable attractors with sufficiently robust and distinguishable basins. If the output is written as $y(t) = f_{\phi}(h(t))$, the readout parameters $\phi$ are trained to identify or use the retained state. Depending on the architecture, the recurrent parameters may be learned using an associative learning rule or gradient-based optimization. In a physical implementation whose internal dynamics are fixed by the device, the attractor landscape may instead be engineered or calibrated, while only the write controls and readout are optimized \cite{hopfield1982neural,amit1985storing}. 

More generally, we use the term persistent-state system for a physical device or architecture in which information is encoded in a stable or metastable state that remains distinguishable over the relevant retention time. A persistent-state system is not necessarily a neural network; however, it can implement or support an attractor-based RNN when its retained states participate in an interconnected recurrent computational architecture. 

\paragraph{Transformers}
In recent years, a paradigm shift has occurred with the introduction of the Transformer architecture.
Unlike RNNs, which must process inputs sequentially to update a hidden state, Transformers ingest the entire data sequence simultaneously.
While technically a feed-forward architecture due to the absence of internal feedback loops, the Transformer utilizes a self-attention mechanism to handle temporal context.
By computing similarity scores between all pairs of elements in the sequence, the system estimates the importance of every input relative to every other input, thereby establishing an explicit, content-dependent memory \cite{vaswani2017attention}.
This enables the identification of dependencies between distant data points directly, allowing access to a global memory of the sequence without the constraints of serial iteration. This architecture underlies modern large language models (LLMs).
Consequently, the attention mechanism shifts the computational burden from arithmetic operations to memory bandwidth and storage \cite{gholami2024ai}.

While research on photonic implementations of standard feed-forward and recurrent architectures is well-established although far from complete, photonic hardware explicitly designed to physically emulate the self-attention mechanism of Transformers is a very new area of research \cite{tian2025photonic}.

\subsection{%
Fading-Memory Characterization }
\label{sec:memory_char}

An advantage of RC is the mathematical and statistical tractability of the system dynamics.
This has facilitated the development of metrics to quantify properties such as memory depth and dimensionality expansion \cite{dambre2012information, jaeger2001echo}.
In practice, these procedures are versatile and can be adapted to evaluate any neural network architecture possessing intrinsic memory \cite{barak2017recurrent, vandoorne2014experimental}.

\paragraph{Response to Constant Input and Edge of Chaos}

Analyzing the response of a reservoir to a constant input reveals the system's autonomous dynamical regime.
Depending on the input magnitude, the system may settle into a fixed point, enter a limit cycle, or exhibit chaotic behavior.
This baseline response is critical because it determines the stability required for reliable computing.
Typically, a stable fixed point is preferred in standard RC to ensure the Echo State Property (ESP), which guarantees that the system's state is a function of the input history rather than its initial conditions \cite{jaeger2001echo, lukovsevivcius2009reservoir, yildiz2012re}.

By varying the constant input level, one can map bifurcation diagrams to identify transitions between stable, oscillatory, and chaotic regimes (\cref{fig:2_level_system}).
The dynamical behavior of a reservoir can be characterized in terms of two opposing tendencies: \emph{contraction}, in which nearby trajectories converge, and \emph{expansion}, in which they diverge. Strong contraction causes the system to forget past inputs rapidly, while strong expansion leads to chaotic behavior that degrades reproducibility.

In some reservoir models, the transition from contractive dynamics to unstable or chaotic dynamics is referred to as the edge of chaos \cite{langton1990computation}. Operation near this boundary can increase sensitivity to past inputs and enrich the generated features. In fact, perturbations may decay more slowly without being amplified uncontrollably \cite{jaeger2001echo,maass2002real}. Such behavior can support a useful balance between memory and nonlinear processing, and operation in this regime has been found empirically favorable for some reservoir-computing tasks \cite{fang2019nonequilibrium}. However, it can also reduce reproducibility and noise tolerance. Its computational benefit therefore depends on the task, input scaling, reservoir architecture, noise level, and readout, and it should not be regarded as a universally optimal operating regime \cite{bertschinger2004edge, legenstein2007edge, boedecker2012information}. 

\paragraph{Echo State Property}

For conventional fading-memory reservoir computing, the ESP is a key condition for reliable operation \cite{jaeger2001echo, yildiz2012re, lukovsevivcius2009reservoir}. The ESP guarantees that the internal state of the reservoir is asymptotically determined solely by the history of the input signal, rather than by its initial conditions. Thus, trajectories initialized in different states must converge when driven by the same input history \cite{manjunath2013echo, buehner2006tighter, gallicchio2017echo}. In photonic hardware intended to operate as a fading-memory reservoir, optical feedback and nonlinearities should provide rich dynamics while remaining sufficiently contractive to ensure this state convergence \cite{vandoorne2014experimental, larger2012photonic, brunner2013parallel}. 

Violation of the ESP through chaotic divergence or uncontrolled initial-state dependence makes a system unsuitable for tasks requiring a unique fading-memory response \cite{wainrib2016local, bertschinger2004edge}. In persistent-state or attractor-based systems, however, controlled dependence on a previously written state is intentional and represents the stored information  \cite{chaudhuri2016computational,khona2022attractor}.

\paragraph{Impulse Response}

The impulse response, or the system's reaction to a single spike, provides a direct window into the memory structure of the reservoir. In purely linear systems, this response corresponds to the Green's function, which fully characterizes the system's temporal behavior. For nonlinear reservoirs, analyzing the relaxation time of the transient response to a spike serves as a proxy for the "fading memory" timescale \cite{ganguli2008memory, dambre2012information, white2004short}.
This measurement physically quantifies the effective memory depth by determining how long the hardware reverberates a single bit of information before returning to its baseline state \cite{vandoorne2014experimental, appeltant2011information}.
If the response does not fade, for instance, if the spike triggers sustained limit cycles or chaotic divergence, the ESP may be violated, rendering the system unstable for standard RC tasks \cite{manjunath2013echo, yildiz2012re, wainrib2016local}.


\paragraph{Memory Capacity and Information Processing Capacity}

Linear Memory Capacity (MC) is a standard metric used to quantify the ability of a reservoir to reconstruct past input information from its current state \cite{white2004short, ganguli2008memory}.
To measure this property, the system is driven by a random input sequence $x(t)$ drawn from a uniform distribution, ensuring that the input values are uncorrelated over time.
The distribution interval and input scaling form part of the measurement specification because they affect the response of nonlinear systems. 
The full vector of reservoir states $\vb{h}(t)$ is recorded at each time step. For each specific time delay $k$, a linear readout is trained to approximate the delayed input $x(t-k)$ using the state vector $\vb{h}(t)$. The reconstruction quality for each delay $k$ is evaluated using the coefficient $\text{MC}_k$, defined as $\text{MC}_k = 1 - \frac{\text{SSE}_k}{\text{SST}_k}$, where $\text{SSE}_k$ is the sum of squared errors between the target and the reconstruction, and $\text{SST}_k$ is the total sum of squares of the mean-centered target signal.

By construction, $\text{MC}_k$ is the coefficient of determination $R^2$ for each evaluated lag $k$. 
For an unregularized ordinary least-squares regression with an intercept, evaluated on the same data used for fitting,  $R^2$ equals the squared Pearson correlation coefficient between the target and its reconstruction. This equality does not generally hold for regularized models or predictions evaluated on an independent test dataset. 
A complete measurement protocol should therefore report the input distribution and range, number of observable states, washout length, training and test lengths, regression method and any regularization method, evaluation dataset and maximum lag $k_{\max}$. 

Formally, the total Memory Capacity (MC) is the sum of $\text{MC}_k$ over all positive delays, $\text{MC} = \sum_{k=1}^{\infty} \text{MC}_k$. In a finite measurement, MC is estimated by summing up to a maximum lag $k_{\max}$, chosen sufficiently large that $\text{MC}_k$ has decayed to zero within the statistical uncertainty or measurement noise floor. The selected $k_{\max}$ therefore forms part of the measurement specification. 
The lag $k$ measures memory in input samples. If $T_s$ is the input sampling or symbol period, the corresponding physical delay is $T_{\mathrm{k}} = kT_s$.
Both quantities should be reported when comparing systems operating at different sampling or symbol rates. 
For binary inputs, the $\mathrm{MEM}_k$ task described in \cref{sec:benchmarks} uses the same delayed target $x(t-k)$ as the lag-resolved memory-capacity measurement. However, BER and $\text{MEM}_k$ quantify performance differently and should be reported as distinct metrics.
Furthermore, for a purely linear system, the profile of $\text{MC}_k$ versus $k$ is related to the squared envelope of the system's impulse response, providing a link between statistical memory capacity and physical transient dynamics.

However, MC does not account for the nonlinear memory capabilities of the network. To address this limitation, MC is extended by training the readout to predict nonlinear functions of the input history.
For example, reconstructing powers of delayed inputs, such as $x(t-k)^2$, yields to nonlinear memory capacities, while predicting products of distinct past inputs, such as $x(t-k)x(t-j)$, assesses cross-term memory capacities. Nevertheless, these specific measures provide only a partial description of the nonlinear memory of the system.
A more comprehensive framework is the Information Processing Capacity (IPC) \cite{dambre2012information, inubushi2017reservoir}.
The total computational capacity of a dynamical system can be decomposed into the sum of capacities for reconstructing a complete set of orthogonal basis functions defined over the input history.

One defines a set of orthogonal basis functions, such as the Legendre polynomials. For every function $f$ in this set and every delay $k$, the target output is generated from the input history. Its normalized reconstruction capacity is $C_{f,k} = 1 - \text{SSE}_{f,k}/\text{SST}_{f,k}$. The capacity associated with a specific basis function is then obtained by summing over all delays, $C_f = \sum_k C_{f,k}$, and the global Information Processing Capacity is obtained by summing over all basis functions, $\text{IPC} = \sum_f C_f = \sum_f \sum_k C_{f,k}$. 

Although theoretically exhaustive, calculating the full IPC is often very computationally expensive. The number of required basis functions grows combinatorially with the polynomial degree and memory depth, making the evaluation of high-order capacities impractical.

In telecom, Pseudo-Random Bit Sequences (PRBSs) are usually employed as test series for characterizing optical links. 
A PRBS is a deterministic binary test sequence, commonly generated by a linear-feedback shift register, that is designed to reproduce the statistical properties of a random binary input. A maximal-length PRBS of order $m$ has a period length of $P=2^m-1$ bits. After mean subtraction or bipolar mapping to (-1,+1), its nonzero-lag cyclic correlations scales as 1/P, allowing it to approximate an independent binary sequence over lags much shorter than its period. Measurements should therefore specify the PRBS order, period, numerical and physical symbol encoding, and maximum evaluated lag, with $k_{\max} \ll P$. Performance must be reported using a metric such as Bit Error Rate (BER), reconstruction error, $\text{MC}_k$ or MC. 

\paragraph{Fading Memory}

To experimentally verify the fading memory property, one can assess the convergence of the system state under identical driving conditions. This procedure involves subjecting the reservoir to distinct random input sequences that switch to a common input sequence at a specific point in time. If the system adheres to the ESP, the state trajectories generated during the common phase must asymptotically synchronize. The divergence between the two trajectories should decay to zero, demonstrating that the system has forgotten the disparate initial histories. This convergence confirms that the current state is determined solely by the recent input history rather than the initial conditions \cite{jaeger2001echo}.

\subsection{%
Persistent-State Memory Metrics }
\label{sec:persistent_state_metrics}

Persistent-memory metrics do not belong to a single standardized framework: their terminology is drawn from information theory \cite{shannon1948mathematical}, attractor-network and stochastic-dynamics literature \cite{khona2022attractor,brinkman2022metastable}, physical memory devices \cite{sebastian2020memory}, and oscillator physics \cite{demir1998phase}. We therefore use operational definitions based on the complete write--retain--read--reset cycle. These metrics are particularly useful for characterizing attractor-based RNNs, in which memory is encoded in stable or metastable states. They describe whether the intended state is reached, how long it remains available, whether it can be identified, and how it responds to perturbations.

\paragraph{%
Write Success Rate }

The write success rate quantifies how reliably a write operation places the system in the requested state. For a target state $\mu$, the operation is repeated and the fraction of trials ending in $\mu$ provides the empirical success rate. This fraction estimates $p_{\mathrm{write},\mu}^{\mathrm{succ}} = P(s_{\mathrm{after}} = \mu \mid \text{write } \mu)$, where $s_{\mathrm{after}}$ is the state observed after writing \cite{worledge2010switching,apalkov2016magnetoresistive}.

\paragraph{%
Read Success Rate }

Once a state has been prepared, the read success rate describes whether the readout recognizes it correctly. A known state $\mu$ is read repeatedly, and the fraction of correct labels estimates $p_{\mathrm{read},\mu}^{\mathrm{succ}} = P(\widehat{s} = \mu \mid s = \mu)$. Here, $s$ denotes the stored state and $\widehat{s}$ the label assigned by the readout \cite{shannon1948mathematical,apalkov2016magnetoresistive}.

When the memory has $M$ available states, the state-specific read outcomes can be collected in a read confusion matrix. Each element records how often a stored state $i$ is identified as state $j$:
\begin{equation}
	C_{j \mid i}^{\mathrm{read}} = P\left( \widehat{s} = j \mid s = i \right).
\end{equation}
The diagonal element $C_{i \mid i}^{\mathrm{read}}$ is the read success rate for state $i$. The off-diagonal elements $C_{j \mid i}^{\mathrm{read}}$, with $j \neq i$, are the probabilities of specific read errors in which state $i$ is identified as state $j$. Their sum gives the total read error rate for state $i$, $1-C_{i \mid i}^{\mathrm{read}}$. The matrix therefore provides a conditional-probability description of the complete readout channel \cite{shannon1948mathematical}.

\paragraph{%
Storage Capacity }

Storage capacity describes how many distinct memories can be encoded in a network and subsequently retrieved. In attractor networks, each memory is represented by a stable attractor, and the capacity is studied by increasing the number of stored patterns until interference between them compromises their stability. It therefore depends on the network size, the learning rule, and the required robustness of the attraction basins \cite{amit1985storing,mceliece1987capacity}. In a physical implementation, the same concept corresponds to the number of distinct memory states that the system can reliably support.

\paragraph{%
Dwell-Time Distribution }

For metastable states, retention is characterized by the dwell time $\tau$, namely the interval during which the system remains in the written state before a spontaneous transition. The same metric applies to nominally stable attractors when noise can induce transitions between their basins. Preparing the same state repeatedly and recording $\tau$ in each trial produces the dwell-time distribution \cite{mazzucato2015dynamics,brinkman2022metastable}.

\paragraph{%
Noise-Induced Transition Rate }

Whereas dwell times describe how long a state survives, the noise-induced transition rate identifies how frequently one state changes into another. During observation without a switching command, the number of transitions from $\mu$ to $\nu$ is divided by the total time spent in $\mu$. The resulting rate is denoted by $\Gamma_{\mu \to \nu}$ \cite{bressloff2010metastable,moreno2007noise}.

\paragraph{%
Basin Robustness }

The stability of a stored state can also be tested directly by perturbing it. After preparing state $\mu$, perturbations of different amplitudes are applied and the subsequent evolution is observed. The fraction of trials that return to the attractor representing $\mu$, considered as a function of perturbation amplitude, characterizes the robustness of its basin \cite{storkey1999basins,ye2023controlling}.

\paragraph{%
Reset Time and Energy }

At the end of the memory cycle, the system is returned to a reference state. The elapsed time between the reset command and arrival at that state defines the reset time. The reset energy is obtained by integrating the total applied power over the reset interval, which at fundamental thermodynamic limits is constrained by the energy dissipation required for state erasure \cite{proesmans2020finite}. Together, these quantities describe the time and energy required to prepare the memory for a subsequent write operation in physical implementations \cite{wong2010phase,apalkov2016magnetoresistive}. The reported value should state whether it includes optical pumps, electrical biasing and control electronics.

\paragraph{%
Phase Diffusion }

When information is encoded in the phase of an oscillator or limit cycle, retention requires a phase-sensitive metric. Starting repeatedly from the same initial phase $\phi(0)$, one records the phase $\phi(t)$ after an elapsed time $t$. Noise causes the phase trajectories to spread, and the growth of their variance defines the phase-diffusion coefficient $D_{\phi}$ through
\begin{equation}
	\operatorname{Var}\left[ \phi(t) - \phi(0) \right] \approx 2 D_{\phi} t.
\end{equation}
In this expression, $\operatorname{Var}$ denotes the variance across repeated trials. Its slope is $2D_{\phi}$, so a larger diffusion coefficient corresponds to a faster loss of the initially stored phase \cite{lax1967classical,demir1998phase}. 

\subsection{Benchmark Tasks}
\label{sec:benchmarks}

The metrics introduced in \cref{sec:memory_char} quantify intrinsic properties of a computing substrate, such as memory depth and fading timescale, in a task-agnostic manner.
While this generality is valuable for characterizing and comparing physical systems, it does not directly predict performance on a specific application.
This distinction is formalized by the \emph{No Free Lunch theorem} \cite{wolpert2002no}, which states that no learning algorithm or computing architecture can outperform any other when performance is averaged uniformly over all possible tasks.
In practice, this implies that a system optimized according to general memory metrics may not excel on a specific target task, and vice versa.
Consequently, task-specific benchmarks serve as an essential complement to general characterization metrics: they probe whether the memory and nonlinearity provided by a physical substrate translate into useful computation on concrete problems.

Benchmark tasks therefore occupy a central role in the evaluation of PNNs \cite{yik2025neurobench}.
They provide reproducible, community-accepted figures of merit that enable direct comparison across implementations, platforms, and computational paradigms.
In what follows, we survey the benchmark tasks most commonly reported.

\paragraph{Benchmark Tasks involving Binary Inputs}

To characterize the computational power of a photonic system, a binary input stream $x(t)$, typically a PRBS is employed \cite{argyris2005chaos}.
A binary input consists of two distinguishable symbols mapped to the logical values false and true, independently of their physical representation. At the output, the readout is decoded into a predicted logical label using an appropriate decision rule, such as a threshold. Binary-task performance is quantified by the BER, defined as the fraction of predicted labels that differ from their target labels. 

In the specific context of optical telecommunications, a primary challenge is the reconstruction of the transmitted sample series following propagation through a dispersive optical fiber. Consequently, the target for channel equalization is defined as the original input $y(t) = x(t)$.

The simplest task is the Memory task ($\text{MEM}_n$), where the objective is to reconstruct the input bit from $n$ steps earlier, such that $y(t) = x(t-n)$. This task strictly evaluates the linear fading memory of the system \cite{jaeger2001echo, white2004short, dambre2012information}.

Another task is the $\text{AND}_n$ operation, which requires the system to output the logical AND of the current bit and the delayed bit $x(t-n)$, formally %
$y(t) = x(t) \land x(t-n)$ . This tests the ability of the reservoir to perform cross-integration of immediate and past information.
The most demanding task is the delayed $\text{XOR}_n$ operation, defined as $y(t) = x(t) \oplus x(t-n)$. Since the XOR function is not linearly separable, solving this task confirms that the reservoir can combine relatively strong nonlinearity and fading memory \cite{duport2012all}.

The \emph{header recognition} task is a standard benchmark that evaluates the ability of a reservoir to identify a specific bit pattern within a continuous input stream.
In this task, a binary sequence is continuously fed into the system, and the target output is a high value whenever the last $n$ bits of the input match a predefined header pattern, and a low value otherwise.
This benchmark simultaneously probes both the memory depth of the system, since the reservoir must retain the last $n$ bits, and its ability to perform nonlinear pattern matching, as it must distinguish the target sequence from all other possible bit combinations \cite{vandoorne2014experimental}.
The task difficulty scales with the header length $n$: longer headers require deeper memory and more expressive nonlinear transformations.
For this reason, header recognition of increasing lengths (e.g., 3-bit, 6-bit) is frequently used to benchmark integrated photonic reservoirs and to compare the effective memory depth across different hardware implementations.
Header recognition is usually formulated as one predefined header against all other $n$-bit patterns. For an approximately independent and balanced binary input, the selected header occurs with probability approximately $2^{-n}$, making the task strongly class-imbalanced. Class-sensitive measures such as precision, recall, F1 score, balanced accuracy, or a complete confusion matrix therefore provide a more informative evaluation than BER alone. 

In experimental photonic realizations, the output signal is typically sampled, digitized, and in many cases thresholded before the final decision is formed. This step can act as an additional nonlinear transformation and may substantially improve classification accuracy beyond what would be achievable from a strictly linear readout of the raw analog state. Accordingly, benchmark results should specify clearly whether thresholding is part of the evaluated physical computing substrate or an external post-processing step \cite{soriano2014delay, van2017advances}.

\paragraph{%
NARMA-N as a Nonlinear Temporal Benchmark }

The Nonlinear AutoRegressive Moving Average (NARMA) task is a widely adopted nonlinear temporal benchmark for evaluating the computational capabilities of recurrent neural networks and RC systems \cite{atiya2000new, rodan2010minimum}. Its performance depends jointly on fading memory, nonlinear transformation, input distribution and scaling, readout dimension, fitting procedure and noise. It therefore provides a useful measure of the combined temporal-processing capability of a system, but it cannot isolate memory capacity from the other contributions. Specifically, the NARMA-N task requires the system to predict the value of a target sequence $y(t)$ based on the current input $x(t)$ and its history over the past $N$ steps. The target is generated by the recursive formula:
\begin{equation}
	y(t+1) = \alpha y(t) + \beta y(t) \left( \sum_{i=0}^{N-1} y(t-i) \right) + \gamma x(t-N+1) x(t) + \delta
\end{equation}
where $\alpha, \beta, \gamma, \delta$ are constants typically set to ensure long-term dependencies and nonlinear interactions. For instance, in the standard NARMA-10 task ($N=10$), the output depends on the product $x(t-9)x(t)$ of the current and delayed inputs, as well as a summation of previous outputs, creating a complex dependency that requires both significant memory depth and nonlinear processing capabilities.

Because the task involves cross-terms between delayed inputs and recursive nonlinearities, NARMA-N is a benchmark task that jointly tests the ability of a system to retain information and perform nonlinear operations on it. Its construction is readily interpretable because the required delayed and nonlinear interactions are explicitly defined. Consequently, although NARMA-N is not a formal capacity measure, its performance provides a useful proxy estimate of nonlinear memory. 
Standard NARMA-N configurations use defined parameter values and a bounded input drawn from a specified distribution and range, improving comparability across systems \cite{wringe2025reservoir}. 
However, practical implementation of the NARMA task requires care. The recursive nature of the target generation means that for certain input sequences or parameter settings, the target value $y(t)$ can diverge or exhibit numerical instability over long sequences. To mitigate this, it is common to use bounded input sequences or to reset the target generation periodically.
Furthermore, a significant efficiency advantage of this benchmark is data reuse. The same random input sequence $x(t)$ used to calculate the linear MC can be used to generate the NARMA target $y(t)$ in post-processing. %
The same experimental trace can therefore support the calculation of linear MC, IPC, and evaluation of NARMA task performance, reducing the data-acquisition time required for hardware characterization. 

\paragraph{Time-Series Prediction}

Time-series prediction tasks require the reservoir to reconstruct the future state of a dynamical system from its past trajectory.
These benchmarks are widely used to assess the joint ability of a computing substrate to retain a sufficient history of past inputs and to apply nonlinear transformations to that history.

The \emph{Santa Fe} dataset consists of a single univariate time series measured from a far-infrared laser operating in a chaotic regime \cite{weigend2018time}.
The task requires the system to predict one of the following samples given a window of past observations.
Because the underlying attractor is low-dimensional yet strongly nonlinear, the Santa Fe task probes the balance between memory depth and nonlinear expressivity.
Performance is typically reported as the Normalized Mean Square Error (NMSE) or as the Normalized Root Mean Square Error (NRMSE).

The \emph{Mackey-Glass} series is generated by the delay-differential equation
\begin{equation}
	\dv{x}{t} = \frac{\beta\, x(t - \tau)}{1 + x(t-\tau)^n} - \gamma x(t),
\end{equation}
where the delay $\tau$ controls the complexity and dimensionality of the resulting attractor \cite{mackey1977oscillation}.
For $\tau = 17$ the dynamics are weakly chaotic, while $\tau = 30$ yields a higher-dimensional, more demanding attractor.
The task is single-step-ahead prediction, reported as NMSE.
Because the Mackey--Glass series is generated deterministically, the dataset can be reproduced exactly, which facilitates reproducible comparison across implementations \cite{jaeger2004harnessing, vandoorne2014experimental, donati2024time}.

The \emph{Lorenz63} system is a canonical three-dimensional chaotic attractor described by the Lorenz equations
\begin{equation}
    \begin{dcases}
        \dv{x}{t} &= \sigma (y - x) \\
        \dv{y}{t} &= x (\rho - z) - y \\
        \dv{z}{t} &= xy - \beta z \\
    \end{dcases}
\end{equation}
where $x$, $y$ and $z$ are the state variables and $\sigma$, $\rho$ and $\beta$ are the system parameters \cite{lorenz1963deterministic}.
The standard choice of $(\sigma, \rho, \beta)$ = $(10, 28, \dfrac{8}{3})$ places the system in a chaotic regime, producing a strange attractor and rendering long-term prediction of the trajectory impossible.

\paragraph{Classification Tasks}

Classification benchmarks evaluate the ability of a system to assign a discrete label to an input pattern, rather than to reconstruct a continuous target trajectory. Different benchmarks are reported at \cite{kelly2024welcome}. Here we list the most commonly used ones.

\emph{Iris-flower classification} is a multiclass classification task on the four-feature Iris dataset \cite{fisher1936use}.
Samples are drawn equally from three species of iris flowers, with 50 samples per class.
Each sample is described by four morphological features: sepal length, sepal width, petal length, and petal width, all measured in centimeters.
Because the three classes are not linearly separable in the full four-dimensional feature space, a linear discriminant applied directly to the raw features achieves a classification accuracy of approximately \qty{96}{\percent} \cite{fisher1936use}.
The residual misclassification originates from the partial overlap between some classes, whose feature distributions are not disjoint.
The motivation for employing a nonlinear computing substrate is precisely to resolve this overlap: a nonlinear transformation of the input features can map the original feature space into a higher-dimensional representation in which the three classes become linearly separable, thereby improving classification accuracy beyond the linear baseline.
For this reason, the Iris task serves as a compact and reproducible benchmark for assessing whether a physical system provides useful nonlinearity, independently of the presence of optical memory.

\emph{MNIST handwritten digit classification} presents grayscale images of handwritten digits from zero to nine.
The dataset comprises \num{70000} grayscale images of size \numproduct{28x28} pixels, quantized to 256 intensity levels, of which \num{10000} are reserved for testing \cite{lecun2002gradient}.
A linear classifier trained with cross-entropy loss typically achieves approximately \qty{92}{\percent} accuracy on this benchmark \cite{foradori2025neuromorphic}.
A standard multilayer perceptron is not well suited to this task, as it lacks the translational invariance and local feature extraction capabilities that are essential for image recognition.
Convolutional Neural Networks (CNNs), by contrast, are specifically designed for structured spatial data and represent the dominant approach for MNIST classification, achieving accuracies well above \qty{99}{\percent} \cite{krizhevsky2012imagenet, ciregan2012multi}.
Nonetheless, MNIST is routinely employed as a benchmark for RNN and RC systems, where the image is unrolled into a flat temporal sequence of pixels and presented to the network row by row or pixel by pixel.
This sequential encoding deliberately ignores the spatial structure of the image and instead stresses the temporal memory of the system.
The primary motivation for adopting this encoding is interpretability and comparability: the benchmark is widely recognized, the baseline performances are well established, and the sequential formulation provides a direct measure of the effective memory depth and nonlinear processing capability of the physical substrate, independently of spatial inductive biases \cite{jaeger2004harnessing, tanaka2019recent}.
Moreover, this dataset is frequently adapted to lower complexity settings by reducing the number of training samples or restricting evaluation to a binary subset of classes, such as the discrimination of digits zero and one.

\emph{Fashion-MNIST} \cite{xiao2017fashion} is a direct drop-in replacement for MNIST that substitutes handwritten digits with grayscale images of clothing articles, drawn from ten categories such as T-shirts, trousers, and ankle boots.
The dataset preserves the same structure as MNIST: \num{70000} images of size $28 \times 28$ pixels, with \num{10000} samples reserved for testing.
However, Fashion-MNIST is considerably more challenging than its digit counterpart.
A linear classifier trained on the raw pixel values achieves approximately \qty{85}{\percent} accuracy.
Moreover, unlike handwritten digit images, whose pixel intensity distributions are nearly binary (dark background and bright stroke), Fashion-MNIST images exploit a much richer grayscale dynamic range.
Fabric textures, shadows, and gradual intensity transitions produce smooth spatial variations across the image, so that the full eight-bit depth of the pixel values carries discriminative information.

Similar classification benchmarks exist for temporal audio data.
\emph{Spoken digit recognition} requires the system to identify isolated spoken digits, typically zero through nine, from a set of speakers.
Unlike static image datasets, each sample is a time-varying acoustic waveform, so the task inherently probes the temporal memory and nonlinear processing capability of the computing substrate.
A widely used dataset for this benchmark is the Free Spoken Digit Dataset (FSDD) \cite{jackson2018jakobovski}, which provides recordings of spoken digits from multiple speakers at a fixed sampling rate.
Because the discriminative information is distributed across time, a system with no memory cannot solve this task by processing a single sample in isolation.

\begin{table}[t]

\caption{Comparison between biological and integrated photonic memory: computational role, learning, and system-level implications.}
\label{tab:bio_photonic_memory_computational}
\renewcommand{\arraystretch}{1.2}
\tablealtrows
\begin{tabular}{@{}p{0.22\textwidth} p{0.35\textwidth} p{0.35\textwidth}@{}}
\toprule
Aspect & Biological memory & Integrated photonic memory \\
\midrule
Computation--memory relation & Strongly co-localized: the same substrate stores history and computes & Often partially co-localized in the photonic substrate, but frequently hybridized with electronics for control, detection, and training \\

Learning mechanism & Local plasticity gated by biochemical and neuromodulatory signals; Hebbian/STDP-like correlation rules & Usually external or offline training, or electronic control; limited demonstrations of local in-situ physical plasticity \\

Nonlinearity source & Neuronal thresholding, synaptic saturation, biochemical cascades, and recurrent collective dynamics & Material nonlinearity, resonant enhancement, gain saturation, injection locking, detector square-law response, and external electronics \\

State update locality & Typically local or quasi-local at synapses and neurons & Often local at the device level in principle, but in practice frequently assisted by global control and readout circuitry \\

Scalability constraint & Metabolic cost, anatomical wiring, reliability, and biological speed & Footprint, optical loss, tuning power, detector overhead, phase stability, and fabrication tolerances \\

Energy-efficiency principle & Extremely efficient due to sparse, event-driven, massively parallel distributed dynamics & Potentially efficient for high-bandwidth analog operations, but total efficiency depends strongly on sources, tuning, detection, and control electronics \\

Bandwidth / operating speed & Slow compared with photonics; biologically relevant dynamics are typically in the Hz--kHz range at spike and network levels & Extremely fast; photonic signals can support several GHz carrier dynamics and very high-throughput \\

Memory organization & Address-free, distributed, associative, and engram-like & Can be distributed and dynamical, but also device-local and programmable; may combine associative dynamics with addressable weight setting \\

Recall / readout & Reactivation of neuronal ensembles and network attractors & Optical or electrical readout of field, transmission, phase shift, resonance shift, or detector current \\

Best conceptual analogue & Memory as distributed adaptive dynamics & Memory as a hierarchy of physical state variables with different retention times \\

\botrule
\end{tabular}
\end{table}

\section{Integrated Photonic Neural Networks with Memory}\label{sec:PNNs}

Before discussing specific photonic neural-network architectures, it is useful to compare biological and integrated photonic memory from the point of view of computation, learning, and system-level constraints. \Cref{tab:bio_photonic_memory_computational} summarizes these aspects and clarifies how differences in locality, nonlinearity, programmability, and readout affect the architectural role of memory in neuromorphic photonics.
\Cref{sec:memory} reviewed the physical mechanisms that provide memory in integrated photonic platforms, ranging from linear delay-based approaches to nonlinear and multistable dynamics, as well as non-volatile material reconfiguration. Here we shift perspective and treat these mechanisms as computational primitives embedded in PNN. Accordingly, this section focuses on integrated photonic implementations that exploit memory to process time-dependent data, with emphasis on the ML viewpoint.

Rather than classifying devices only by their underlying physics, we discuss how memory contributes to task performance, including accuracy, energy efficiency, and area efficiency, across representative benchmarks and application-driven workloads. This viewpoint highlights the design trade-offs that are most relevant to learning systems, such as the balance between memory depth and signal bandwidth, the placement of nonlinearity inside or outside the photonic network, and the extent to which a system is trainable. A further distinction used throughout this section concerns system completeness. We differentiate between fully integrated architectures, hybrid systems that rely on external optical or electronic support, and proof-of-principle demonstrators in which part of the relevant memory, nonlinearity, or stabilization is located outside the chip. This distinction is essential when assessing scalability, energy efficiency, and architectural relevance.

A comparison of the different memory mechanisms discussed in this chapter is provided in \cref{tab:memory_comparison}. To distinguish between optical propagation losses from the energy required to preserve a memory state, \cref{tab:memory_comparison} reports both the dominant loss mechanism and the corresponding state-maintenance condition. The latter refers specifically to the power or feedback required after the writing has ended and should not be confused with the energy required for signal generation, tuning, programming, readout, or detection.

\begin{sidewaystable}[p]
	\caption{Comparison of working memory mechanisms used in PNNs. \emph{$\lambda$-sensitive} indicates sensitivity to changes in input optical wavelength.
	 The \emph{Loss/state maintenance} column separately reports the main optical loss (\emph{Loss}) and the condition required to retain the relevant memory state (\emph{Hold}). \emph{Hold: none} indicates that no continuous state-holding power is required beyond the propagating or initially excited optical signal; it does not imply zero energy for signal generation, tuning, programming, readout, or detection. The \emph{Evidence} column indicates the breadth of experimental support for each physical mechanism: \emph{S},  numerical studies only; \emph{E}, experimental demonstration by one research group; \emph{M}, experimental demonstrations by multiple groups. This classification does not represent system-level integration maturity, which is specified in the text for individual implementations by distinguishing device experiments, integrated optical cores with offline readout, hybrid real-time processors, complete integrated inference systems, and systems with integrated or in-situ training.
	Abbreviations: \emph{opt.}, optical; \emph{temp.}, temperature; \emph{fab.}, fabrication; \emph{mem.}, memory; \emph{pow.}, power; \emph{amorph.}, amorphization; and \emph{cryst.}, crystallization.}
	\label{tab:memory_comparison}
	\renewcommand{\arraystretch}{1.2}
	{\footnotesize
	\setlength{\tabcolsep}{3pt}
	\tolerance=9999 \hyphenpenalty=0 \exhyphenpenalty=0 \emergencystretch=1em
	\tablealtrows
	\begin{tabular}{@{}
		>{\centering \arraybackslash}m{.051\textheight}
		>{\centering \arraybackslash}m{.076\textheight}
		>{\centering \arraybackslash}m{.066\textheight}
		>{\centering \arraybackslash}m{.101\textheight}
		>{\centering \arraybackslash}m{.197\textheight}
		>{\centering \arraybackslash}m{.188\textheight}
		>{\centering \arraybackslash}m{.159\textheight}
		>{\centering \arraybackslash}m{.086\textheight}
		@{}}
		\toprule
		Ref & Mech. & Mem. scale & Input type & Loss / state maintenance & Pros/cons & Type & Evidence \\
		\midrule
		\cite{vandoorne2014experimental,sackesyn2021experimental,sunada2021photonic,nakajima2021scalable,mancinelli2022photonic,staffoli2024silicon,wang2024ultrafast,chen2024efficient}
		& Inte\-grated delay lines
		& \unit{\nano\second}--\unit{\pico\second}
		& High-speed opt. signal
		& Loss: transmission. Hold: none beyond signal propagation.
		& High-speed, broadband / chip area, short mem.
		& Volatile state: finite-window delay
		& M \\
		\cite{laporte2018numerical,heinsalu2024silicon,takano2018compact,harkhoe2020demonstrating,li2023packet,ren2024photonic,nakajima2021scalable}
		& Optical cavities (linear)
		& \unit{\nano\second}--\unit{\pico\second}
		& High-speed opt. signal
		& Loss: cavity and insertion loss. Hold: none for the photon state; heater power may be required for resonance alignment
		& High-speed, $\lambda$-sensitive, compact / narrow band, short mem., temp. and fab. sensitive
		& Volatile state: fading, photon lifetime
		& M \\
		\cite{donati2024time}
		& Exter\-nal opt. fiber
		& \unit{\micro\second}--\unit{\nano\second}
		& High-speed opt. signal
		& Loss: transmission and insertion loss. Hold: none beyond signal propagation
		& Broadband, \unit{\micro\second}-scale mem. / bulky, phase noise, temp. sensitive
		& Volatile state: finite-window delay
		& E \\
		\cite{borghi2021reservoir,donati2024time,donati2021microring,bazzanella2022microring,giron2024effects,gretter2025dynamic,lugnan2025emergent}
		& MRR nonlinearity
		& \unit{\micro\second}--\unit{\nano\second}
		& sub-GHz, opt. signal \(\gtrsim\) \unit{\milli\watt}
		& Loss: cavity and nonlinear absorption. Hold: none for the fading state; optical drive creates the state and heater power may be required for alignment
		& $\lambda$-sensitive, compact, nonlinear, multiscale mem. / fab. and temp. sensitive, nonlinear distortion
		& Volatile state: fading nonlinear relaxation
		& M \\
		\cite{donati2024time,lugnan2025reservoir,donati2025all}
		& Self-Pulsing MRR
		& Non-\newline fading
		& sub-GHz, opt. signal \(\gtrsim\) \unit{\milli\watt}
		& Loss: cavity and nonlinear absorption.Hold: continuous resonant optical pump
		& $\lambda$-sensitive, compact, nonlinear, persistent mem. / fab. and temp. sensitive, nonlinear distortion, needs resetting
		& Volatile, optically sustained state: persistent oscillatory or multistable dynamics
		& E \\
		\cite{gao2022thin}
		& Opto\-fluidics, thermocapillary
		& ms
		& Opt. pulse sequence, sub-\unit{\milli\watt}
		& Loss: gold-patch absorption. Hold: optical heating for the bistable operating state; otherwise passive thermal relaxation
		& Nonlocal, multiphysical, high index contrast / slow, complex fabrication
		& Volatile state: thermocapillary relaxation or bistability
		& S \\
		\cite{tait2017neuromorphic,wang2022multi}
		& O-E-O Feedback
		& \unit{\nano\second}--tens \unit{\pico\second} (projected)
		& High-speed opt. signal
		& Loss: optoelectronic conversion and transmission loss. Hold: closed feedback loop with continuous optical and electrical operation
		& Compact, tunable, nonlinear, amplification / latency, energy consumption, wiring, complex fabrication
		& Volatile state: recurrent feedback dynamics
		& E \\
		\cite{feldmann2019all,lugnan2025emergent}
		& PCM (GST) on a waveguide
		& Non-vol\-atile
		& Opt. pulses \qty{10}{\nano\second} \unit{\milli\watt} to amorph., lower pow. to cryst.
		& Loss: absorption and scattering from the GST patch. Hold: none after programming; optical write and reset pulses required
		& Physical plasticity, no active pow., permanent, compact, broadband, \unit{\mega\hertz} switching / non self-resetting, high opt. absorption, material fatigue
		& Non-volatile parameter / configuration memory
		& M \\
		\botrule
	\end{tabular}}
\end{sidewaystable}

\subsection{Reservoir computing based on optical delays}

RC architectures (\cref{sec:nn_architectures}) based on optical delays leverage the finite propagation velocity of light to implement memory. By routing optical signals through structures such as spiral waveguides, multimode fibers, or resonant cavities, the network maintains a volatile memory of past inputs, allowing them to interfere with the current state.
Within this broad family, we distinguish recurrent reservoirs containing a closed feedback path from passive feed-forward random-feature processors and tapped-delay or next-generation reservoir-computing architectures. Although all these systems process delayed observations, only the first class possesses autonomous recurrent dynamics.

\subsubsection{Spatial RC}

\begin{figure}[!htb]
	\centering
	\includegraphics[width=1\linewidth]{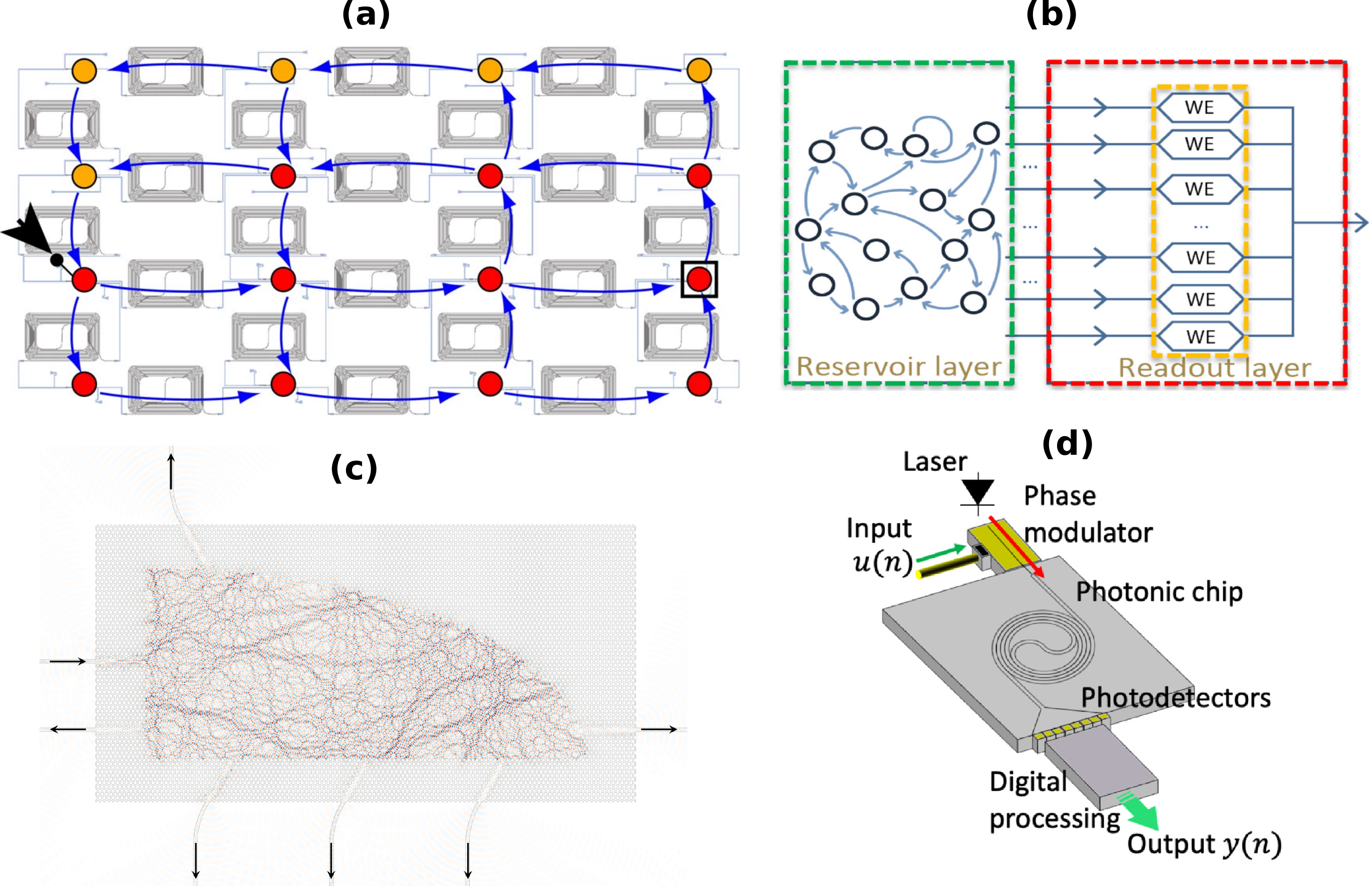}

	\begin{subcaptiongroup}
		\phantomcaption\label{fig:delay_lines-a}
		\phantomcaption\label{fig:delay_lines-b}
		\phantomcaption\label{fig:delay_lines-c}
		\phantomcaption\label{fig:delay_lines-d}
	\end{subcaptiongroup}

	\caption{
		Implementations of spatial RC in PICs.  \subref{fig:delay_lines-a} Integrated spatial RC circuit composed of \numproduct{2x2} combiners connected by delay lines (waveguides curled up into spirals) providing memory for high-speed signal processing (image reprinted with permission from \cite{vandoorne2014experimental}, licensed under \href{https://creativecommons.org/licenses/by-nc-nd/3.0/}{CC BY-NC-ND 3.0}).
        \subref{fig:delay_lines-b} Schematic of an all-optical trainable readout for a photonic spatial RC system, where balanced MZIs and heater-based phase shifters implement the readout weights optically (image reprinted with permission from \cite{ma2023integrated} © Optica Publishing Group).
        \subref{fig:delay_lines-c} A photonic crystal cavity as a compact on-chip spatial reservoir. The memory is provided by the travel time of optical information through the intricate and randomly intialized optical paths connecting the input and the output waveguides (image reprinted with permission from \cite{laporte2018numerical} © Optical Society of America).
        \subref{fig:delay_lines-d} RC system based on a silicon multimode waveguide, where the different delays caused by modal dispersion are exploited as a source of memory (image reprinted with permission from \cite{sunada2021photonic} © Optical Society of America).}
	\label{fig:delay_lines}
\end{figure}

The first experimental demonstration of an on chip photonic reservoir appears in 2014 \cite{vandoorne2014experimental}.
The implementation uses a simple PIC that comprises a \numproduct{4 x 4} matrix of grating couplers, each of which acts as an input port or an output port, and is interconnected by a network of single mode waveguides and splitters (\cref{fig:delay_lines-a}).
In particular, an input laser signal at a wavelength near \qty{1550}{\nm}, injected through a grating coupler, can reach the output ports through multiple optical pathways, including pathways that form recurrent connections.
Moreover, each direct connection between neighbouring grating couplers comprises a delay line consisting of a long waveguide curled up into a spiral.
Therefore, the input coherent signal optically interferes with multiple delayed \textquote{copies} of itself at the output ports, producing multiple representations of the input time series with different memory characteristics.
The resulting signals are read out by photodetectors, which are the only source of the nonlinearity required by the RC scheme, since the optical circuit is linear.
The corresponding electrical output is then sampled and fed into a linear classifier implemented in software, following the RC scheme.

In this implementation, each connection between neighbouring nodes is a delay line providing a finite-window delay of around \qty{280}{\ps} (a \qty{2}{\cm}-long waveguide); the fading memory of the reservoir arises from the recurrent interconnection of these delay lines, through which each input contributes via multiple paths with progressively attenuated weights.

Simple time-dependent ML tasks are carried out at high-speed (up to \qty{12.5}{\Gbps}), such as Boolean logic operations with memory, 5-bit pattern recognition and classification of spoken digits.
An important takeaway of this work is that tuning the ratio between interconnection delay and bit duration is crucial to obtain good ML performance.
Therefore, the length of the optical delay lines should be carefully designed to address a target input bit rate (or input signal bandwidth in general) and ML tasks. The RC system achieves good accuracy in performing the XOR Boolean operation between the \nth{7} and the \nth{11} bit of its input bit stream, provided that the ratio of the interconnection delay over the bit duration is optimized, showing that information on these two lags remains jointly accessible.
The speed of processable signals can be easily increased just by shortening the delay lines, making this type of optical reservoir suitable for the high-speed processing required in optical telecommunication, as it will be further discussed in \cref{subsec:PNNs_telecom}.
On the other hand, the maximum length of on-chip delay lines is limited by the available footprint area and by the waveguide propagation losses, preventing applications to sub-\unit{\GHz} input signals.
For example, this type of optical reservoir is therefore not suited to process the time-dependent output of most optical sensors, where the measured time variations are usually much slower than \qty{1}{\ns}.
Moreover, scaling up this type of system to tackle much more complex tasks is not trivial and likely requires the introduction of nonlinear and/or amplifying elements within the reservoir, such as SOAs, to improve nonlinear processing and compensate for unavoidable optical losses \cite{vandoorne2008toward, van2017advances}.
Recent improvements of the photonic reservoir architecture achieved lower losses, a better mixing of information and an all-optical integrated trainable readout \cite{katumba2018low,sackesyn2021experimental,ma2023integrated} (\cref{fig:delay_lines-b}).

\subsubsection{Multimode and complex waveguide geometries}

Numerical simulations in \cite{laporte2018numerical} present a silicon photonic crystal cavity featuring multiple input and output waveguides (\cref{fig:delay_lines-c}). This architecture improves the energy and area efficiency of integrated spatial RC.
In this case, the delay lines correspond to the optical paths through a large slab waveguide connecting the input and output waveguides.
Employing photonic crystals to implement the optical cavity, while fabrication is challenging in practice, allowed the authors to achieve a strong enough light confinement (high Q factor), and thus to exploit multiple reflections within the cavity to efficiently delay and mix optical information.
The photonic reservoir performs Boolean operations and header recognition with headers up to 6 bits long, across a range of bitrates roughly between \qtylist{25;60}{\Gbps}, which requires retaining information over the corresponding six symbol periods.
Subsequent studies integrate the mixing of optical information and delay-based memory \cite{sunada2021photonic,heinsalu2024silicon}.

Experiments in \cite{sunada2021photonic} demonstrate that silicon multimode waveguide spirals (\cref{fig:delay_lines-d}) provide an optical memory of approximately \qty{200}{\ps} when phase modulation is applied to the input. This memory originates from modal dispersion as the optical field propagates through the device. The authors demonstrate chaotic time series prediction on the Santa Fe dataset (see \cref{sec:benchmarks}) with a NMSE of 0.039 at a speed of \qty{12.5}{\GS\per\s}. This performance relies on the integration of spatial and temporal information from the output field within a single readout time sample.
In particular, the optical power is acquired at 65 different locations along the output interference pattern, and for each of these, 4 time samples (also called \emph{virtual nodes} of the reservoir) are measured for each input sample used to train the linear readout.
In other words, the photonic reservoir provides a total of 260 output features, by expanding the spatial feature set via time multiplexing.

Temporal multiplexing is often employed to increase the dimensionality and memory of RC systems at the cost of:
(i) increased sampling rate at the reservoir readout for a fixed input sample rate,
(ii) application of time-dependent trainable readout weights, as opposed to constant ones.
This last requirement often hinders the photonic hardware implementation of the readout linear combination, with related advantages in throughput and energy efficiency, which is otherwise much more feasible\cite{mancinelli2022photonic, ma2023integrated}.

Authors of \cite{sunada2021photonic} also demonstrate a further increase in the dimensionality of the reservoir output representation by exploiting wavelength multiplexing on top of the aforementioned space and time multiplexing, and leveraging the high sensitivity of the multimode waveguide's response to optical wavelength in order to get non-redundant output features.
In particular, they consider up to 5 different wavelengths with \qty{1}{\nm} steps around \qty{1550}{\nm}, thus expanding the reservoir's output dimensionality up to 1300 features and improving the NMSE up to 0.018 for the Santa Fe prediction task.
The employment of multiple wavelength in parallel is emulated by a sequential measurement of the reservoir output at different locations and wavelength, using a tunable-wavelength laser source and a single movable optical fiber to scan the multiple output locations across the width of the multimode waveguide.
A truly parallel implementation of this multiwavelength readout would require the use of one photodetector per spatial and wavelength output and additional optical circuitry (e.g. based on a series of MRRs) \cite{tait2016microring}) to separate and route the different output wavelengths to different detectors.
Moreover, laser sources with multiple wavelengths would be needed concurrently, which can be efficiently provided by a frequency comb source \cite{feldmann2021parallel, xu202111}.
Finally, the RC system could learn to infer the optical phase shift applied to the input laser beam with a NMSE of 0.029.

In \cite{heinsalu2024silicon}, a large SOI multimode waveguide forming a loop (\cref{fig:delay_lines-d}) is experimentally characterized.
It provides optical memory up to \qty{800}{\ps}, and the corresponding RC system could solve the NARMA3 task and presented a MC of about 11 (\cref{sec:memory_char}).

\subsubsection{Delay-based RC}

As discussed in the previous paragraph, time multiplexing can enhance a reservoir's dimensionality.
This approach is taken to the extreme in  \emph{delay-based RC} \cite{appeltant2011information}, which is mainly based on a single nonlinear node connected to a delayed feedback loop (\cref{sec:nn_architectures}).
Several photonic implementations were demonstrated so far, motivated by the high bandwidth and the potential energy efficiency of optical data transmission.

Photonic delay-based RC systems usually require adding to the original input signal an additional faster modulation, where each original time sample is modulated by the same time sequence, called \emph{mask}.
Masking the reservoir input allows to break the correlation between the reservoir's output features (or virtual nodes), thus providing non-redundant dimensionality expansion.
When masking is performed digitally, it imposes the additional requirement of a modulating signal at a higher speed than the original input signal, although some hardware-based masking has been proposed \cite{nakayama2016laser}.
In general, delay-based RC allows the use of a single-node reservoir at the cost of applying time-dependent weights to the input and output signals at a higher speed than the speed of the processed input signal.
Therefore, if long enough delay lines are used, for a fixed available modulation speed, the reservoir dimensionality can be expanded by slowing down the processing speed.

\begin{figure}[!htb]
	\centering
	\includegraphics[width=1\linewidth]{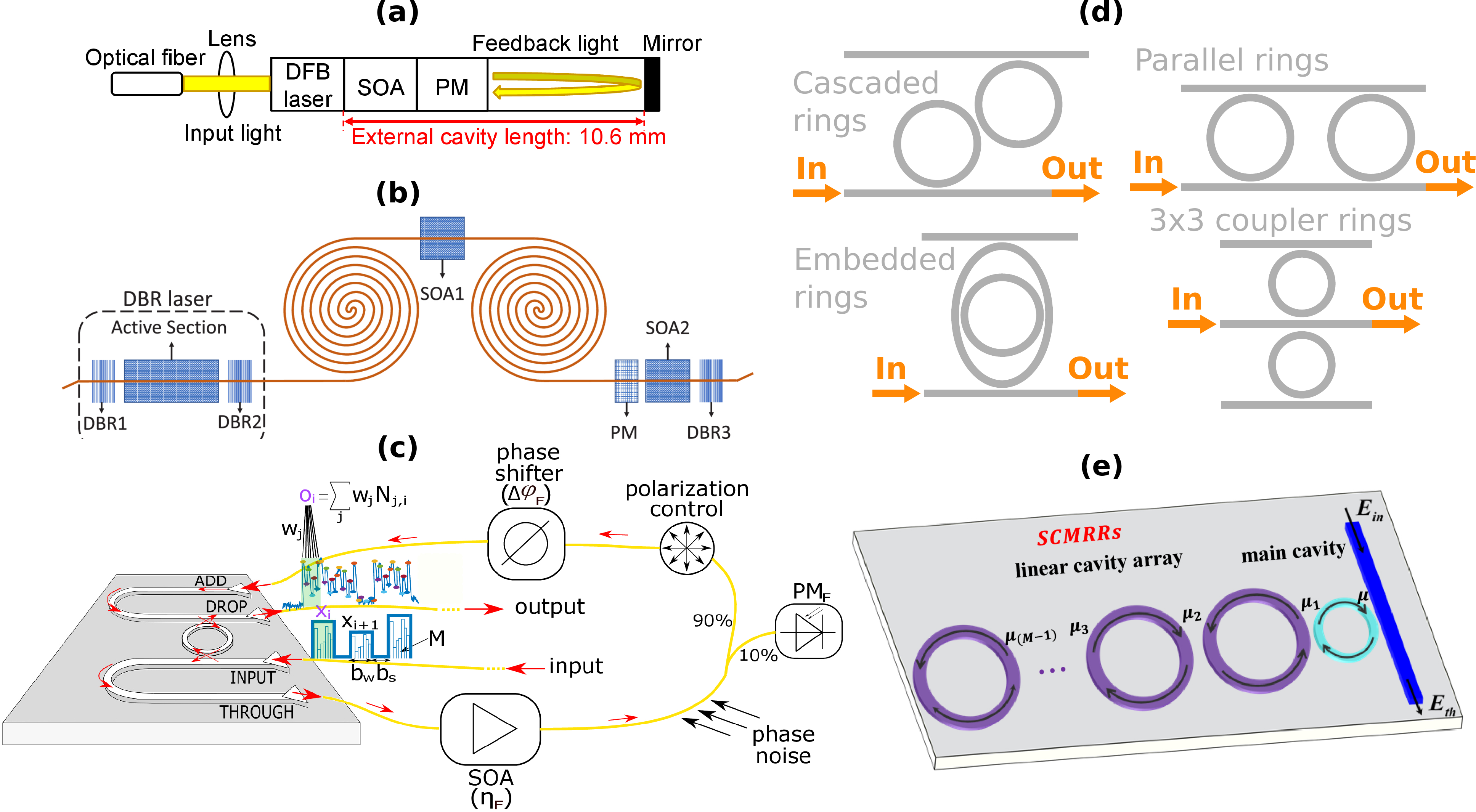}

	\begin{subcaptiongroup}
		\phantomcaption\label{fig:linear_cavities-a}
		\phantomcaption\label{fig:linear_cavities-b}
		\phantomcaption\label{fig:linear_cavities-c}
		\phantomcaption\label{fig:linear_cavities-d}
		\phantomcaption\label{fig:linear_cavities-e}
	\end{subcaptiongroup}

	\caption{%
		Photonic implementations of delay-based reservoir computing.
        \subref{fig:linear_cavities-a}, \subref{fig:linear_cavities-b} Integrated delay-based RC systems based on a distributed-feedback (DFB) laser, whose external cavities host optical delay lines (memory source) and amplifiers (images reprinted with permission respectively from \cite{takano2018compact} and \cite{harkhoe2020demonstrating} © Optical Society of America).
		\subref{fig:linear_cavities-c} Delay-based photonic reservoir combining the nonlinear memory of an integrated silicon MRR and the linear memory of an external optical fiber, which requires amplification and real-time phase noise compensation (image reprinted with permission from \cite{donati2024time} © Optica Publishing Group).
		\subref{fig:linear_cavities-d} Four different configurations of two coupled MRRs are simulated, and their linear memory optimized, to carry out RC tasks \cite{li2023packet}.
		\subref{fig:linear_cavities-e} A series of coupled MRRs is simulated, providing delay-based memory and, only in the last MRR, there is a nonlinearity (image reprinted with permission from \cite{ren2024photonic} © Optica Publishing Group).}
	\label{fig:linear_cavities}
\end{figure}

A representative delay based RC implementation is reported in \cite{takano2018compact}.
The system uses an integrated distributed feedback (DFB) semiconductor laser as the nonlinear node, and an external cavity that includes a SOA, a phase modulator, and a passive waveguide terminated by an external mirror (\cref{fig:linear_cavities-a}).
The cavity provides a round-trip time of \qty{254}{\ps}.
Because the obtained memory is limited, the reservoir provides only a small number of features, and a high error on the Santa Fe prediction task is observed (NMSE=0.423).
The error is reduced to NMSE=0.089 by using virtual nodes, i.e. applying a mask on the optical input and tightly sampling the output (5 virtual nodes per input sample).
This is a hybrid approach that exploits both the reservoir's analog memory and the digital memory provided by the readout.
Therefore, in such cases, it is preferable to compare the achieved results with a baseline result, obtained by only employing the digital memory without the use of a reservoir, in order to evaluate how much of the achieved performance level is actually due to the use of the photonic reservoir.

A similar approach is presented in \cite{harkhoe2020demonstrating}, which employs a distributed Bragg reflector laser integrated on an \ce{InP} chip and an external cavity that includes two spiral delay lines, two SOAs, and a phase modulator (\cref{fig:linear_cavities-b}).
The on-chip external cavity presented a round-trip time of \qty{1170}{\ps} and a minimum NMSE of 0.049 is obtained for the Santa Fe prediction task.

\cite{donati2024time} demonstrates a delay-based RC system composed of a single MRR, which takes the role of a nonlinear node while also providing fading short-time memory, thanks to silicon nonlinear effects.
This memory mechanism is discussed in \cref{subsubsec:response_mem} and \cref{subsec:RCwithMRRs}.
Additional memory is provided by an amplified external optical delay line of \qty{88}{\ns}, consisting of an optical fiber creating a feedback loop (\cref{fig:linear_cavities-c}).
A phase shifter autonomously driven by an electronic controller is employed in order to stabilize the optical phase of signals through the fiber against environmental noise (thermal drift/vibrations), which is typically a major source of instability in such systems.
This additional source of cost and complexity is the price to pay for extending the optical delay-based memory beyond the limits imposed by the PIC.
This fiber delayed feedback enables delayed Boolean operations with up to 3 bits of memory at a few Mbps, whereas related MRR based implementations without an optical delay line, and relying on silicon nonlinear effects as the memory source, remain limited to 1 bit of memory. The PNN is benchmarked with the Santa Fe and Mackey Glass time series prediction tasks, and show that low errors are achieved without the optical delay provided by the fiber, relying only on silicon nonlinear dynamics and on the resulting self-pulsing mechanism.

\subsubsection{Resonator-based delay lines}

An alternative approach for optical delay based memory exploits the photon lifetime of a MRR, rather than the propagation time in a spiral waveguide or an optical fiber, and it therefore requires a much smaller chip area.
While one resonator is unlikely to provide long enough memory, multiple coupled MRRs can be employed, as proposed in two numerical studies \cite{li2023packet, ren2024photonic}.
The first focuses on optimizing 4 different configurations of coupled MRRs (\cref{fig:linear_cavities-d}) to maximize the delay-bandwidth product, finding the same maximum value of \qty{1395}{\ps} \(\times\) \unit{\giga\hertz} for all the considered configurations.
This quantity estimates the trade-off between the group delay (i.e. the memory duration) of a linear MRR and the bandwidth of the signal it can process.

Indeed, for a single MRR, the higher the Q factor, the larger the group delay and the smaller the resonance width (or bandwidth), making it difficult to obtain long memory on a high speed input signal.
Coupling two MRRs, though, allows the authors to mitigate this limitation by achieving a flat-top delay spectrum.
The optimized systems are then employed to solve the 3 and 6 bit header recognition tasks.
In the latter work \cite{ren2024photonic}, multiple coupled MRRs provide a source of linear memory, while the final ring in the cascade operates in a nonlinear regime and supplies the nonlinearity (\cref{fig:linear_cavities-e}).
The study addresses multiple time-dependent benchmark tasks, including MC, NARMA10, Mackey Glass, and Santa Fe, with input rates in the \unit{\giga\hertz} range.

It should be stressed that, in a real system, the resonance wavelength of the coupled MRRs should be somehow aligned (e.g., via heaters) to compensate for the variability due to fabrication errors.
Moreover, the characteristic time of the considered free carriers-based nonlinearity in the last MRR might be too long to match the \unit{\GHz} processing speed. In addition, it varies significantly depending on the fabrication procedure and related inaccuracies.

\subsubsection{Feed-forward and \textquote{next-generation} architectures}

\begin{figure}[!htb]
	\centering
	\includegraphics[width=0.8\linewidth]{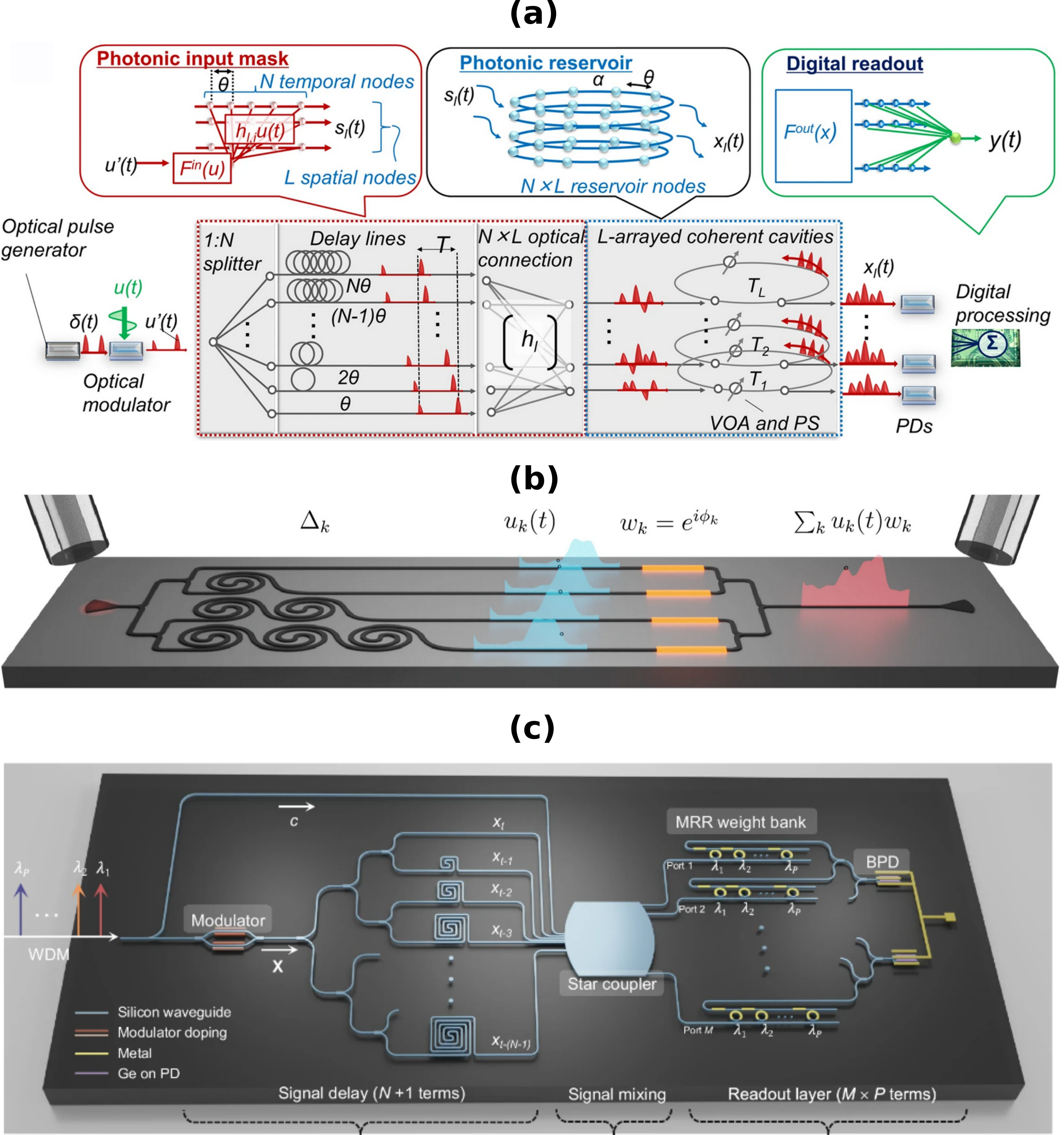}

	\begin{subcaptiongroup}
		\phantomcaption\label{fig:next_genRC-a}
		\phantomcaption\label{fig:next_genRC-b}
		\phantomcaption\label{fig:next_genRC-c}
	\end{subcaptiongroup}

	\caption{
        Hybrid spatial and delay-based reservoir computing architectures.
		\subref{fig:next_genRC-a} A passive PIC hosts a combination of spatial and delay-based RC.
		The input is encoded as optical pulses, masking is performed on-chip exploiting a programmable feed-forward network with multiple delay lines and a reservoir is implemented through coupled programmable optical cavities formed by waveguide loops (image from \cite{nakajima2021scalable}, licensed under \href{https://creativecommons.org/licenses/by/4.0/}{CC BY 4.0}).
		\subref{fig:next_genRC-b} An integrated silicon perceptron is implemented by splitting the input optical signal, each split is subjected to different delay lines, is optically weighted (with complex values) and recombined with the others to provide an output signal (image from \cite{mancinelli2022photonic}, licensed under \href{https://creativecommons.org/licenses/by/4.0/}{CC BY 4.0}).
		\subref{fig:next_genRC-c} An on-chip implementation of the \emph{next generation RC} approach: the input signal is split, each split fed into increasingly longer delay lines and recombined by a star coupler (image from \cite{wang2024ultrafast}, licensed under \href{https://creativecommons.org/licenses/by/4.0/}{CC BY 4.0}).
		All the systems in this figure rely on photodetection as a source of nonlinearity.}
	\label{fig:next_genRC}
\end{figure}

\cite{nakajima2021scalable} combines time-delay RC and spatial RC to generate 512 output features at a rate of \qty{3.75}{\GS\per\s}, with the rate limited by the oscilloscope bandwidth, using a linear PIC on a low loss silica platform (\cref{fig:next_genRC-a}).
Furthermore, the processing dimensionality can be further expanded via multi-wavelength operations.
The system input is an electrical signal that modulates the amplitude of optical pulses, with the pulse repetition rate equal to the input sample rate.
The modulated input pulses are split, delayed and recombined by reconfigurable optical circuit based on phase shifters, delay lines and MZIs, thus applying a spatiotemporal mask to the input.
This PIC applies a mask to the input. Another PIC acts as the reservoir and processes the output of the first PIC. The reservoir is based on coupled passive ring cavities, where variable optical attenuators and phase shifters are employed to control the inter-cavity couplings.
Its output is finally read out by 16 photodetectors, which also provide the nonlinearity required by RC systems.
Finally, the trainable readout layer is implemented in software.

A complex-valued optical perceptron based on delay lines and the intrinsic nonlinearity of photodetection is presented in \cite{mancinelli2022photonic} (\cref{fig:next_genRC-b}).
The same architecture is applied to all-optical equalization of optical telecom signals in \cite{staffoli2023equalization}.
In \cref{subsec:PNNs_telecom} we discuss this type of application in more details.
The input optical signal is split into four parallel waveguides, each introducing a fixed differential time delay relative to the others (\qty{50}{\ps}).
This structure acts as a passive optical tapped-delay line, effectively creating a \textquote{sliding window} over the input symbol sequence in the complex domain.
Each delayed copy is subjected to a trainable complex weight, adjusted via thermal phase shifters, before all branches are coherently recombined.
Experimentally, this four channel architecture supports binary pattern recognition, delayed XOR operations, and equalization of a \qty{10}{\Gbps} intensity modulation direct detection (IM/DD) signal distorted by chromatic dispersion.

Regarding the considered equalization task, the system is tested over standard single-mode fiber links ranging up to \qty{125}{\km} in length.
By training the on chip complex weights, using a particle swarm optimization algorithm that maximizes the eye diagram opening, the device reconstructs the transmitted bits from the distorted signal and reopens the closed eye induced by inter symbol interference.
Subsequent work, based on an improved 8 channel version of the optical perceptron, experimentally demonstrates the equalization of
(i) 4-level pulse amplitude modulated signals at \qty{20}{\Gbps} \cite{staffoli2024silicon}, and of
(ii) nonlinear distortion up to \qty{450}{\km} \cite{staffoli2025nonlinear}.
These all-optical processing solutions aim to offer much improved energy efficiency and latency compared to conventional digital signal processing systems, by skipping the optical-to-electrical, and vice versa, conversions.

In \cite{wang2024ultrafast}, a PNN based on optical delay lines and a star coupler (\cref{fig:next_genRC-c}) realizes an all optical and high speed implementation of a next generation reservoir computing (\cref{sec:nn_architectures}) \cite{gauthier2021next}.
A \qty{1550}{\nm} input laser modulated at \qty{60}{\GHz} is split into 8 waveguides, each going through an optical delay line whose delay is incrementally increased by \qty{16.7}{\ps}.
This delay is the inverse of the input modulation rate, to enable the coupling between subsequent time steps of the input signal.
The delayed optical replicas are then coherently recombined, together with an unmodulated split of the input laser, by a star coupler with 9 inputs and 45 outputs.
The output signals are read by high-speed photodetectors, providing quadratic nonlinearity.
Then, as in conventional RC, the 45 output features are fed into a linear regressor or classifier, implemented offline in software.

The authors also demonstrate the extension of the system's throughput via wavelength division multiplexing, employing 5 wavelengths and reading 9 spatial outputs.
Competitive performances are achieved in different ML benchmark tasks, namely Santa Fe, Lorentz63, nonlinear channel equalization, NARMA 10 and COVID-19 X-ray image classification.
The proposed system can be seen as a more tidy, and perhaps more efficient version of the spatial RC system described in \cite{vandoorne2014experimental}.

\subsection{RC based on nonlinear light-matter interactions}\label{subsec:RCwithMRRs}

Photons do not naturally interact with one another.
However, nonlinearity and memory can be introduced through light-matter interactions, where an intense optical signal temporarily alters a material's physical properties (such as refractive index), effectively controlling the transmission of other optical signals.
These nonlinear effects are useful in photonic RC, since they provide local nonlinearity, as opposed to photodetection nonlinearity, which is applied only at the output of the PIC.
This way, nonlinearity directly affects the network recurrency and dynamics as in RNNs, allowing for higher expressivity and more powerful dimensionality expansion, which are key to strong RC performances.
However, nonlinear effects usually require much higher optical power than just linear transmission, where it is sufficient that the output power level is enough to be readable.
This makes the realization of large nonlinear optical networks challenging, especially considering that energy efficiency is one of the main motivation for optical computing.

\subsubsection{Microring resonators as nonlinear nodes}

One way to achieve energy efficient and compact nonlinear nodes, especially in silicon photonics where second-order nonlinearity is absent, is to rely on optical cavities, such as MRRs.
Their resonant behaviour enhances nonlinearity in different ways:
(i) since the resonant optical field travels more than once along the ring waveguide, the sensitivity to small changes in refractive index or absorption (e.g. due to light-driven free carrier generation or heat) is enhanced;
(ii) for the same reason, input resonant light is accumulated by the ring, thus enhancing optical power;
(iii) the optical transmission of MRRs depends on optical interference, which efficiently translates optical phase shifts due to refractive index perturbations into variations in transmitted power, which can be easily read by a photodetector.

\begin{figure}[!htb]
	\centering

	\begin{subcaptiongroup}
		\phantomcaption\label{fig:nonlinMRRs-a}
		\phantomcaption\label{fig:nonlinMRRs-b}
		\phantomcaption\label{fig:nonlinMRRs-c}
		\phantomcaption\label{fig:nonlinMRRs-d}
		\phantomcaption\label{fig:nonlinMRRs-e}
	\end{subcaptiongroup}

	\includegraphics[width=0.85\linewidth]{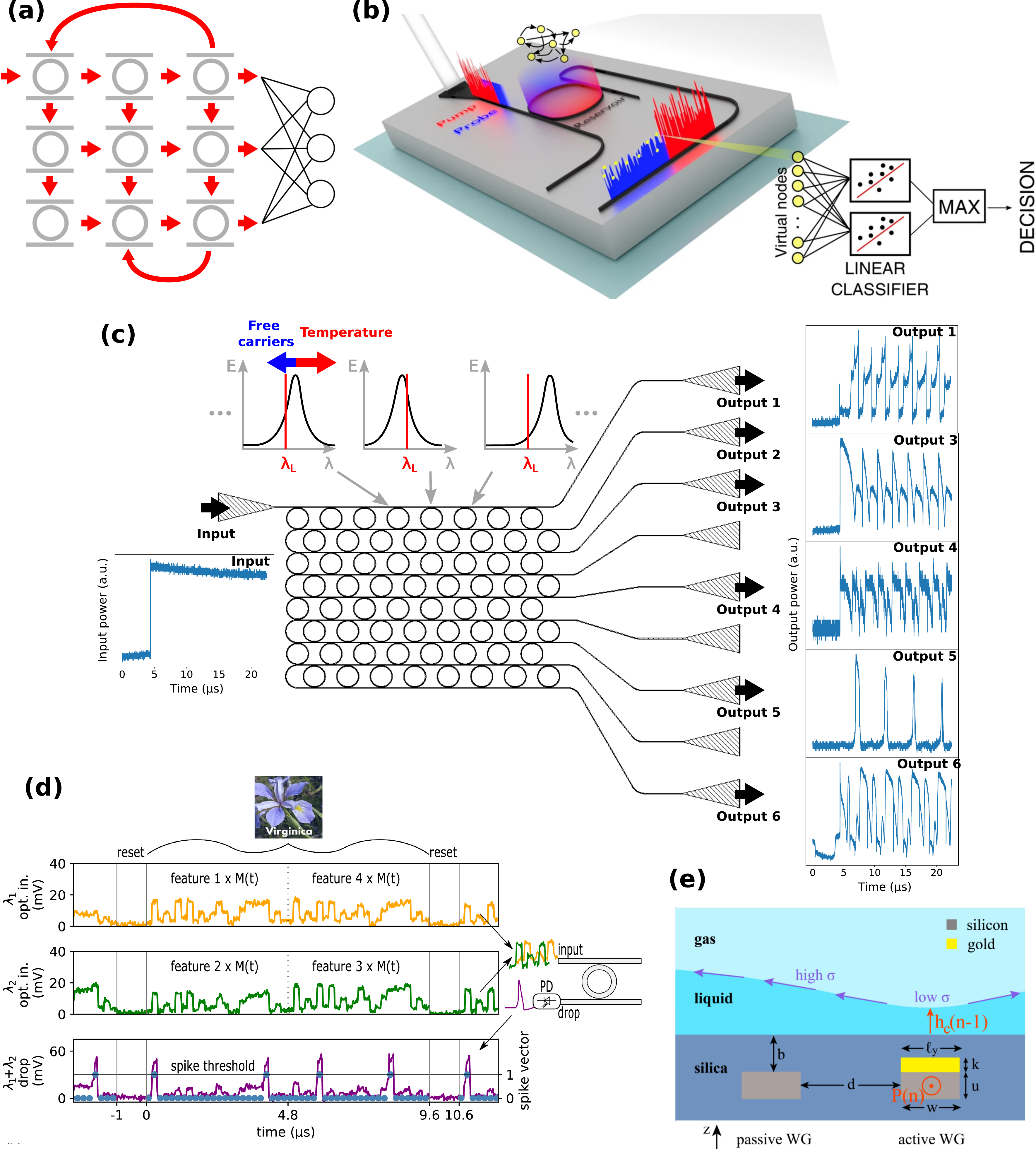}
	\caption{%
        Response-induced nonlinear reservoir computing implementations in PICs.
		\subref{fig:nonlinMRRs-a} Example representation of the photonic reservoir topology, based nonlinear coupled MRRs, employed in \cite{mesaritakis2013micro}.
		\subref{fig:nonlinMRRs-b} A single nonlinear silicon MRR is exploited for delay-based RC. A pump and probe mechanism is employed, via two input optical wavelengths that are nonlinearly coupled by the MRR's nonlinear dynamics (image from \cite{borghi2021reservoir}, licensed under \href{https://creativecommons.org/licenses/by/4.0/}{CC BY 4.0}).
		\subref{fig:nonlinMRRs-c} An \numproduct{8x8} matrix of coupled silicon MRRs hosts complex self-pulsing dynamics, producing different complex output signals. This enables dimensionality expansions employable for RC, by combining the space, wavelength and time dimensions (image from \cite{lugnan2025reservoir}, licensed under \href{https://creativecommons.org/licenses/by/4.0/}{CC BY 4.0}).
		\subref{fig:nonlinMRRs-d} Input and output data encodings enabling the use of a self-pulsing MRR as an artificial spiking neuron for multi-wavelength delay-based RC (image from \cite{donati2025all}, licensed under \href{https://creativecommons.org/licenses/by/4.0/}{CC BY 4.0}).
		\subref{fig:nonlinMRRs-e} Cross section of a simulated on-chip optofluidic system, where a nonlocal fading memory effect is obtained by the interplay between optical pulses in a waveguide, a gold patch and a liquid-gas interface on top of the PIC (image from \cite{gao2022thin}, licensed under \href{https://creativecommons.org/licenses/by/4.0/}{CC BY 4.0}).}
	\label{fig:nonlinMRRs}
\end{figure}

Because of their complex multiscale dynamic behaviour and sensitivity to fabrication variability, nonlinear silicon MRRs are difficult to model accurately, especially in a network of coupled resonators.
Moreover, heaters or other electrical controls are usually needed to precisely align their resonance wavelengths.
Since the application of standard backpropagation-based training methods usually require accurate modeling, controls and observability of the internal states in a photonic ANN, nonlinear MRRs are much more easily employable in RC approaches, where these requirements are removed for a physical reservoir.

In \cite{mesaritakis2013micro}, a network of nonlinear \ce{InGaAsP}/\ce{InP} MRRs is proposed for RC and investigated via numerical simulations.
A \numproduct{5x5} matrix of randomly interconnected MRRs is employed as a reservoir (\cref{fig:nonlinMRRs-a}, note that the depicted scheme is only \numproduct{3x3}) to classify 8-bit time sequences encoded in the amplitude of an input laser beam.
A similar approach is reported in \cite{denis2018all}, based on a network of silicon MRRs and focusing on the delayed XOR Boolean task.
Both works relied on optical delays for memory and do not demonstrate a direct advantage in terms of ML performances with respect to linear PICs.

In \cite{borghi2021reservoir}, the nonlinear effects in a single silicon MRR are exploited as a source of both fading memory and nonlinearity for delay-based RC (\cref{fig:nonlinMRRs-b}). Experiments were performed in a pump-probe configuration.
The input pump signal consists of a modulated light with wavelength around \qty{1550}{\nm}, which modifies the MRR's resonance via the silicon nonlinear effects based on two-photon absorption, free carrier density and temperature. A second input probe beam that has constant power and a different wavelength, with the wavelength separation set to one MRR free spectral range relative to the pump, is nonlinearly coupled by the MRR internal dynamics to the pump beam.
Exploiting only this probe signal as a source of virtual nodesthe 1-bit delayed XOR and the Iris-Flower classification tasks are successfully solved. This demonstrates that the MRR's nonlinearity provided both memory and nonlinearity required by RC.
The signal processing speed is a crucial parameter to exploit either free carrier-based or thermo-optic effects for computation.
Indeed, the XOR task is carried out at \qty{30}{\mega\hertz} by only exploiting the nonlinear free carrier dynamics, while the Iris-Flower classification is carried out at \qty{0.38}{\mega\hertz}, a regime characterized by the coupling between thermal and free carrier dynamics.
This work shows that material nonlinearities in a MRR provide both energy efficient nonlinearity and compact on chip memory (referred to as \emph{nonlinear relaxation–induced memory} in \cref{subsubsec:response_mem}).
This type of memory is best viewed as complementary to integrated delay based memory, rather than as its replacement.
Indeed, they cover different, barely overlapping, timescales: from microseconds to nanoseconds for silicon MRRs and below nanoseconds for integrated delay lines.

The two memory mechanisms also differ in how they relate with variations of input signal timescale.
Indeed, a nonlinear memory becomes negligible when its characteristic relaxation time is much longer than the input variation time, provided that the system evolves through an initial warm up interval, since the input induced perturbations effectively average out.
Conversely, when the nonlinear relaxation dynamics is much faster than the input variation, the same mechanism acts as an effective instantaneous nonlinearity.
These relaxation-based effects do not apply to delay based memory, since propagation delays reshape the effective network connectivity and determine how the propagating field coherently interferes with delayed replicas of itself, while the underlying operations remain linear in the complex optical domain. Delay based memory nonetheless requires matching the delay to the symbol duration, as discussed, e.g., in \cite{vandoorne2014experimental}.

Another important difference arising from the use of optical cavities, such as MRRs, is wavelength specificity.
A simple non-resonant optical delay line is typically broadband and provides a memory duration that depends only weakly on wavelength, as in SOI waveguides operating in the infrared C band (\qtyrange{1530}{1565}{\nano\meter}).
On the other hand, the nonlinearity of MRRs primarily acts on input light that is close to resonance, at least for optical powers in the \unit{\milli\watt} range, and it depends strongly on the detuning between the input wavelength and the resonance wavelength.
While this property enables wavelength selective operations and wavelength dependent effective network topologies, it also makes MRR-based systems sensitive to fabrication variability.
In particular, the resonance wavelength varies across devices and is difficult to predict before fabrication. Therefore precise resonance alignment often requires external electrical tuning, for example via integrated phase shifters, although RC operation typically tolerates imperfect alignment.

In \cite{donati2024time}, which is already discussed in the previous section, the authors show that benchmark time series prediction tasks, namely Santa Fe and Mackey Glass, can be addressed by exploiting the nonlinear dynamics of a silicon MRR, in close analogy with \cite{borghi2021reservoir}.
In this approach, the experiment employs a single input wavelength at a time, which removes the pump and probe configuration based on two wavelengths.
Numerical simulations further extend the analysis of a single nonlinear MRR as a time delay reservoir in
\cite{donati2021microring,bazzanella2022microring,giron2024effects,gretter2025dynamic}.

\subsubsection{Self-pulsing MRRs and spiking neural networks}

Nonlinear effects in silicon MRRs give rise to self sustained dynamics, namely self-pulsing (discussed in \cref{subsubsec:multistable_mem}), when the MRR is continuously driven by resonant optical power that exceeds a threshold, which typically lies in the milliwatt range.
While this effect is originated by the interplay between the aforementioned nonlinear effects, based on free-carriers and temperature, it can provide memory effects lasting much longer than the relaxation times of both.
An input signal can perturb the dynamics of this self pulsation. The perturbation either delays the pulsation by altering its temporal phase or forces the system to transition into a different static or dynamic equilibrium state. Such a transition between attractors typically occurs in systems of multiple coupled MRRs.

These effects can outlast the carrier and thermal relaxation times by orders of magnitude, but with different retention limits (\cref{subsubsec:multistable}): a transition between attractors persists until noise induces an escape from the basin or the optical power sustaining the self-pulsing is switched off, and is thus non-fading, whereas a phase offset along a single limit cycle is not protected by an attractor boundary and is progressively lost through phase diffusion (\cref{sec:persistent_state_metrics}).
For example, memory is erased when the self-pulsing is switched off for a duration longer than the thermal relaxation time.
Self-pulsing-based memory is experimentally demonstrated in single and multiple coupled MRRs \cite{biasi2024exploring}.
In particular, the self-pulsing in two or three coupled MRRs is sensitive to the timing of a perturbing optical pulse and to the pulse rate of a perturbing 20-pulse sequence.
These effects are relevant for the development of photonic spiking neural networks and of optical smart sensors, where the information carried by a fast and short optical signal can be transferred to slower and long-lasting network dynamics.

This research direction continues in \cite{lugnan2025reservoir}, which experimentally demonstrates, in a matrix of 64 coupled MRRs (\cref{fig:nonlinMRRs-c}), that ML can exploit this long lasting memory effect at least up to \SI{75}{\micro\second} after the end of the perturbing signal.
In an RC scheme, a linear readout, implemented in software, is trained on the optical signals measured at the five output ports of the MRR network.
The system allows determination of the timing and the pulse rate of input perturbing pulses and trains of pulses respectively, by processing the downsampled self-pulsing dynamics at the outputs much after the end of the input perturbation.
This is accomplished at two different timescales of the input perturbation: with pulse widths of \SI{60}{\nano\second} and \SI{200}{\nano\second}.
While the investigated PIC is fixed, a large variety of self-pulsing dynamics can be obtained depending on the measured output port, the input laser wavelength and power.
This flexibility enables exploration of many operating points, so the configuration that best matches a target task can be selected. Moreover, this ML approach differs fundamentally from conventional RC, since it relies on long-lasting memory sustained by the self pulsation dynamics (based on self-pulsing phase or on transitions between attractors, the latter referred to as \emph{passive multistable-induced memory} in \cref{subsubsec:multistable_mem}), rather than on the short-lived fading memory that characterizes standard reservoirs.
A similar concept, where non-fading network dynamics are exploited for working memory, is found in a recent paper about \emph{non-dissipative RC}  \cite{gallicchio2024euler}, where improved performances in several time-dependent ML tasks are demonstrated in a software implementation.
The same MRR network and RC approach is further studied in \cite{foradori2025neuromorphic}, and applied to the classification of images (from MNIST and Fashion-MNIST).
The authors investigate how the ML performances vary with different configurations, e.g. the encoding of the input or the sample rate or the number of output ports.
Accuracies up to \qty{96.49}{\percent} and \qty{85.81}{\percent} are achieved respectively for the MNIST and Fashion-MNIST tasks, against corresponding baseline accuracies of \qty{92.12}{\percent} and \qty{83.84}{\percent}.

As evidenced in \cite{van2012cascadable,xiang2022all,donati2025all}, self-pulsing silicon MRRs present different computational properties resembling the ones of biological neurons and therefore they can be employed as artificial spiking neurons.
In \cite{donati2025all}, these properties are exploited to demonstrate delay-based RC computation using a spiking silicon MRR.
In particular, the Iris-flower classification task is encoded by using two input wavelengths (two features per wavelength) and the total power at the output is measured (\cref{fig:nonlinMRRs-d}).
The system relies on the ability of a MRR to nonlinearly couple optical signals at multiple wavelengths and at different times, demonstrating the integration of different inputs as in biological neurons.
Moreover, a digital linear readout is applied with a threshold to the output spikes (generated via self-pulsing).
This way, the authors obtained an accuracy of \qty{92}{\percent} with a sparse readout, where the digital linear classifier looks at a few on-off events to carry out the task.
The high computational efficiency of a ML inference process based on a sparse set of spikes is one of the main motivations driving research on artificial spiking neural networks.

\subsubsection{Optofluidics as an alternative nonlinear platform}

\cite{gao2022thin} introduces, through numerical simulations, an optofluidic approach that provides optical nonlinearity and volatile memory in a PIC, where a thin liquid film sits above a waveguide and a gold patch is deposited on top of the film (\cref{fig:nonlinMRRs-e} and \cref{subsubsec:multistable}).
The optical field that propagates in the waveguide heats up the gold patch, inducing a deformation in the liquid gas interface of the liquid layer above the waveguide, due to thermo-capillary.
In turn, such a deformation interacts with the evanescent optical field thus causing a phase shift in the signals transmitted by the same waveguide and by other nearby waveguides.
The authors propose delay-based RC systems implemented by adding the optofluidic cell on one arm of an integrated MZI and on top of a directional coupler.
Benchmark ML tasks include a delayed 2-bit XOR task and a binary classification task on handwritten digits 0 and 1 from the MNIST dataset.
The proposed nonlinear node allows for high refractive index changes within a time scale of 10 ms and energy consumption in the order of \qty{10}{\nano\joule} per bit.
This optofluidic platform and similar multiphysics systems can be particularly interesting for smart sensing applications, where any source of perturbation for the system can be sensed (e.g., chemicals or vibrations) and, in principle, interact in a two-ways manner with a photonic computing circuit.

\subsection{Trainable integrated photonic neural networks with optical memory}

While RC enables the usage of many different types of physical hardware for computing without requiring these systems to be fully observable and controllable, highly engineered PICs can host photonic implementations of more powerful ML models such as RNNs, where the internal parameters of the network (hidden layers) can be precisely tuned by a training algorithm.
Fabricating, controlling and scaling up this type of system is still very challenging, hence the scarcity of related works compared to RC implementations and feed-forward photonic processors.

\subsubsection{Optoelectronic recurrent architecture}

This type of trainable photonic RNNs rely on a hybrid approach, exploiting optoelectronic conversion to achieve non-coherent summation and nonlinear activation of weighted signals at different wavelengths. These are core operations for an artificial neuron based on wavelength division multiplexing, and are difficult to implement purely in the optical domain.

\begin{figure}[!htb]
	\centering
	\includegraphics[width=0.99\linewidth]{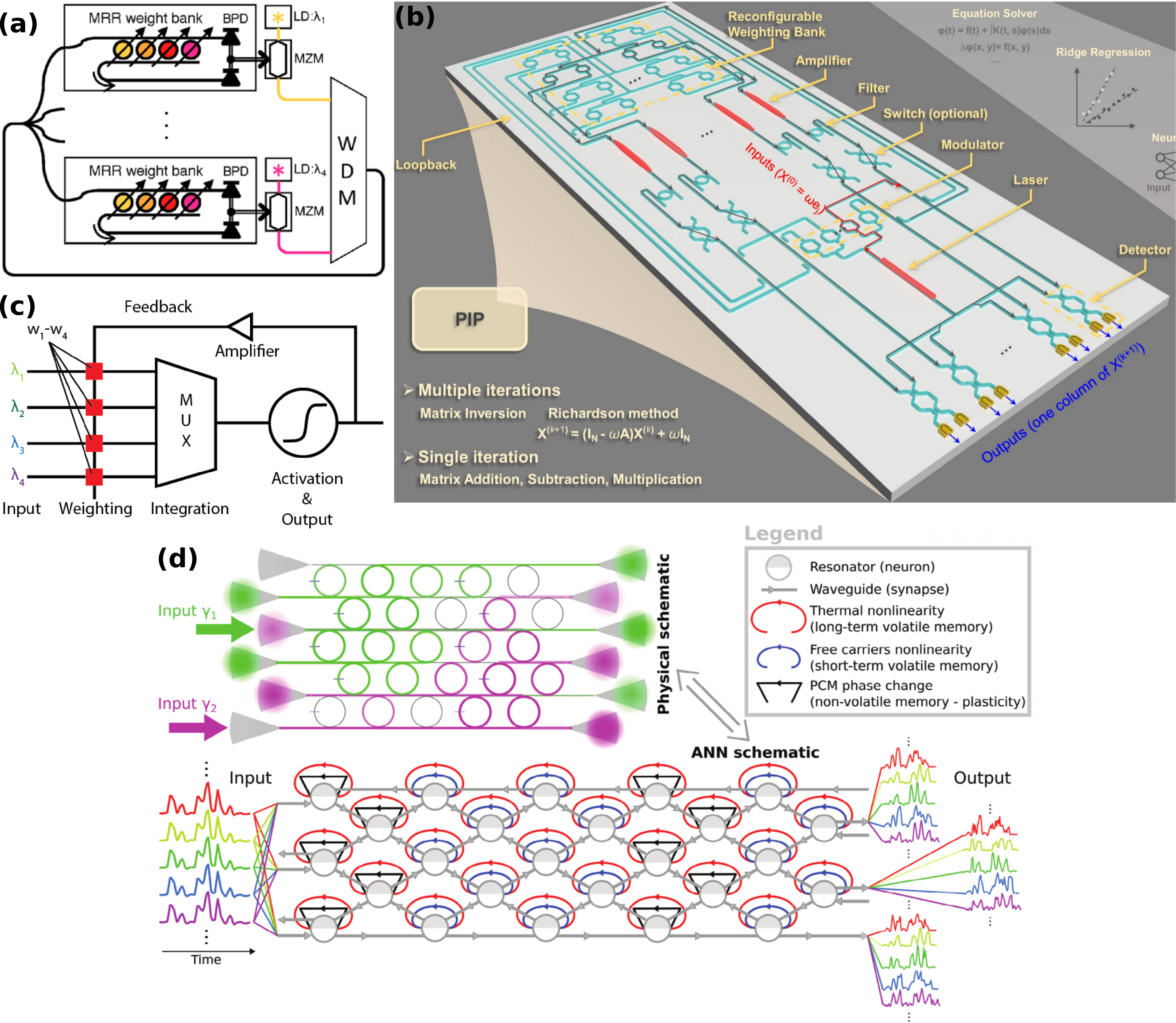}

	\begin{subcaptiongroup}
		\phantomcaption\label{fig:RNNs-a}
		\phantomcaption\label{fig:RNNs-b}
		\phantomcaption\label{fig:RNNs-c}
		\phantomcaption\label{fig:RNNs-d}
	\end{subcaptiongroup}

	\caption{%
        Implementation of PNNs with recurrent dynamics and adaptive memory.
		\subref{fig:RNNs-a} A photonic RNN where on optical-electrical-optical (O-E-O) conversion is employed for non-coherent summation and nonlinearity in the artificial neurons. A single feedback waveguide takes the role of multiple recurrent connections by exploiting WDM. Synaptic weights are applied to specific wavelength channels via tunable MRRs (image from \cite{tait2017neuromorphic}, licensed under \href{https://creativecommons.org/licenses/by/4.0/}{CC BY 4.0}).
		\subref{fig:RNNs-b} Iterative photonic network for matrix inversion. Recurrent connections are implemented via feedback optical delay lines (image from \cite{chen2024efficient}, licensed under \href{https://creativecommons.org/licenses/by/4.0/}{CC BY 4.0}).
		\subref{fig:RNNs-c} Schematic representation of an integrated self-adaptive photonic neuron based on WDM and PCM (GST) cells featuring Hebbian plasticity. An optical feedback loop connecting the neuron's output and the PCM-based synaptic weights allows the strengthening of those synapses that contribute to the neuron activation (image from \cite{feldmann2019all}, licensed under \href{https://creativecommons.org/licenses/by/4.0/}{CC BY 4.0}).
		\subref{fig:RNNs-d} A self-adapting nonlinear MRR network with PCM cells and its corresponding ANN schematic. The system integrates non-volatile plastic memory (MRRs with PCM) on top of multi-scale fading nonlinear memory (nonlinear MRRs without PCM). Image from \cite{lugnan2025emergent}, licensed under \href{https://creativecommons.org/licenses/by/4.0/}{CC BY 4.0}.}
	\label{fig:RNNs}
\end{figure}

In \cite{tait2017neuromorphic}, a silicon PIC is employed to implement a recurrent neural network, where the nonlinearity is provided by the Optical-Electrical-Optical (O-E-O) conversion and the memory by the electrical dynamics within the feedback loop, which allows to control how an optical signal interacts with another by means of simple auxiliary electric circuits.
The architecture is based on the broadcast-and-weight protocol, utilizing WDM to enable parallel interconnectivity (\cref{fig:RNNs-a}).
Each neuron's state is encoded as the optical power of a specific wavelength on a common bus waveguide.
The weighting mechanism is implemented using MRR weight banks, able to extract and weight multiple wavelengths traveling through a single waveguide.
By tuning the MRRs in or out of resonance, the network performs channel-specific multiplication.
Summation is achieved via incoherent photodetection using balanced photodetectors, which allow for both positive and negative (inhibitory) synaptic weights.
The memory that implements network recurrence is realized through an optoelectronic feedback loop: the summed photocurrent from the balanced detectors drives an electro optic modulator that acts as the nonlinear activation function, which re modulates the WDM carrier signals and injects them back into the network. The state variable of each neuron is the voltage applied to the modulator, held at the output of a low-pass filtered transimpedance amplifier, so that the neuron time constant is set by the electrical filter rather than by the optical path. In the reported experiment the detected signals are low-pass filtered at \qty{10}{\kilo\hertz}.
This creates a closed-loop programmable dynamical system.
A mathematical isomorphism between the integrated photonic circuit and the equations of continuous time RNNs is demonstrated.
In the simulated fully integrated version, the optical round trip of the feedback path amounts to \qty{47.8}{\ps}, and the operating timescale is set about two orders of magnitude above it, at \qty{12.5}{\nano\second}, with the signal bandwidth restricted to \qty{1}{\GHz} in order to suppress delay induced dynamics. The optical propagation delay is therefore not used as a memory resource, but kept negligible with respect to the electrically defined state time constant.
By tuning the MRR weight banks, the study shows, supported by simulations, that this PNN implements programmable dynamical systems, including the solution of ordinary differential equation systems at sub \unit{\GHz} rates, with substantial acceleration relative to conventional electronic processors.

\cite{wang2022multi} presents, via simulation, a similar approach and system as in \cite{tait2017neuromorphic}, where the recurrent feedback is provided by a waveguide connecting all the WDM channels. The work focuses on applying the system to intra and inter-channel impairments compensation in WDM systems, with a throughput up to \qty{56}{\GS\per\s}.

\subsubsection{All-optical recirculation via delay lines}

To overcome the latency and energy overhead associated with O-E-O conversion, the memory mechanism is kept entirely within the optical domain.
By utilizing low-loss waveguides as physical delay lines, these architectures perform iterative computations where the memory state is encoded in the time-of-flight of photons circulating in a closed loop.

\cite{chen2024efficient} demonstrates a PIC implementing an iterative network based on on-chip delay lines as memory mechanism (\cref{fig:RNNs-b}), to perform matrix inversion.
The core hardware is a photonic iterative processor constructed from a mesh of MZIs, where the weight matrix is encoded into the static phase settings of the mesh.
Instead of detecting the output after each matrix multiplication, the optical signal representing the current solution estimation is routed back to the input ports via on-chip optical waveguides acting as delay lines.
This creates a closed optical loop where the light pulse continuously circulates, evolving to the next improved estimation through physical propagation alone, effectively turning the iterative computation time into the simple time-of-flight of the photon.
The all-optical recirculation strategy achieves a net inversion time of \qty{1.2}{\nano\second} for \numproduct{2x2} complex-valued matrices. This figure corresponds to the core processing time defined by the authors as $t_{core}=t_{loop}\cdot P\cdot N$, with a single-iteration time $t_{loop}=\qty{300}{\ps}$ and $P=2$ iterations, and does not include the input/output access time, which the same work defines separately as the time required to load the weight matrix, fetch the input data and store the output data.
The system's versatility was verified across three distinct problem types.
By eliminating the intermediate electronic processing steps, the authors estimate an improvement in input/output efficiency (i.e. the cost of generating input signals and detecting output signals) of at least one order of magnitude compared to electronic processors and nonrecurrent photonic accelerators.

\subsubsection{Plasticity via phase change materials}

A fundamentally different approach to memory involves the integration of PCMs (see \cref{subsec:non-volatile}) directly onto the photonic waveguides.
Unlike previously discussed architectures that rely on feedback loops or volatile nonlinear effects, PCMs provide non-volatile weights.
This allows the hardware to exhibit physical plasticity, where the network's physical properties evolve in response to the input optical signal.

In the past years, PCMs have been employed to efficiently set non-volatile weights in PNN \cite{feldmann2019all,feldmann2021parallel}.
One example appears in \cite{feldmann2019all}, which develops a SiN PIC with four neurons and sixty synapses for pattern recognition, using PCMs and WDM.
The synaptic weights and the neuron functions rely on GST cells, which provide nonvolatile storage and a nonlinear activation, and on MRRs that act as wavelength selective elements for multiwavelength processing.
In this system, the optical pulses that encode the neuron output originate from a dedicated optical source that is distinct from the optical pulses that enter the synapses and address the neuron PCM cell.
Moreover, since the neuron's activation function is implemented via the nonvolatile PCM memory, after each activation it requires a reset pulse to bring the PCM cell back to its initial state.
Similarly to Hebbian learning, self-adaptation to input photonic signals is demonstrated for unsupervised learning by means of a feedback waveguide connecting the neuron's output to the PCM-based synaptic weights (\cref{fig:RNNs-c}).
If a synapse transmits an optical pulse and the downstream photonic neuron fires within the same clock cycle, the superposition of the input pulse and the feedback pulse strengthens the synapse, since the PCM cell reaches a fully amorphized state and its transmission is maximized.
If a synapse does not transmit a pulse while the downstream neuron fires within the same clock cycle, the synaptic weight is weakened, since the crystallization level of the PCM region on the synaptic waveguide increases and the waveguide transmission is reduced.
The experiment uses optical pulses with a duration of \qty{200}{\nano\second} and peak powers up to a few milliwatts, with a maximum pulse energy of \qty{710}{\pico\joule}.

In \cite{lugnan2025emergent}, the volatile memory due to nonlinear effects in silicon MRRs and the non-volatile memory of PCMs are combined in a \numproduct{13x14} matrix of coupled MRRs.
The system can be seen as a nonlinear reservoir similar to the one studied in \cite{lugnan2025reservoir,foradori2025neuromorphic}, with the addition of plastic self-adaptation provided by PCM cells, which were added to the waveguide of some of the MRRs (one every three).
In this case, the MRRs with PCM act as a sort of plastic synapse, directing the light propagation through the network depending on the PCM state, which modifies the MRR resonance.
Therefore, an input optical signal that is powerful enough to modify the PCM states, and presents fast enough transients, will permanently modify the dynamical properties of the MRR network, which in turn determine how and which PCM weight may be changed by another input signal (\cref{fig:RNNs-d}).

Allowing plastic adaptation to the input improves the RC performance in the classification of five time series, each defined by a distinct sequence of four pulses with \qty{5}{\nano\second} duration and peak power on the order of \qty{10}{\milli\watt} (pulse energy is of the order of \qty{100}{\pico\joule}).
While a classification error reduction is not achieved after each plastic adaptation, it is more likely to happen than an error increase.
Moreover, the MNIST classification task is addressed by preprocessing the images, mainly via downsampling, and encoding them as time sequences injected into the MRR network.
Multiple laser wavelengths are used, and the output signals are measured at multiple output ports.
A classification accuracy of \qty{98.23}{\percent} is obtained by building an ensemble of RC systems, using a method termed \emph{chaining}, from the nonlinear representations of the input data available at the MRR network outputs. The reported experimental baseline accuracy, obtained by applying a linear classifier directly on the acquired optical pixel sequences, is approximately 86\%.
More specifically, each representation corresponds to a temporal output signal generated by the network during the injection of the input sequence, recorded at different output ports and optical wavelengths, and, in principle, acquired in parallel (sequentially in the reported experiment), while also varying the input optical power.
Each RC system in the chain is obtained by training a linear classifier on a distinct output representation, where trainable weights act on time samples within a fixed time window, and on the output of the preceding RC system in the chain.
This ensemble is built via greedy optimization, exploring different combinations of RC systems in a growing chain.

Integrating a PCM cell into a MRR brings advantages beyond wavelength selection, enabling faster and more energy efficient access to the non-volatile memory functionality, as it is shown via simulations in \cite{lugnan2022rigorous}.
A GST patch deposited on a waveguide introduces high optical loss through absorption, with the loss typically larger in the crystalline phase.
This behaviour is expected because physical plasticity relies on optical pulses in the waveguide that heat the PCM and raise its temperature close to, or above, the melting point.
The power enhancement in a MRR allows to modify the PCM solid state with lower input pulse power.
Moreover, because of the enhanced sensitivity to optical changes on the ring waveguide, shorter GST patches can produce strong enough changes in light transmission.
The shorter the GST patch along the waveguide, the lower the optical loss due to absorption and the shorter the pulse duration required to switch the memory.
Therefore, combining MRRs and PCM cells enable direct cascadability of multiple non-volatile components, as in \cite{lugnan2025emergent}.

\subsection{Applications to channel equalization in telecom}
\label{subsec:PNNs_telecom}

Fiber optic communications represent the backbone of the Internet, enabling fast and efficient data transfer in different scenarios, from short-reach links in intra-data center applications to ultra-long-haul submarine transmission \cite{mukherjee2020springer}.
The constantly increasing data traffic pushes the market towards transmission at larger bitrates, which in turn requires peak performance in the traffic handling.
A problem is the fact that modulated optical signals propagating in fiber are affected by distortions induced both by linear effects, such as Chromatic Dispersion (CD), and nonlinear effects, such as Self-Phase Modulation (SPM) \cite{agrawal2013nonlinearCh2}.
These create impairments in the signal, which become more severe with the increase of modulation frequency, propagation distance, and input power in the fiber.
The associated distortions introduce the need for recovering the signal's integrity after the transmission, namely, operating an equalization process. Nowadays this is mainly provided via Digital Signal Processing (DSP) operated on board of the transceivers, requiring huge computational effort, power consumption, and latency \cite{zhong2018digital, zhou2019recent}.
In this context, PNNs are emerging as a more efficient solution to this problem, since they can exploit all the properties offered by light-based computation in terms of bandwidth, speed of propagation, latency reduction, and parallelization.

Many different solutions have been experimentally proven to be effective against distortions induced on propagating optical signals \cite{silva2025integrated}.
The various implementations span different architectures, from FFNN to RC, with an increasing level of complexity for more demanding transmission scenarios.
Common features among the different implementations are the tunable weights, which can operate directly on the optical signal or at the readout level, and the presence of memory within the PNN.
Memory is fundamental in counteracting distortions generated by CD and SPM, which indeed arise from the interference between temporally adjacent symbols during propagation and thus require accessing optical samples close in time to reconstruct the original symbol shape.
In the following paragraphs, we present some examples of implementations, starting from those featuring the largest amount of memory and proceeding to shorter temporal dynamics. The implementations are grouped into nonlinear reservoirs based on optical feedback and laser dynamics, passive integrated reservoirs in which the only nonlinearity is provided by photodetection and whose readout is either digital or all-optical, linear optical FIR filters, lattice filters, and resonant filters based on cascaded MRRs.

\begin{figure}[!htb]
	\centering
	\includegraphics[width=\linewidth]{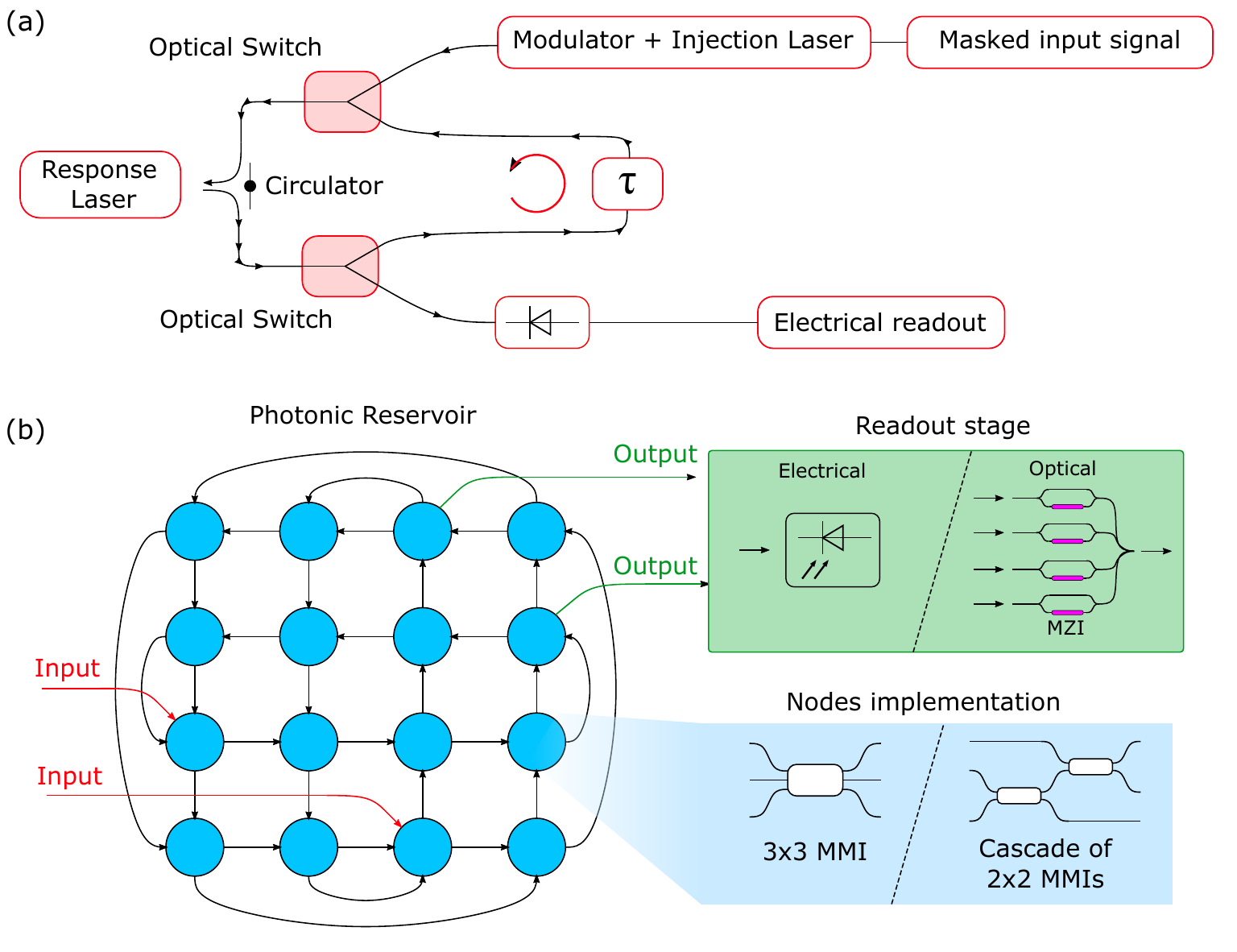}
	\begin{subcaptiongroup}
		\phantomcaption\label{fig:rc_equalization-a}
		\phantomcaption\label{fig:rc_equalization-b}
	\end{subcaptiongroup}

	\caption{%
		Implementations of a photonic Reservoir Computing for optical signal equalization.
		\subref{fig:rc_equalization-a} Semiconductor laser with time-delayed optical feedback \cite{argyris2018photonic}. The simulated signal after fiber-optic transmission is randomly masked offline and applied to the injection laser. Then, this is directed to the response laser, which receives delayed optical feedback from a fiber loop providing a delay $\tau = \qty{66}{\ns}$. Optical switches control the loop status, finally directing the output of the reservoir to the readout stage.
		\subref{fig:rc_equalization-b} Integrated 16-node photonic reservoir featuring the same building blocks used in \cite{sackesyn2021experimental, gooskens2023experimental, masaad2024experimental, vanassche2025real} for nonlinear equalization. }
	\label{fig:rc_equalization}
\end{figure}

\paragraph{Reservoir computing with optical feedback and laser nonlinearity}

PNNs based on RC exploits randomness and recursivity.
In \cite{argyris2018photonic}, the reservoir consists of a semiconductor laser operating in the C-band wavelength range and an optical fiber delay loop from which the laser receives delayed feedback, as illustrated in \cref{fig:rc_equalization-a}. The optical processing provided by the reservoir enriches the signal's dynamics and projects its samples to a higher-dimensional space, where the action of a linear classifier completes the equalization.
This approach is applied to CD and SPM equalization for a \qty{25}{\Gbps} 2-level Pulse Amplitude Modulated (PAM2) signal after \qty{45}{\km} of propagation.
The fiber loop acts as the main source of memory for the system, providing up to $\tau = \qty{66}{\ns}$ of delay, corresponding to about 1650 times the symbol period $T_0 = \qty{40}{\ps}$ of the \qty{25}{\Gb\per\s} PAM2 signal considered.
Not only does this prevent a real-time processing of the data, due to its discrepancy with the time scale imposed by the signal's baud rate, but it is also significantly longer than what is actually needed for the selected equalization task, which indeed requires on the order of hundreds of ps (a few $T_0$).
A more detailed treatment of the role of RC memory is provided in \cite{estebanez202256, skontranis2023multimode}, which compare the equalization performances of the system with the feedback loop (time-delayed RC configuration) and without it (Extreme Learning Machine, ELM approach).
In this last case, the memory is solely provided by the nonlinear dynamics of the response laser, which, however, is sufficient for the equalization of \qty{112}{\Gbps} PAM4  signal after \qty{100}{\km} propagation (with CD removed via DSP).

\paragraph{Passive integrated reservoirs with hybrid and all-optical readout}

Within the RC approach, \cite{sackesyn2021experimental, gooskens2023experimental, masaad2024experimental, vanassche2025real} propose the PIC
schematized in \cref{fig:rc_equalization-b}.
Each circle represents a node, which is implemented as a \numproduct{3x3} MMI, and the arrows indicate the inter-node connections, which are realized via spiralized waveguides.
The input signal can be fed to the reservoir at different input nodes, as well as the output can be extracted from many ports at the readout stage. Notice that no intrinsic nonlinearity is incorporated in the reservoir at each node and the square nonlinear response of the photodetector is used.

In a first implementation \cite{sackesyn2021experimental}, the output of a 32-node reservoir is post-processed via software to maximize the similarity with a digital target sequence.
This allows reaching nonlinear equalization of a \qty{32}{\Gb\per\s} PAM2 signal after \qty{25}{\km} of fiber.
The same hybrid approach has also proven effective in a WDM scenario \cite{gooskens2023experimental}, as well as for nonlinear equalization in coherent systems \cite{masaad2024experimental}.
Ultimately, the combination of a reservoir structure with a weighted summation tree allows for all-optical signal processing  \cite{vanassche2025real}.
The DSP has thus been removed, even at the readout stage.
In these examples, the memory is set by the length of the internode connections and their lengths span from fractions to a few multiples of the input signal baud duration $T_0$.
Thus, the overall memory is compatible with the typical time scale of the system, which is set by the modulation frequency of the input signal. The critical role of the internode connections length is addressed by \cite{zuo2023integrated}.
Channel equalization using a 32-node integrated RC scheme, featuring an all-optical trainable readout as in \cite{vanassche2025real}, is simulated.
Three lengths $\Delta L$, $2\times\Delta L$, and $3\times\Delta L$ are considered for the internode connections
Multiple training sessions for equalization after 25 km of fiber at \qty{25}{\Gbps} PAM2 are performed while varying $\Delta L$ so that it provides a delay between $\Delta T = 0$ and $\Delta T = 2\times T_0$.
System's performance degrades for $\Delta T = 0$ and $\Delta T > T_0$, namely when memory is absent or is too large compared with the requirements of the task.

\paragraph{Linear optical FIR filters}

An alternative to the RC approach is provided by integrated Finite Impulse Response (FIR) filters.
\cite{mancinelli2022photonic} presents a simple perceptron with integrated delay lines, i.e. a Time-Delayed Complex Perceptron (TDCP).

\begin{figure}[!htb]
	\centering
	\includegraphics[width=\linewidth]{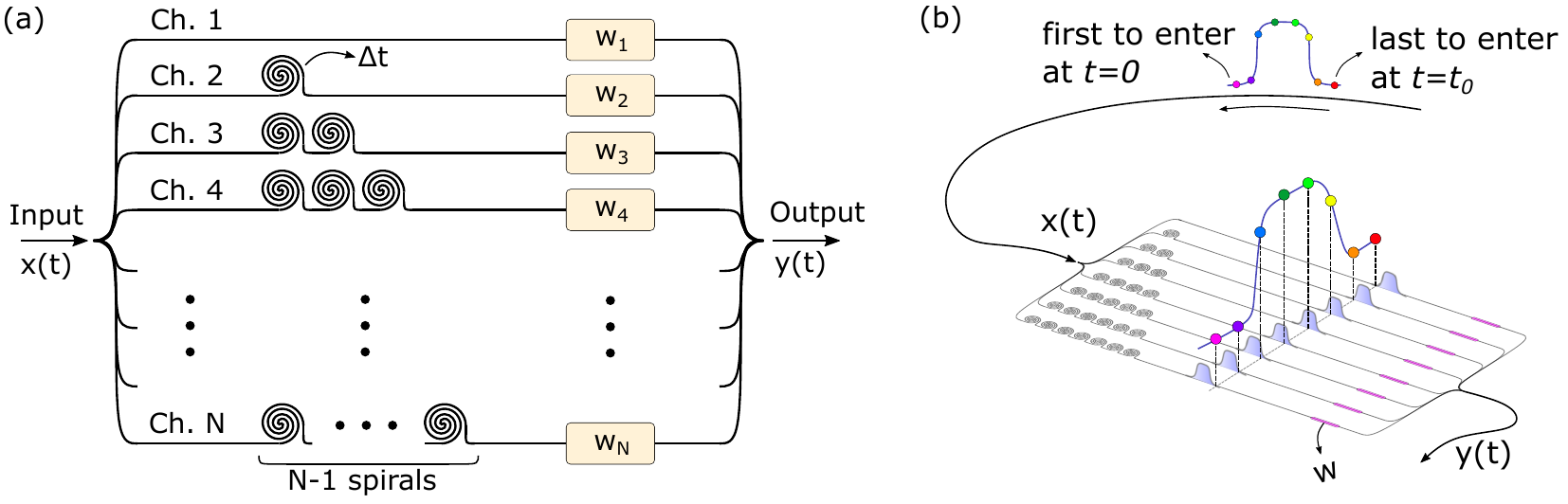}
	\begin{subcaptiongroup}
		\phantomcaption\label{fig:tdcp_generic-a}
		\phantomcaption\label{fig:tdcp_generic-b}
	\end{subcaptiongroup}

	\caption{
        Time-delayed complex perceptron.
		\subref{fig:tdcp_generic-a} Schematics of an $N$-channel Time-Delayed Complex Perceptron \cite{staffoli2023equalization, staffoli2024chromatic, staffoli2024silicon, staffoli2025nonlinear, marciano2024chromatic}. The optical input time sequence $x(t)$ is split into $N$ copies, which then propagate through spiralized waveguides to accumulate a relative delay multiple of $\Delta t$. Copies traveling in each channel are then applied with independent complex weights $w_i$ ($i=1,\dots,N$). Finally, they reach the recombination stage where they are complex-summed to produce the optical output sequence $y(t)$.
		\subref{fig:tdcp_generic-b} Illustration of the working principle upon the arrival of a 100-ps optical pulse when $\Delta t = 25$ ps. The optical samples (colored dots) from the input enter the chip in sequence, from $t=0$ (pink dot) to $t_0 = 175$ ps (red dot), and then they are parallelized in time by the combined action of the splitting stage and the delay lines.}
	\label{fig:tdcp_generic}
\end{figure}

The TDCP design is composed of an $N$-line FIR filter in the optical domain, followed by a nonlinear activation function due to the detection.
The FIR filter operates the linear processing as illustrated in \cref{fig:tdcp_generic}, acting on the input optical signal $x(t)$ (complex field) to generate an output $y(t)$ according to
\begin{equation}\label{eq:cperc}
	y(t) = \sum_{i=1}^{N} x\left[ t - (i-1)\Delta t \right] w_i.
\end{equation}
The input signal $x(t)$ is first split into $N$ separate waveguides (channels, or taps) of different length, where each duplicate undergoes a delay that is an integer multiple of $\Delta t$.
Each copy is then modified by applying a trainable complex weight $w_i$ and all of them are then summed together to produce the optical output $y(t)$ (complex field), subsequently sent to the nonlinear stage.
Overall, the optical samples serially inserted in the TDCP switch to a parallel propagation in the corresponding channels, finally reaching the summation stage simultaneously, as illustrated in \cref{fig:tdcp_generic-b}.
In other words, the delay lines parallelize the input information, allowing for the simultaneous processing of pieces of optical information initially separated in time.
In the photonic implementation, crucial design parameters are the unitary delay $\Delta t$, which regulates the density of the optical samples reaching the recombination stage simultaneously, and the number of taps $N$, determining the total observation window of the device as $(N-1)\times \Delta t$.

The design specifics in terms of the number of taps $N$ and unitary delay $\Delta t$ are closely related to the target baud rate and the amount of distortions to be compensated.
As reported in \cite{staffoli2023equalization}, for a given $\Delta t$, larger $N$ increases the memory, but also the insertion losses, and the number of parameters to train.
On the other hand, for fixed $N$, a smaller $\Delta t$ provides a higher density of optical samples reaching the recombination stage simultaneously, thus increasing, in principle, the quality of symbol reconstruction.
Overall, the design parameters result from a trade-off between a sufficiently dense optical sampling (at least 2 samples per baud), insertion loss minimization, and a reduced number of trainable parameters.
In this sense, a useful comparison can be performed between the two implementations presented in \cite{staffoli2023equalization} and \cite{staffoli2024chromatic, staffoli2024silicon}, featuring $N=4$ with $\Delta t = \qty{50}{\pico\second}$ and $N=8$ with $\Delta t = \qty{25}{\pico\second}$, respectively , i.e. observation windows of \qty{150}{\pico\second} and \qty{175}{\pico\second}, corresponding to $1.5\,T_0$ and $1.75\,T_0$ since both devices operate at a symbol rate of \qty{10}Gbaud ($T_0 = \qty{100}{\pico\second}$).
Despite the devices are characterize by a similar memory amount, better equalization results on a target 10G applications are obtained with the 8-tap version, equipped with thermally-tuned amplitude and phase weights on each tap.
Overall, the device shows excellent equalization capabilities in different scenarios, obtained by varying the transmission rate (\qtylist{10;20}{\Gb\per\s}), the modulation format (PAM2, PAM4, OFDM \cite{marciano2024chromatic}), and the transmission reach (up to \qty{200}{\km}).
In a multi-span nonlinear propagation regime, SPM is equalized up to \qty{450}{\km} (with CD removed and signal amplification every \qty{50}{km} provided by external instruments) \cite{staffoli2025nonlinear}.

\begin{figure}[!htb]
	\centering
	\includegraphics[width=\linewidth]{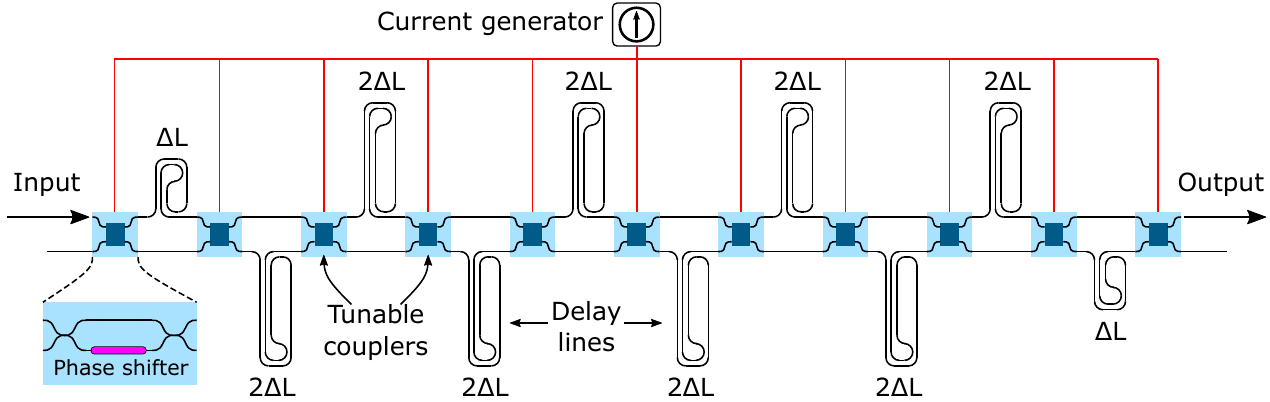}
	\caption{Schematics of a tunable lattice filter implemented in SiN and used in \cite{moreira2016programmable,brodnik2018extended} for CD equalization. The input signal propagates through subunits constituted by a tunable coupler and an unbalanced MZI. The imbalance is provided by delay lines implemented via spirals of lengths $\Delta L$ (first and last units) or $2\Delta L$. The inset shows the internal structure of a tunable coupler, which is represented by a balanced MZI with a tunable PS. The couplers' response is driven by a common external current generator, which controls all of them simultaneously.}
	\label{fig:lattice_filter}
\end{figure}

\paragraph{Lattice filters}

A different approach based on integrated lattice filter is presented in \cite{moreira2016programmable,brodnik2018extended} and schematized in \cref{fig:lattice_filter}.
It is based on a cascade of repeated units constituted by a tunable coupler and an unbalanced MZI. The couplers, in turn, are constituted by balanced MZIs with one of the arms equipped with a phase shifter driven by an external current generator.
At each tunable coupler, spectral components are sent respectively to the upper and lower arms of the subsequent unbalanced MZI, thus imposing a longer propagation path to specific wavelengths.
These accumulate a delay with respect to those traveling in the straight waveguide, balancing the effect generated by CD.
The spiral unit length $\Delta L$ and the number of stages determine the whole structure's behavior in terms of memory (still provided by optical delay lines), Free Spectral Range, maximum achievable dispersion, and bandwidth.
The \ce{SiN} version of \cite{brodnik2018extended} manages to generate dispersion up to \qty{\pm 500}{\ps\per\nm}, sufficient to compensate for \qty{40}{\km} propagation of a \qty{53}{\Gbps} PAM4 signal.
This device implementation, based on a multi-stage approach, promises low insertion loss, since the optical power is preserved at every stage but the last.

\begin{figure}[!htb]
	\centering
	\includegraphics[width=\linewidth]{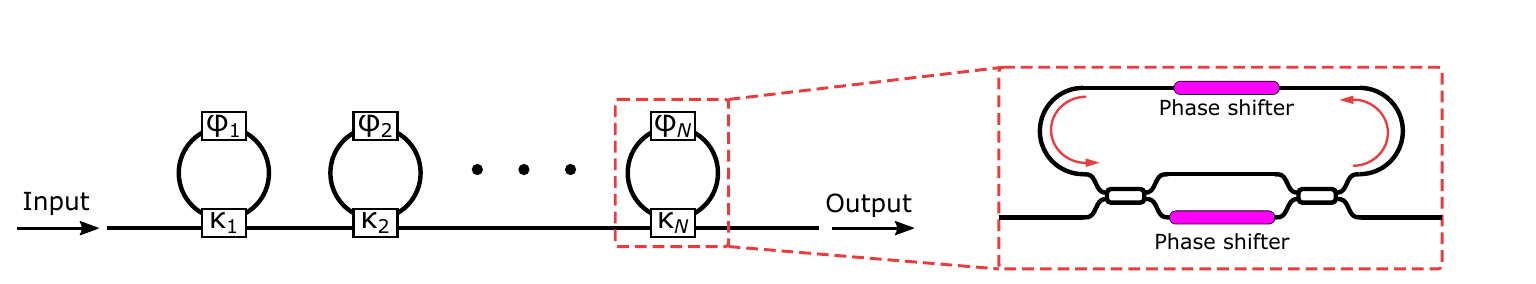}
	\caption{%
		Schematics of a cascade of $N$ MRRs used in \cite{sorianello2018polmux,liu2022silicon} for CD equalization. As shown in the inset, within each subunit, the coupling between the bus waveguide and the resonant structure is provided by a MZI, which can be thermally tuned to regulate the corresponding coupling coefficient $\kappa_i$ with $i=1,\dots,N$. Each MRR is then equipped with another phase shifter that drives the phase weight $\phi_i$.}
	\label{fig:mrr_sequence}
\end{figure}

\paragraph{Resonant filters based on cascaded MRRs}

Other filter implementations are based on MRR architectures, which can be arranged in different topologies to mimic the inverse transfer function of CD.
\cite{liu2022silicon} and \cite{sorianello2018polmux} propose sequences of, respectively, 7 and up to 9 MRRs coupled to the same bus waveguide, following the scheme proposed in \cref{fig:mrr_sequence}.
The overall frequency response of the structure is determined by the waveguide-MRR coupling coefficients $\kappa_i$, and the phase shifts $\phi_i$ imposed in each structure, setting the maximum achievable dispersion and the bandwidth.
In this setting, the nonlinear dynamics related to the free-carriers inside the MRR cannot be directly exploited for a real-time processing of the input data. Indeed, the typical free-carrier lifetime in silicon is on the order of nanoseconds, which is not compatible with the much faster time scale set by the symbol duration in telecom applications, which lowers to \qty{<10}{\ps}.
In \cite{dong2025augmented}, the free carrier nonlinear dynamics is simulated by setting the free-carrier lifetime to \qty{30}{\ps}, much lower than its physical value, to make it compatible with the signal's bandwidth.
On the contrary, in the linear regime, as is the case for \cite{liu2022silicon, sorianello2018polmux}, memory is provided by the resonant nature of the MRR, which allows storing the optical information in the signal coupled to the rings. In \cite{sorianello2018polmux}, a 9-stage architecture was proved to equalize a \qty{100}{\Gbps} polarization multiplexed PAM2 signal over \qty{30}{\km} of optical fiber, corresponding to a maximum dispersion of \qty{-540}{\ps\per\nm}. In \cite{liu2022silicon}, the number of stages is set to 7, allowing the equalization of a \qty{25}{\Gbps} PAM2 signal in \qty{40}{\km} fiber, corresponding to \qty{-720}{\ps\per\nm}.
The periodic nature of these filters' frequency response allows operating in a WDM scenario, which, for the considered cases, corresponds to a spacing of \qty{100}{GHz} between the carriers. In general, the maximum dispersion that cascade filters can compensate grows with the number of stages, which, however, leads to a larger insertion loss.

\paragraph{Programmable filters as linear stages of PNNs}

The approaches presented so far as programmable optical filters can be interpreted as the linear stages of PNNs, whose activation function is the square modulus applied at the end-of-line photodetector.
Indeed, the parameters related to each structure require a proper adjustment to provide the desired amount of dispersion. When these PNNs are inserted in a transmission line, a training procedure provides for the tuning, driven, for example, by maximizing the signal quality at the receiver (BER, OSNR, eye diagram aperture, \dots).

\Cref{tab:comparison_cd} collects and compares the most relevant works in this context \cite{argyris2022photonic}, giving priority to experimental realizations over simulated ones.
From this study, photonics platforms emerge as integrated, scalable, and tunable solutions that perform signal processing with minimized latency and power consumption.
They appear as key-enabling technologies for future applications in telecommunications, removing (or at least alleviating) the computational effort required by traditional DSP.

\begin{table}[!htb]
	\caption{Comparison between different techniques for channel equalization. Maximum dispersion is the maximum amount provided by the device under test. We add NL (Nonlinearities) when the equalization includes nonlinear distortions as well. ELM: Extreme Learning Machine, O-FIR: Optical Finite Impulse Response filter, MRR: Microring Resonator, RC: Reservoir Computing, SiN: Silicon Nitride, Si: Silicon, PLC: Planar Lightwave Circuit, LNOI: Lithium Niobate On Insulator, OF: Optical Feedback, WDM: Wavelength Division Multiplexing, MZM: MZI Modulator. WDM column refers to the spacing between channels that the device in a static configuration manages to compensate for. All the reported measurements were performed at 1550 nm, except for \cite{ran20254}, which worked at 1270 nm.}
	\label{tab:comparison_cd}
	\renewcommand{\arraystretch}{1.2}
	\tablealtrows
	\begin{tabular}{@{}
		>{\centering \arraybackslash}m{.05\linewidth}
		>{\centering \arraybackslash}m{.14\linewidth}
		>{\centering \arraybackslash}m{.08\linewidth}
		>{\centering \arraybackslash}m{.07\linewidth}
		>{\centering \arraybackslash}m{.10\linewidth}
		>{\centering \arraybackslash}m{.07\linewidth}
		>{\centering \arraybackslash}m{.12\linewidth}
		>{\centering \arraybackslash}m{.10\linewidth}
		@{}}
		\toprule
		Ref.                                                                     & Layout                    & Platf.     & Bitrate (Gbps) & Encoding    & Fiber (\unit{\km})  & Maximum dispersion (\unit{\ps\per\nm}) & WDM (\unit{\GHz}) \\
		\midrule
		\cite{argyris2018photonic}                                               & O-RC                      & LASER + OF & 25                    & PAM2        & 45                  & \num{-810}+NL                          & -                 \\
		\cite{estebanez202256}                                                   & ELM                       & LASER + OF & 112                   & PAM4        & 100                 & \num{0} + NL                           & -                 \\
		\cite{sackesyn2021experimental}                                          & O-RC                      & Si         & 32                    & PAM2        & 25                  & \num{-450}+NL                          & -                 \\
		\cite{vanassche2025real}                                                 & O-RC + optical readout    & Si         & 28                    & PAM2        & 50                  & \num{-900}+NL                          & -                 \\
		\cite{staffoli2023equalization}                                          & 4-tap O-FIR               & Si         & 10                    & PAM2        & 125                 & \num{\pm2200}                          & 80 (theo.)        \\
		\cite{marciano2024chromatic, staffoli2024silicon, staffoli2024chromatic} & 8-tap O-FIR               & Si         & 20                    & PAM4, OFDM  & 125                 & \num{\pm 2200}                         & 40 (theo.)        \\
		\cite{staffoli2025nonlinear}                                             & 8-tap O-FIR               & Si         & 10                    & PAM2        & 200 (CD), 450 (SPM) & \num{\pm 3600}                         & 40 (theo.)        \\
        \cite{wang2026all} &  3 node RC + 8-tap O-FIR               & Si         &  224                    &  PAM4        & 5 & \num{\pm  85}           + NL              &  3200 (exp.) \\
		\cite{liu2023adaptive}                                                   & 4-tap O-FIR + DSP         & SiN / LNOI & 28                    & QPSK      & 30                  & \num{-570}                             & 28 (theo.)        \\
		\cite{ran20254}                                                          & MZM-based transm. + DSP   & Si         & 256                   & PAM4        & 5                   & \num{\pm 16}                           & 3660 (exp.)       \\
		\cite{sorianello2018polmux}                                              & MRR-based 9 stages filter  & Si         & 100                   & PolMux PAM2 & 30                  & \num{- 540}                            & 100 (sim.)        \\
		\cite{liu2022silicon}                                                    & MRR-based 7 stages filter  & Si         & 25                    & PAM2        & 40                  & \num{- 720}                            & 100 (sim.)        \\
		\cite{brodnik2018extended}                                               & MZI-based 10 stages O-FIR & SiN        & 53                    & PAM4        & 40                  & \num{\pm 500}                          & 100 (exp.)        \\
		\botrule
	\end{tabular}

\end{table}

\section{Conclusion}
\label{sec:conclusion}

This review has examined memory in integrated PNNs from the perspective of physical mechanism, dynamical behavior, and computational function. The central conclusion is that no single physical memory mechanism is optimal. Integrated photonic platforms provide complementary state variables based on propagation delay, resonant storage, nonlinear relaxation, attractor dynamics, and persistent material or internal-field reconfiguration. Their practical usefulness depends on trade-offs among retention, bandwidth, insertion loss, holding power, write and read energy, observability, stability, drift, programmability, footprint, and electronic-control overhead. The biological discussion in \cref{sec:bio} provides a qualitative analogy for the coexistence of short-lived dynamical states and longer-lived adaptive parameters, but the operational classification developed here is grounded in device physics and nonlinear dynamics.

The classification of integrated photonic memory in \cref{sec:memory} organizes available mechanisms into volatile and nonvolatile, with volatile memory further divided into response-induced and multistable-induced classes.
Response-induced mechanisms include delay-based memory in waveguide spirals and coupled resonators, slow-light effects in photonic crystal and resonator structures, and nonlinear relaxation dynamics arising from free-carrier and thermo-optic effects in silicon MRRs. Volatile memory mechanisms include finite propagation and resonant-storage processes, nonlinear relaxation dynamics, and attractor-based states maintained under specified operating conditions. Passive optical bistability and gain-mediated bistability can retain a selected state while the required drive or bias is maintained. Self-pulsing systems require a more restricted interpretation: phase or attractor selection can provide long-lasting dynamical memory, whereas a single stable limit cycle is neither a multistable memory nor, by itself, a spiking-neuron implementation.
Non-volatile mechanisms, including phase-change materials, ferroelectric films, ionic resistive switching, charge-trapping structures, including internal-field readout in lithium niobate,, can retain persistent parameter or configuration states after the writing stimulus has been removed. These states may require no continuous state-holding power; however, complete operation still consumes energy for programming, erasing, addressing, optical readout, detection, and, where applicable, stabilization.
This classification connects device physics directly to the operational role of memory in a learning system: the retention time, volatility, energetics, and programmability of each mechanism determine where in a temporal processing hierarchy it is most appropriately placed.

The theoretical framework in \cref{sec:theo} provides complementary tools for characterizing fading and persistent memory. For fading-memory systems, linear memory capacity, information processing capacity, and the echo-state property quantify how previous inputs affect the present response. Persistent-state devices instead require characterization of the complete write-retain-read-reset cycle, including retention, switching time and energy, state distinguishability, drift, endurance, and reset conditions. Benchmark tasks, including Boolean operations, time-series prediction, and classification, complement these physical metrics by measuring the computational usefulness of the available memory.

The architectural survey in \cref{sec:PNNs} demonstrates how the physical mechanisms reviewed in \cref{sec:memory} have been used as computational primitives in photonic machine-learning systems.
Spatial and delay-based RC implementations exploit integrated waveguide spirals, coupled resonators, and multimode structures to realize volatile fading memory compatible with gigahertz-rate signal processing.
Nonlinear silicon MRRs provide both memory and local nonlinearity through free-carrier and thermo-optic dynamics, while experiments on self-pulsing MRRs provide evidence for long-lasting volatile memory extending beyond the individual carrier and thermal relaxation times.
Optoelectronic recurrent architectures and all-optical delay line loops realize programmable dynamical systems, and PCM integration enables physical plasticity and non-volatile weight storage on chip.
The application to channel equalization in \cref{subsec:PNNs_telecom} illustrates a concrete use case in which the temporal structure of optical fiber distortions is directly matched by the memory depth of photonic filter and reservoir architectures, achieving low-latency signal recovery at line rates. The surveyed evidence nevertheless spans different levels of maturity, including numerical simulations, device-level experiments, hybrid photonic-electronic implementations, and fully integrated demonstrations. These categories should not be interpreted as providing equivalent evidence of system-level scalability.

These results establish that integrated photonic platforms offer multiple memory mechanisms of different kinds, spanning time scales from picoseconds to microseconds in the volatile domain and enabling retention from hours to projected multi-year intervals in the non-volatile domain, depending on the physical mechanism, device design and environmental conditions. Experimental demonstrations show that selected implementations of these mechanisms can support reservoir computing, individual or small-scale spiking-neuron functions, and feed-forward temporal processing in silicon and hybrid SOI circuits.
Nevertheless, several challenges remain before scalable neuromorphic photonic systems can be realized.
The alignment and stabilization of resonant elements across fabrication-induced variability requires precise thermal tuning and represents a significant control overhead. Moreover, while PNN with memory (photonic RC) and with precisely tuned synaptic weights (feed forward neural network accelerators) are separately demonstrated and available, the union of the two (e.g. in photonic recurrent neural networks) remains very challenging and relatively unexplored.

The near-term development of integrated PNNs with memory will rely on the continued integration of low-loss photonic circuits with active, tunable, and nonlinear elements, A longer-term objective is to co-integrate complementary mechanisms with different retention times and computational roles and, ultimately, to establish controlled interactions between transient dynamical states and persistent parameters. At the same time, these systems must be designed in close synergy with their control electronics, so that calibration, thermal stabilization, and phase control are treated as intrinsic aspects of the overall architecture rather than external constraints. In parallel, learning strategies need to be adapted to the specific characteristics of photonic hardware, including analog variability, limited precision, and non-ideal dynamics, to exploit the useful dynamics of volatile memory while maintaining stable and controllable persistent parameters.

Overall, the strongest conclusion supported by the available evidence is that no single photonic memory mechanism simultaneously optimizes retention, bandwidth, loss, holding power, write energy, stability, and programmability. The practical potential of neuromorphic photonics is likely to emerge in hybrid photonic–electronic systems, particularly in applications where data are inherently optical, low latency is critical, and processing close to the signal source provides a clear advantage.

\backmatter

\section*{Acknowledgements}
This work was supported by the European Union under NextGenerationEU project PRIN2022AEEKNC and from the European Union’s Horizon Europe research and innovation programme under the Marie Sklodowska-Curie grant agreement No 101226674 — MINDnet.  S.B. acknowledges the co-financing from European Union-Next GenerationEU, Missione 4 Componente 2 – CUPE53D23002230006. A.F. acknowledges funding by the European Union under GA n°101070238-NEUROPULS.

\printbibliography

\end{document}